\documentclass[floatfix,twocolumn]{revtex4-1}

\usepackage{graphicx} 
\usepackage{dcolumn} 
\usepackage{amsmath}
\usepackage{amssymb}
\usepackage{bm}
\usepackage[dvips]{color}
\usepackage{times}
\usepackage{ulem}
\usepackage[justification=raggedright]{caption}

\usepackage{subcaption}
\usepackage{setspace}

\usepackage{tabularx}
\usepackage{appendix}

\usepackage{amstext}
\usepackage{array}

\usepackage{bbm}

\definecolor{wine-stain}{rgb}{0.5,0,0} 
\definecolor{bblue}{rgb}{0,0.0,0.5} 

\usepackage[colorlinks=true
,urlcolor=blue
,anchorcolor=blue
,citecolor=wine-stain
,filecolor=blue
,linkcolor=blue
,menucolor=blue
,pagecolor=blue
,linktocpage=true
,pdfproducer=medialab
,linktoc=all 
]{hyperref}

\usepackage{multirow}

\usepackage{booktabs}  

\newcommand{\tabitem}{~~\llap{\textbullet}~~}
 
\allowdisplaybreaks

\newcommand{\ncmd}{\newcommand}
\renewcommand{\vee}{\mathsf{v}}
\ncmd{\lt}{\left}
\ncmd{\rt}{\right}
\ncmd{\tr}[1]{\mbox{Tr}\lt({#1}\rt)}
\ncmd{\half}{\frac{1}{2}}
\ncmd{\eps}{\epsilon}
\ncmd{\veps}{\varepsilon}
\ncmd{\dgr}{\dagger}
\ncmd{\sig}{\sigma}
\ncmd{\gam}{\gamma}
\ncmd{\rtarw}{\rightarrow}
\ncmd{\Rt}{\Rightarrow}
\ncmd{\abs}[1]{\lt|{#1}\rt|}
\ncmd{\avg}[1]{\lt<{#1}\rt>}
\ncmd{\dl}{\delta}
\ncmd{\Dl}{\Delta}
\ncmd{\sgn}[1]{\mbox{sgn}\lt(#1\rt)}
\ncmd{\kap}{\kappa}
\ncmd{\wtil}[1]{\widetilde{#1}}
\ncmd{\thrfr}{\therefore}
\ncmd{\eq}[1]{Eq. \eqref{#1}}
\ncmd{\fig}[1]{Fig. \ref{#1}}
\ncmd{\Lam}{\Lambda}
\ncmd{\lam}{\lambda}
\ncmd{\dow}{\partial}
\ncmd{\ordr}[1]{\mathcal{O}\lt(#1\rt)}
\ncmd{\dsty}{\displaystyle}
\ncmd{\alert}[1]{\color{red}{#1}}
\ncmd{\mc}{\mathcal}
\ncmd{\mbf}[1]{\mathbf{#1}}
\ncmd{\derv}[2]{\frac{d{#1}}{d{#2}}}
\ncmd{\pderv}[2]{\frac{\partial{#1}}{\partial{#2}}}
\ncmd{\sub}[1]{{\,\dsty{\!_{#1}}}}
\ncmd{\step}[1]{\Theta\lt(#1\rt)}
\ncmd{\td}{\tilde} 
\ncmd{\what}{\widehat}
\ncmd{\om}{\omega}
\ncmd{\Om}{\Omega}
\ncmd{\vrho}{\varrho}
\ncmd{\vsig}{\varsigma}
\ncmd{\vkap}{\varkappa}
\ncmd{\bqa}{\begin{eqnarray}} 
\ncmd{\eqa}{\end{eqnarray}}
\ncmd{\nn}{\nonumber \\}
\ncmd{\nnum}{\nonumber}
\ncmd{\comment}[1]{{\color{red}{#1}}}
\definecolor{new_color}{RGB}{150,0,150}

\newcommand\redout{\bgroup\markoverwith
{\textcolor{red}{\rule[.5ex]{2pt}{0.8pt}}}\ULon}

\makeatletter
\newcommand*{\rom}[1]{\expandafter\@slowromancap\romannumeral #1@}
\makeatother

\makeatletter
\def\l@subsection#1#2{}
\def\l@subsubsection#1#2{}
\makeatother

\begin{document}

\title{
Anisotropic Non-Fermi Liquids
}

\author{Shouvik Sur$^{1,*}$
 and Sung-Sik Lee$^{1,2}$\\
\vspace{0.3cm}
{\normalsize{$^1$Department of Physics $\&$ Astronomy, 
McMaster University,}}\\
{\normalsize{1280 Main St. W., Hamilton ON L8S 4M1, Canada}}
\vspace{0.2cm}\\
{\normalsize{$^2$Perimeter Institute for Theoretical 
Physics,}}\\
{\normalsize{31 Caroline St. N., Waterloo ON N2L 2Y5, 
Canada}}
}

\date{\today}


\begin{abstract}
We study non-Fermi liquid states that arise at the quantum critical points 
associated with the spin density wave (SDW) and  charge density wave (CDW) transitions 
in metals with  twofold rotational  symmetry.
We use the dimensional regularization scheme, 
where a one-dimensional Fermi surface is embedded in $3-\epsilon$ dimensional momentum space.
In three dimensions, quasilocal marginal Fermi liquids arise 
both at the SDW and CDW critical points :
the speed of the collective mode along the ordering wavevector is logarithmically renormalized to zero compared to that of Fermi velocity.
Below three dimensions, however, the SDW and CDW critical points 
exhibit drastically different behaviors.
At the SDW critical point, 
a stable anisotropic non-Fermi liquid state is realized for small $\epsilon$,
where not only time but also different spatial coordinates develop distinct anomalous dimensions.
The non-Fermi liquid exhibits an emergent algebraic nesting
as the patches of Fermi surface are deformed into a universal power-law shape near the hot spots.   
Due to the anisotropic scaling, the energy of incoherent spin fluctuations disperse with different power laws in different momentum directions.  
At the CDW critical point, on the other hand,
the perturbative expansion breaks down 
immediately below three dimensions
as the interaction renormalizes the speed of charge fluctuations 
to zero within a finite renormalization group scale
through a two-loop effect. 
The  difference originates from the fact 
that the vertex correction anti-screens the coupling at the SDW critical point
whereas it screens at the CDW critical point.\\

$^*$ Present address : National High Magnetic Field Laboratory and Department of Physics, Florida State University, Tallahassee, Florida 32306, USA.

\end{abstract}

\maketitle

\hypersetup{linkcolor=bblue}
\tableofcontents

\section{Introduction}
Quantum phase transitions commonly arise   
in a wide range of strongly correlated metals 
such as high $T_c$ cuprates, iron pnictides, 
and heavy fermion compounds    \cite{Gegenwart, Lohneysen, Stewart, Armitage,Helm,Hashimoto,Park}.
Proximity of metals to symmetry broken phases creates  non-Fermi liquid states 
near quantum critical points
through the coupling between soft particle-hole excitations and the order parameter fluctuations. 
At the critical point, the low-energy excitations near the Fermi surface strongly damp the order parameter fluctuations which, in turn, feed back to the dynamics of low energy fermions 
 \cite{Moriya-1,Moriya-2, Hertz, Millis, Holstein, Reizer, PLee1989, PLee1992, Polchinski, Althshuler, Kim, Halperin, Chubukov, Abanov1, Abanov2}.
The theoretical challenge is to understand the intricate interplay 
between the electronic degrees of freedom and the critical fluctuations of order parameter.
In two space dimensions, the metallic quantum critical points remain largely ill-understood due to strong coupling between itinerant electrons and the collective modes.

In chiral non-Fermi liquids, strong kinematic constraints protect critical exponents
from quantum corrections beyond one-loop, 
even though it is a strongly coupled theory in two space dimensions  \cite{Sur1}.
However, such non-perturbative constraints are unavailable for non-chiral systems in general. 
Therefore, it is of interest to find perturbatively accessible  non-Fermi liquids which can be understood in a controlled way.
Various deformations of  theoretical models have been considered to obtain perturbative control over quantum fluctuations. 
An introduction of a large number of species of fermions fails to   weaken the strong quantum fluctuations in the presence of a  Fermi surface\cite{SSL1, Metlitski1, Metlitski2,Holder15}.
To tame quantum fluctuations, one can use a dimensional regularization scheme where the dimension of space is increased with the co-dimension of the Fermi surface fixed to be one  \cite{Chakravarty,Fitzpatrick1,Fitzpatrick2}. 
This scheme has the merit of preserving a non-vanishing density of states at the Fermi surface. 
However, the increase in the dimension of Fermi surface beyond one results in
a  loss of emergent locality in the momentum space  \cite{SLEE08},
which is an example of ultraviolet/infrared (UV/IR) mixing\cite{Mandal}. 
Consequently, the size of the Fermi surface enters in the low-energy scaling of physical quantities
which are insensitive to the size of Fermi surface in the original two-dimensional theory.
An alternative strategy is to reduce the density of states of  the collective mode  \cite{NayakNFL, Mross}, 
or the fermions  \cite{Senthil, Dalidovich, Sur2}.
This is achieved either by modifying the dispersion,  
or embedding the one-dimensional Fermi surface in a higher dimensional space.
In the latter `co-dimensional' regularization scheme, 
one can preserve  locality and avoid UV/IR mixing 
by introducing a nodal gap,
which leaves behind a one-dimensional Fermi surface 
embedded in general $d$ dimensions  \cite{Dalidovich, Sur2}.
Weakly interacting non-Fermi liquids become accessible near the upper critical dimension, 
where the deviation from the upper critical dimension, $\epsilon$,  becomes a small parameter.

In a recent work   \cite{Sur2}, the spin-density wave critical point was  studied in metals with four-fold rotational ($C_4$) symmetry based on the co-dimensional regularization scheme. 
From  one-loop renormalization group (RG) analysis, 
a non-Fermi liquid state was found 
at the infrared (IR) fixed point below three dimensions. 
Although interactions are renormalized to zero at low energies,
an emergent nesting of Fermi surface and the boson velocity that flows to zero in the low energy limit
enhance quantum fluctuations.
A balance between the vanishing coupling and the IR singularity caused by the dynamically generated quasilocality
results in a stable non-Fermi liquid for small $\epsilon$.
Here quasilocality is different from a completely dispersionless spectrum of the collective mode\cite{Patel,Maier16}.
Instead it refers to the fact that the velocity of the collective mode 
measured in the unit of the Fermi velocity
flows to zero in the low energy limit.

\begin{figure}[!]
\centering
\includegraphics[width=0.9\columnwidth]{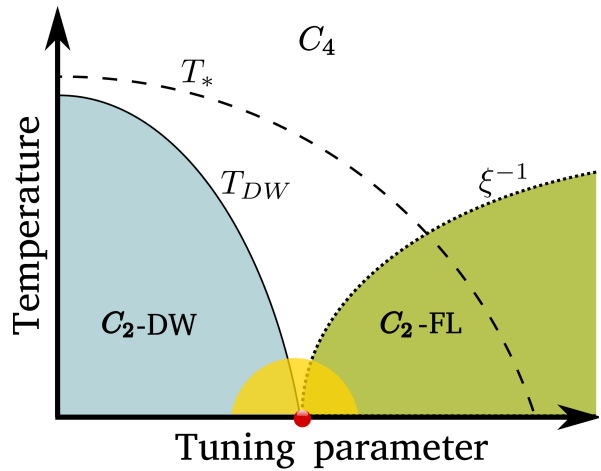}
\caption{
A schematic phase diagram for a density wave transition in metals with the $C_2$ symmetry. 
Here `DW' = density wave, and `FL' = Fermi liquid. 
$T_*$ (dashed line) is a temperature scale which separates the $C_2$ symmetric phase at low temperatures 
from the $C_4$ symmetric phase at high temperatures.
$T_*$ is a crossover when $C_4$ is explicitly broken,  
whereas it becomes a phase transition line when $C_4$ is spontaneously broken.
Either way, the quantum critical point for the density wave transition  
is described by the same theory that respects only $C_2$ symmetry.
$T_{DW}$ (solid line) is the temperature scale below which the system develops a long-range density wave order. 
$\xi^{-1}$ (dotted line) is the inverse correlation length of the density wave fluctuations in the paramagnetic Fermi liquid. 
The dome around the critical point represents 
a potential secondary ordered phase that can appear 
if the critical point is unstable.
}
\label{fig: phase-diag}
\end{figure}
The emergent nesting is a consequence of interaction 
which tends to localize particles in certain directions in real space.
However, the effect of the interactions is rather limited in the presence of the $C_4$ symmetry,
which constrains the $x$ and $y$ components of momentum to scale identically.
Because the deviation from perfect nesting flows to zero only logarithmically in length scale  \cite{Chubukov, Abanov1,Metlitski2,Sur2,Maier16}, 
the Fermi surface nesting becomes noticeable only when the momentum is exponentially close to the hot spots. 
The situation is different when
the $C_4$ symmetry is explicitly or spontaneously broken to  two-fold rotational ($C_2$) symmetry  \cite{Taillefer1, Chu2009, Chuang, Lawler, Ando, Hinkov, Fink, Kivelson,Chubukov2015}. 
If the system undergoes a continuous density wave transition in metals with the $C_2$ symmetry  \cite{Zhou,Chu,Kasahara2012,Lu,Thewalt2015}, 
a new type of non-Fermi can emerge at the quantum critical point.
Because different components of momentum receive different quantum corrections,
the system can exhibit a stronger dynamical nesting.
In this paper, we study the scaling properties of the quantum critical points associated with the  spin density wave (SDW) and  charge density wave (CDW) transitions in metals with the $C_2$ symmetry (see \fig{fig: phase-diag}).

The paper is organized as follows. 
In section \ref{sec: model}, we introduce the low energy effective theory that describes the density wave critical points in metals with the $C_2$ symmetry. 
We take advantage of the formal similarities between the SDW and  CDW critical points to formulate a unified approach to both cases.
Here we employ the co-dimensional regularization scheme, where the one-dimensional Fermi surface is embedded in $3-\epsilon$ space dimensions.
In section \ref{sec: RG}, we outline the RG procedure, and derive the general expressions for the critical exponents and the beta functions.
In section \ref{sec: SDW}, we show that a stable non-Fermi liquid fixed point is realized at the SDW critical point  slightly below three dimensions.
In the low energy limit, not only frequency but also different momentum components acquire anomalous dimensions, 
resulting in an anisotropic non-Fermi liquid.
We compute the critical exponents that govern the anisotropy, and other critical exponents to the leading order in $\epsilon$. 
In the low energy limit, the energy of the collective mode disperses with different powers in different momentum directions.
Furthermore, the Fermi surface near the hot spots connected by the SDW vector is deformed to a universal power-law shape. 
The algebraic nesting is stronger compared to the $C_4$ symmetric case
where the Fermi surface is deformed only logarithmically. 
It is also shown that a component of the boson velocity, which flows to zero at the one-loop order, 
flows to a nonzero value which is order of $\epsilon^{1/3}$ due to a two-loop correction.
The non-zero but small velocity enhances higher-loop diagrams.
Despite the enhancement, higher loop corrections are systematically suppressed 
in the small $\epsilon$ limit, and the $\epsilon$-expansion is controlled.
Section \ref{sec: CDW} is devoted to the CDW critical point.
Although the system flows to a stable marginal Fermi liquid in three dimensions,
it flows out of the perturbative window in the low energy limit for any nonzero $\epsilon$.
In section \ref{sec: conclusion}, we conclude with a summary.


\section{The Model} \label{sec: model}

\begin{figure*}[!]
\centering
\begin{subfigure}{0.28\linewidth}
\centering
\includegraphics[width = 0.7\columnwidth]{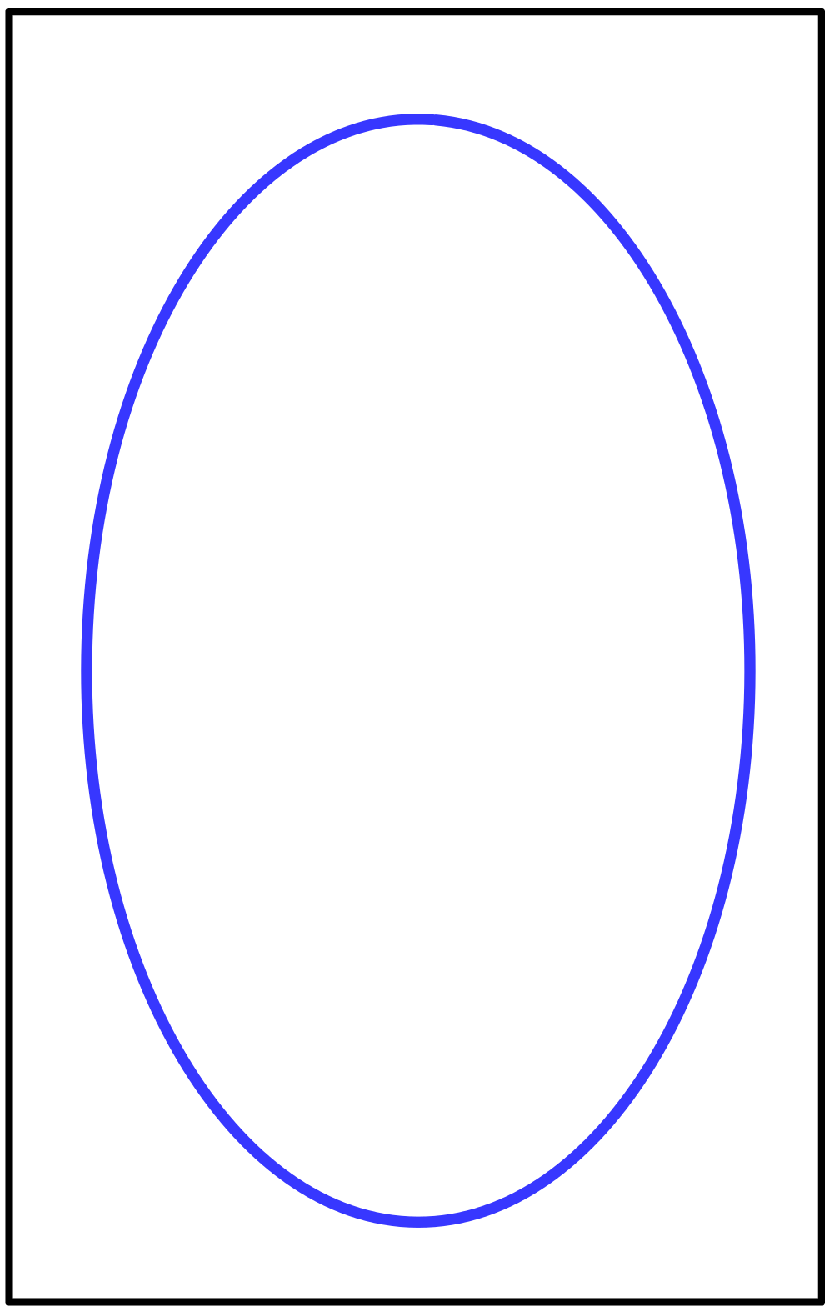}
\caption{}
\label{fig: FS-FL}
\end{subfigure}
\hfill
\begin{subfigure}{0.28\linewidth}
\centering
\includegraphics[width = 0.7\columnwidth]{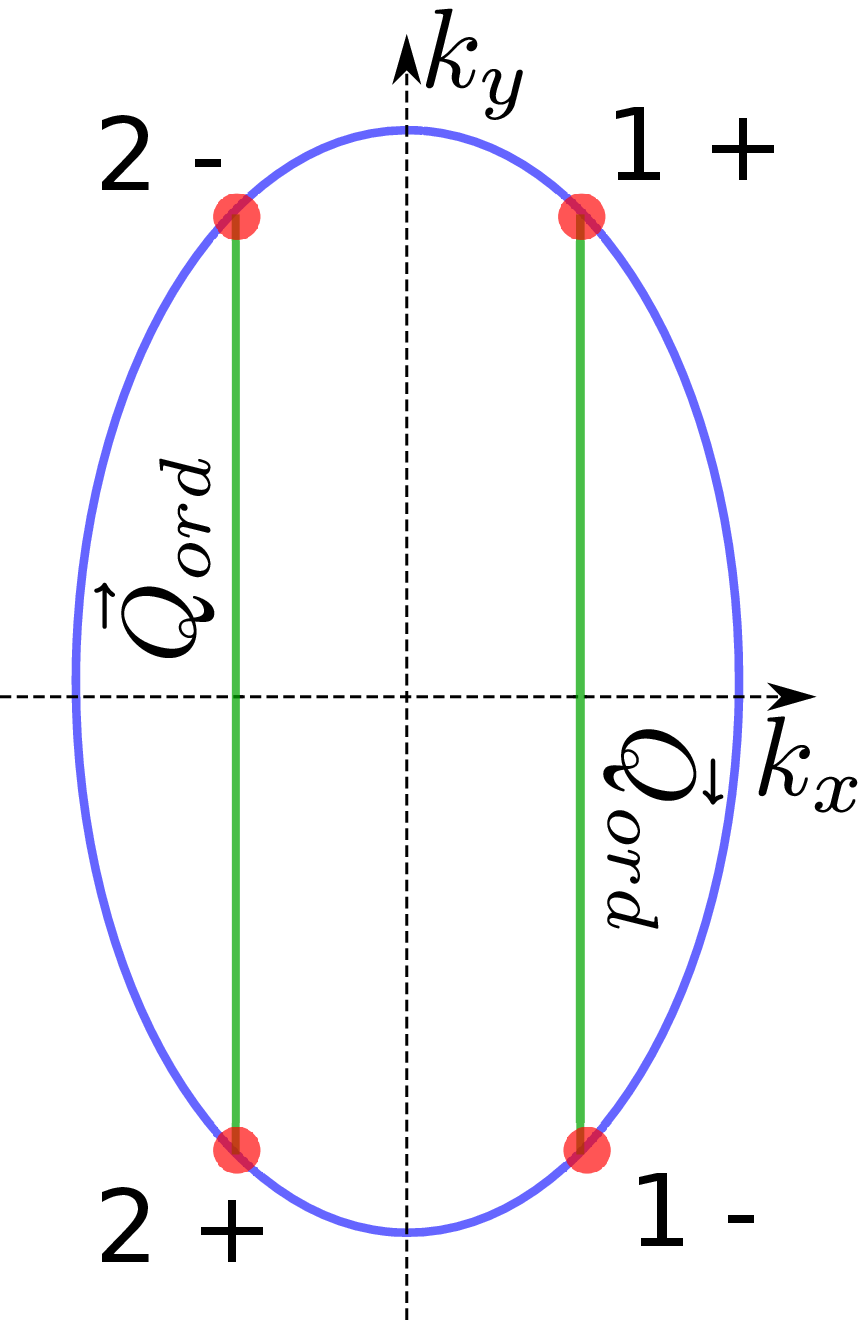}
\caption{}
\label{fig: FS-scheme}
\end{subfigure} 
\hfill
\begin{subfigure}{0.28\linewidth}
\centering
\includegraphics[width = 0.7\columnwidth]{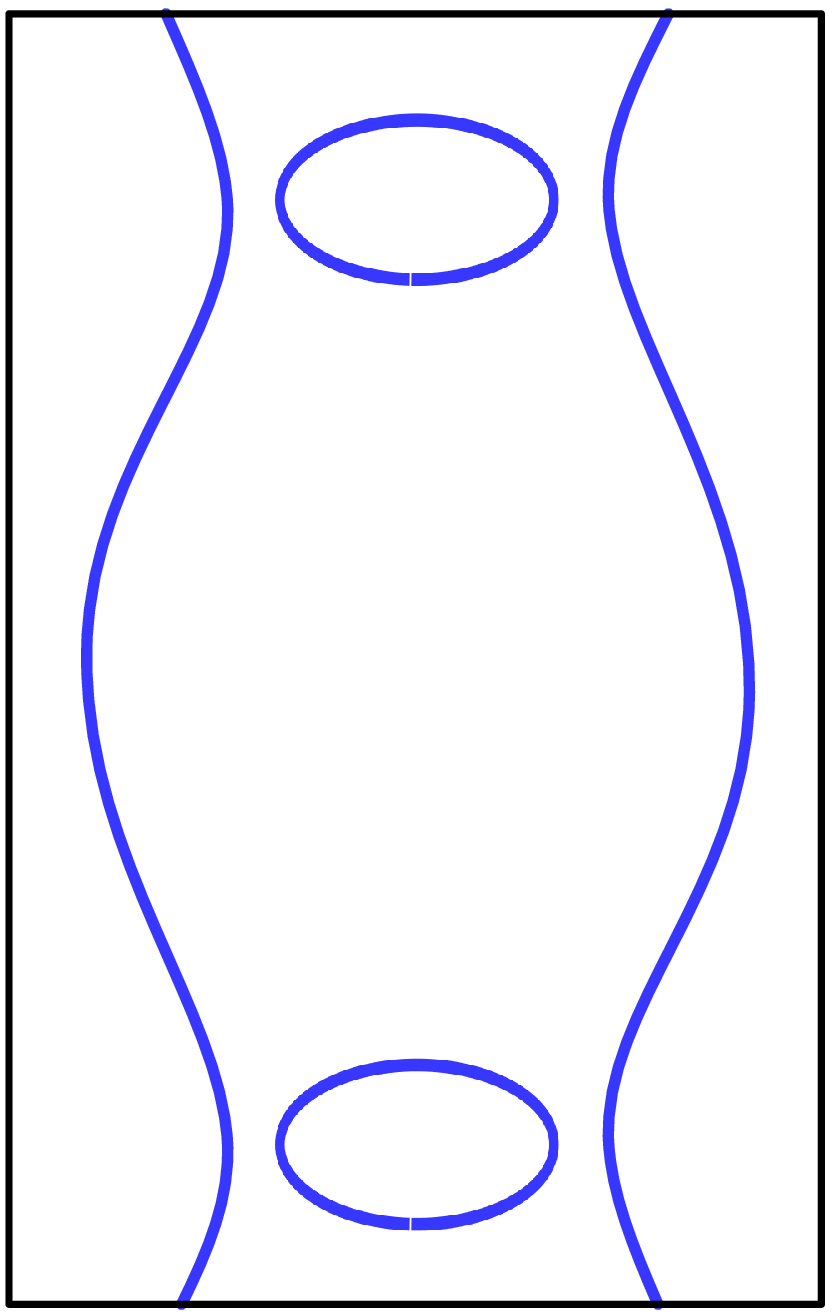}
\caption{}
\label{fig: FS-DW}
\end{subfigure}
%
\caption{
(a) Anisotropic Fermi surface in two space dimensions. 
(b) At the critical point, density wave fluctuations induce strong scatterings 
between electrons near the hot spots denoted by the (red) dots. 
(c) Reconstructed Fermi surface in the ordered phase.
}
\end{figure*}

In this section we introduce the minimal model for the quantum critical point associated 
with the spin and charge density wave transitions in metals with the $C_2$  symmetry. 
A rectangular lattice with anisotropic hoppings in the $\hat x$ and $\hat{y}$  directions
gives an anisotropic Fermi surface as is shown in \fig{fig: FS-FL}.
At a generic filling, the Fermi surface is not nested,
and weak interactions do not produce density wave instabilities. 
Here we assume that there exists a microscopic Hamiltonian with a finite strength of interaction 
that drives a spin or charge density wave transition in the $C_2$ symmetric metal.
We consider a commensurate density wave with wave vector $\vec Q_{ord}$ 
which satisfies $2 \vec Q_{ord} = 0$ modulo the reciprocal vectors.
The specific choice of $\vec Q_{ord}$ and the shape of the Fermi surface is unimportant for the low energy description 
of the quantum critical point.
The order parameter fluctuations are strongly coupled with electrons near a finite number of hot spots 
which are connected to each other through the primary wave vector $\vec Q_{ord}$. 
The hot spots are represented as (red) dots in \fig{fig: FS-scheme}.
In the ordered state, the Fermi surface is reconstructed (see \fig{fig: FS-DW}) 
due to a gap that opens up in the single particle excitation spectrum near the hot spots.

We study the universal properties of the critical points
within the framework of low energy effective field theory 
that is independent of the microscopic details. 
A spin-fermion model is the minimal theory that describes the interaction between the collective mode and the itinerant electrons   \cite{Chubukov, Abanov1}.
In the minimal model, we focus on the vicinity of the hot spots 
and consider interactions of the electrons near the hot spots with long wavelength fluctuations of the order parameter. 
At low energies, we can ignore the Fermi surface curvature and use linearized  electronic dispersions around the hot spots.
We emphasize that linearizing the dispersion is not equivalent to taking the one-dimensional limit 
because the collective modes scatter electrons across the hot spots whose Fermi velocities are not parallel to each other.
Due to the similarities between the SDW and CDW critical points, 
we introduce a general action which is applicable to both cases,
\begin{widetext}
\begin{align}
S &= \sum_{j=1}^{N_f} \sum_{s=1}^{N_c} \sum_{l=1}^{2} \sum_{m = \pm} \int \frac{d^3 k}{(2\pi)^3} ~ \psi_{l,m,j,s}^{*}(k)
\lt( ik_0 + \vec{\vee}_{l,m} \cdot \vec k \rt) \psi_{l,m,j,s}(k)  
+ \frac{1}{4} \int \frac{d^3 q}{(2 \pi)^3} ~ 
\lt( q_0^2 + c_x^2 q_x^2 + c_y^2 q_y^2 \rt) \tr{\Phi(-q) \Phi(q)} \nn
& \quad + \frac{\wtil{g}}{\sqrt{N_f}} ~\sum_{j=1}^{N_f} \sum_{l=1}^{2} \sum_{s,s' = 1}^{N_c} \int \frac{d^3 k}{(2 \pi)^3} 
\frac{d^3 q}{(2 \pi)^3}  
\lt[ \psi_{l,+,j,s}^{*}(k+q)~ \Phi_{s,s'}(q)  ~\psi_{l,-,j,s'}(k) + \mbox{h.c.} \rt] \nn
& \quad + \frac{1}{4} \int \frac{d^3 q_1}{(2\pi)^3} \frac{d^3 q_3}{(2\pi)^3} \frac{d^3 q_3}{(2\pi)^3}
~ \Bigl[ \wtil{u}_1 ~ \tr{\Phi(-q_1 + q_2) \Phi(q_1)} ~ \tr{ \Phi(-q_3 - q_2) \Phi(q_3)} 
\nn &\qquad \qquad
+  \wtil{u}_2 ~ \tr{\Phi(-q_1 + q_2) \Phi(q_1) \Phi(-q_3 - q_2) \Phi(q_3)} \Bigr].
\label{eq: original-action}
\end{align}
\end{widetext}
Here $ \psi_{l,m,j,s}(k)$ describe electrons with momenta near the hot spots, 
where $(l,m)$ with $l=1,2$ and $m=\pm$ labels the four hot spots as shown in \fig{fig: FS-scheme}.
$j = 1,2,..,N_f$ and $s=1,2,..,N_c$ represent a flavor index and the spin, respectively.
The $SU(2)$ spin is generalized to $SU(N_c)$. 
The parameter $N_f$ is an extra flavor which can arise from degenerate bands with the $SU(N_f)$ symmetry. 
$\vec k$ is the two-dimensional momentum that measures a deviation from the hot spots. 
$\vec{\vee}_{l,m}$ is the Fermi velocity at each hot spot : 
$\vec{\vee}_{1,+} \equiv (v_x, v_y) = - \vec{\vee}_{2,+}$, $\vec{\vee}_{1,-} \equiv (v_x, - v_y) = - \vec{\vee}_{2,-}$.
If the ordering wave vector happens to coincide with $2 \vec K_F$ ($\vec K_F$ being a Fermi vector),
$v_x$ vanishes and one needs to include the local curvature of the Fermi surface\cite{Holder14,Schafer16}.
In this paper, we consider the generic case with $v_x, v_y > 0$,
where the hot spots connected by the order vector are not pre-nested.
The $N_c \times N_c$ matrix field $\Phi(q)$ represents the density wave mode of frequency $q_0$ and momentum $\vec Q_{ord}+ \vec q$.
The boson field satisfies $\Phi^{\dagger}(q) = \Phi(-q)$ because $2 \vec Q_{ord} = 0$    \cite{NayakDW}.
The matrix field can be written as
\begin{align}
\Phi(q) = 
\begin{cases}
\vec \phi(q) \cdot \vec \tau & \mbox{for ~ SDW} \\[.1em] 
\sqrt{\frac{2}{N_c}} ~ \phi(q) ~ \mc I_{N_c} & \mbox{for ~ CDW}
\end{cases}
\label{eq: Phi}
\end{align}
where $\tau^{(\alpha)}$ is the $\alpha$-th generator of $SU(N_c)$ in the fundamental representation, and
$\mc I_{N_c}$ is the $N_c \times N_c$ identity matrix.
$\tau^{(\alpha)}$ and $\mc I_{N_c}$ represent the spin and charge vertices, respectively. 
We choose the normalization  $\tr{\tau^{(\alpha)} \tau^{(\beta)}} = 2 \dl^{\alpha \beta}$ for the $\tau$-matrices. 
For $N_c = 2$ and $3$ in the SDW case and for any $N_c$ in the CDW case, $\wtil{u}_1$ and $\wtil{u}_2$ are equivalent, and we can set $\wtil{u}_2 = 0$ without loss of generality.
For the CDW critical point, both $N_f$ and $N_c$ play the same role, and the physics depends only on the total number of electron species, $\wtil{N}_f = N_c N_f$.

Some parameters in \eq{eq: original-action} can be absorbed into scales of momentum and fields.
We scale $(k_x, k_y) \mapsto \lt(\frac{k_x}{c_x}, \frac{k_y}{v_y} \rt)$ and $(\Phi, \psi) \mapsto \sqrt{c_x v_y}(\Phi, \psi)$ to rewrite the action as 
\begin{widetext}
\begin{align}
& S = \sum_{j=1}^{N_f} \sum_{s=1}^{N_c} \sum_{l=1}^{2} \sum_{m = \pm} \int \frac{d^3 k}{(2\pi)^3} ~ \psi_{l,m,j,s}^{*}(k)
\lt( ik_0 + \mc{E}_{l,m}(\vec k) \rt) \psi_{l,m,j,s}(k) 
+ \frac{1}{4} \int \frac{d^3 q}{(2 \pi)^3} ~ 
\lt( q_0^2 + q_x^2 + c^2 q_y^2 \rt) \tr{\Phi(-q) \Phi(q)} \nn
& \quad + \frac{g_{0}}{\sqrt{N_f}} \sum_{j=1}^{N_f} \sum_{l=1}^{2} \sum_{s,s' = 1}^{N_c} \int \frac{d^3 k}{(2 \pi)^3} 
\frac{d^3 q}{(2 \pi)^3}  
\lt[ \psi_{l,+,j,s}^{*}(k+q)~ \Phi_{s,s'}(q)  ~\psi_{l,-,j,s'}(k) + \mbox{h.c.} \rt] \nn
&  \quad + \frac{1}{4} \int \frac{d^3 q_1}{(2\pi)^3} \frac{d^3 q_2}{(2\pi)^3} \frac{d^3 q_3}{(2\pi)^3}
~ \Bigl[ u_{1;0} ~ \tr{\Phi(-q_1 + q_2) \Phi(q_1)} ~ \tr{ \Phi(-q_3 - q_2) \Phi(q_3)} 
\nn &\qquad \qquad
+ u_{2;0} ~ \tr{\Phi(-q_1 + q_2) \Phi(q_1) \Phi(-q_3 - q_2) \Phi(q_3)} \Bigr].
\label{eq: dw-action-2d}
\end{align}
\end{widetext}
The rescaled dispersions are  $\mc{E}_{1,+}(\vec k) = - \mc{E}_{2,+}(\vec k) = v k_x + k_y$, and $\mc{E}_{1,-}(\vec k) = - \mc{E}_{2,-}(\vec k) = v k_x - k_y$,
where $v \equiv \dfrac{v_x}{c_x}$ and $c \equiv \dfrac{c_y}{v_y}$ represent the relative velocities between electron and boson in the two directions.
The couplings are also rescaled to $g_{0} \equiv  \dfrac{\wtil{g}}{\sqrt{c_x v_y}}$ and $u_{i;0} \equiv \dfrac{\wtil{u}_i}{c_x v_y}$.

The $(2+1)$-dimensional theory is now generalized to a $(d+1)$-dimensional theory which describes the one-dimensional Fermi surface embedded in d-dimensional momentum space. 
Following the formalism in Ref.     \cite{Sur2}, we express  \eq{eq: dw-action-2d} in the basis of spinors $\Psi_{+,j,s}(k) = \lt( \psi_{1,+,j,s}(k), \quad \psi_{2,+,j,s}(k) \rt)^T$ and $\Psi_{-;j;s}(k) = \lt( \psi_{1,-;j;s}(k), \quad - \psi_{2,-;j;s}(k) \rt)^T$, and add $(d-2)$ extra  co-dimensions to the Fermi surface, 
\begin{widetext}
\begin{align}
& S = \sum_{j=1}^{N_f} \sum_{s=1}^{N_c} \sum_{n = \pm} \int dk ~ \bar \Psi_{n,j,s}(k) \lt( i \mbf{K} \cdot \mbf{\Gamma} + i \veps_{n}(\vec k) ~ \gamma_{d-1} \rt) \Psi_{n,j,s}(k) 
 + \frac{1}{4} \int dq ~ \lt( |\mbf{Q}|^2 + q_x^2 + c^2 q_y^2 \rt) ~ \tr{\Phi(-q) \Phi(q)} \nn
&  + i ~ \frac{g_{0}}{\sqrt{N_f}} \sum_{j=1}^{N_f}  \sum_{s,s' = 1}^{N_c} \int dk ~ dq ~  
\lt[ \bar \Psi_{+,j,s}(k+q)~ \gam_{d-1} ~ \Phi_{s,s'}(q) ~\Psi_{-,j,s'}(k) - \mbox{h.c.} \rt] \nn
&   + \frac{1}{4} \int dq_1  dq_2  dq_3 ~ \Bigl[ u_{1;0} ~ \tr{\Phi(-q_1 + q_2) \Phi(q_1)} ~ \tr{ \Phi(-q_3 - q_2) \Phi(q_3)} 
+ u_{2;0} ~ \tr{\Phi(-q_1 + q_2) \Phi(q_1) \Phi(-q_3 - q_2) \Phi(q_3)} \Bigr],
\label{eq: gen-d}
\end{align}
\end{widetext}
where $k \equiv (\mbf{K}, \vec k)$ and $dk \equiv \dfrac{d^{d+1} k}{(2\pi)^{d+1}}$. 
The two dimensional vectors on the plane of the Fermi surface are denoted as $\vec k = (k_x, k_y)$, while $\mbf{K} = (k_0, k_1, \ldots, k_{d-2})$ denotes $(d-1)$ dimensional vectors with $k_1, \ldots, k_{d-2}$ being the newly added co-dimensions. 
We collect the first $(d-1)$ $\gamma$-matrices in $\mbf{\Gamma} = (\gam_0, \gam_1, \ldots, \gam_{d-2})$. 
The conjugate  spinor is defined by $\bar \Psi_{n,j,s} = \Psi_{n,j,s}^{\dag} \gam_0$. 
The dispersions of the spinors along the $\vec k$ direction are inherited from the two dimensional dispersion, $\veps_{\pm}(\vec k) = v k_x \pm k_y$. 
It is easy to check that  we recover \eq{eq: dw-action-2d} in  $d = 2$ 
with $\gamma_0 = \sigma_y$ and  $\gamma_1 = \sigma_x$, where $\sigma_i$ are Pauli matrices.
The theory in general dimensions interpolate between the two-dimensional metal and a semi-metal with a line node in three dimensions \cite{Burkov2011, Bian2016}. 
The action is invariant under $U(1) \times SU(N_c) \times SU(N_f)$, which are associated with the particle number, spin and flavor conservations, respectively.  
The theory is also invariant under time reversal, inversion, and $SO(d-1)$ rotations in $\mbf{K}$.

The engineering scaling dimensions of the $(d+1)$-momentum, the fields and the couplings  are
\begin{align}
& [\mbf{K}] = 1, \quad [k_x] = 1, \quad [k_y] = 1, \nn
& [\Psi_{n,j,s}] = -\half (d+2), \quad [\Phi] = -\half (d+3), \quad [v] = 0 \nn  
& [c] = 0, \quad [g_0] = \half (3-d), \quad \mbox{and} \quad [u_0] = 3 - d.
\label{eq: GaussianScaling}
\end{align}
Classically, frequency and all momentum components have the same scaling dimension.
The upper critical dimension is $d=3$ at which all the couplings in the theory are dimensionless at the Gaussian fixed point.
We apply the field theoretic RG  
based on a perturbative expansion in $\eps \equiv 3 - d$.

\section{Renormalization Group} 
\label{sec: RG}

In this section we outline our RG scheme, and derive the general expressions for the beta functions and the critical exponents.
The readers who wish to skip the details can jump to Eqs.  \eqref{eq: zTau} - \eqref{eq: uvBetaChi2} which are the main results of this section.

Starting with the action in \eq{eq: gen-d}, 
we define dimensionless couplings 
\begin{align}
g = \mu^{-(3-d)/2} ~ g_0, 
\qquad
u_i = \mu^{-(3-d)} ~ u_{i;0},
\label{eq: dimlss}
\end{align}
where $\mu$ is a scale at which the renormalized couplings are to be defined.
From the action in \eq{eq: gen-d}, the quantum effective action is computed perturbatively in the couplings. 
The logarithmic divergences that arise at the upper critical dimension manifest themselves as poles in $\epsilon$.
Requiring the renormalized quantum effective action to be analytic in $\epsilon$, 
we add counter terms of the form,
\begin{widetext}
\begin{align}
S_{CT} &= \sum_{j=1}^{N_f} \sum_{s=1}^{N_c} \sum_{n = \pm} \int dk ~ \bar \Psi_{n,j,s}(k) \lt( i \mc{A}_1 \mbf{K} \cdot \mbf{\Gamma} + i (\mc{A}_2 v k_x + n \mc{A}_3 k_y) ~ \gamma_{d-1} \rt) \Psi_{n,j,s}(k) \nn 
& \quad + \frac{1}{4}  \int dq ~ 
\lt( \mc{A}_4 |\mbf{Q}|^2 + \mc{A}_5 q_x^2 + \mc{A}_6 c^2 q_y^2 \rt) \tr{\Phi(-q) \Phi(q)} \nn
& \quad + \mc{A}_7 \mu^{(3-d)/2} \dfrac{i g}{\sqrt{N_f}}  \sum_{j=1}^{N_f} \sum_{s,s' = 1}^{N_c} \int dk ~ dq ~  
\lt[\bar \Psi_{+,j,s}(k+q)~ \gam_{d-1} ~\Phi_{s,s'}(q) ~\Psi_{-,j,s'}(k) - \mbox{h.c.} \rt] \nn
&  \quad + \frac{\mu^{(3-d)}}{4} \int dq_1 ~ dq_2 ~ dq_3 ~ \Bigl[ \mc{A}_8 u_{1} ~ \tr{\Phi(-q_1 + q_2) \Phi(q_1)} ~ \tr{ \Phi(-q_3 - q_2) \Phi(q_3)} \Bigr. \nn 
& \hspace{2in} + \Bigl. \mc{A}_9 u_{2} ~ \tr{\Phi(-q_1 + q_2) \Phi(q_1) \Phi(-q_3 - q_2) \Phi(q_3)} \Bigr],
\label{eq: dw-action-CT}
\end{align}
\end{widetext}
with $
\mc{A}_i \equiv \mc{A}_i(v, c, g, u, \eps) 
= \sum_{m=1}^{\infty} Z_{i,m}(v, c, g, u)~ \eps^{-m}$.
The counter terms are chosen to cancel the poles in $\epsilon$ 
based on the minimal subtraction scheme. 
Due to the lack of full rotational  symmetry in the $(\mbf{K}, \vec k)$-space and the $C_4$ symmetry in the $(k_x, k_y)$-plane, 
the kinetic terms are renormalized differently in the $\mbf K$, $k_x$, and $k_y$ directions, respectively.
Therefore, $\mbf K$,  $k_x$ and $k_y$ can have different quantum scaling dimensions. 
The sum of the original action and the counter terms gives the bare action,
\begin{widetext}
\begin{align}
& S_{B} = \sum_{j=1}^{N_f} \sum_{s=1}^{N_c} \sum_{n = \pm} \int dk_{{B}} ~ \bar \Psi_{B; n,j,s}(k_{{B}}) \lt( i \mbf{K}_B \cdot \mbf{\Gamma} + i  (v_B k_{B;x} + n k_{B;y}) ~ \gamma_{d-1} \rt) \Psi_{B; n,j,s}(k_{{B}}) \nn 
& \quad + \frac{1}{4}  \int dq_B ~  \lt( |\mbf{Q}_B|^2 + q_{B; x}^2 + c_{{B}}^2 q_{B; y}^2 \rt) \tr{\Phi_{B}(-q_B) \Phi_{B}(q_B)} \nn
& \quad + i ~ \frac{g_{B}}{\sqrt{N_f}} \sum_{j=1}^{N_f} \sum_{s,s' = 1}^{N_c} \int dk_{{B}} ~ dq_B ~   \lt[\bar \Psi_{B; +,j,s}(k_{{B}} + q_B)~ \gam_{d-1} ~ \Phi_{B;s,s'}(q_B)  ~\Psi_{B; -,j,s'}(k_{{B}}) - \mbox{h.c.} \rt] \nn
&  \quad + \frac{1}{4} \int dq_{1;B} ~ dq_{2;B} ~ dq_{3;B} ~ \Bigl[ u_{1;B} ~ \tr{\Phi_B(-q_{1;B} + q_{2;B}) \Phi_B(q_{1;B})} ~ \tr{ \Phi_B(-q_{3;B} - q_{2;B}) \Phi_B(q_{3;B})} \Bigr. \nn 
& \hspace{2in} + \Bigl. u_{2;B} ~ \tr{\Phi_B(-q_{1;B} + q_{2;B}) \Phi_B(q_{1;B}) \Phi_B(-q_{3;B} - q_{2;B}) \Phi_B(q_{3;B})} \Bigr].
\label{eq: sdw-action-ren}
\end{align}
\end{widetext}
Here the bare quantities are related to their renormalized counterparts through the multiplicative factors,
\begin{align}
& \mbf{K} = \mc{Z}_{\tau}^{-1} ~\mbf{K}_B, 
 \qquad \quad
 k_x = \mc{Z}_{x}^{-1} ~k_{B; x}, & \nn
& k_y = k_{B; y}, 
 \qquad \quad
 \Psi_{n,j,s} = \mc{Z}_{\psi}^{-\half} ~\Psi_{B; n, j, s}, & \nn
& \Phi = \mc{Z}_{\phi}^{-\half} ~\Phi_{B},
 \qquad \quad
 v = \frac{\mc{Z}_x^2 ~ \mc{Z}_{\tau}^{d-1} ~ \mc{Z}_{\psi}}{\mc{Z}_2} ~ v_{B}, & \nn
& c = \lt[ \dfrac{\mc{Z}_x ~ \mc{Z}_{\phi} ~\mc{Z}_{\tau}^{d-1}}{\mc{Z}_6} \rt]^{\half} ~ c_{{B}}, \nn
& g = \mu^{-(3-d)/2}~ \frac{\mc{Z}_{x}^2 ~ \mc{Z}_{\tau}^{2(d-1)} ~ \mc{Z}_{\psi} ~ \mc{Z}_{\phi}^{\half}}{\mc{Z}_{7}} ~ g_{B}, & \nn
& u_1 = \mu^{-(3-d)}~ \dfrac{\mc{Z}_{x}^3 ~ \mc{Z}_{\tau}^{3(d-1)} ~ \mc{Z}_{\phi}^{2}}{\mc{Z}_{8}} ~ u_{1;B}, 
\nn
& u_2 = \mu^{-(3-d)}~ \dfrac{\mc{Z}_{x}^3 ~ \mc{Z}_{\tau}^{3(d-1)} ~ \mc{Z}_{\phi}^{2}}{\mc{Z}_{9}} ~ u_{2;B}, & 
\label{eq: bare-ren}
\end{align}
where
\begin{align}
& \mc{Z}_{\tau} = \dfrac{\mc{Z}_1}{\mc{Z}_3}, 
\quad
\mc{Z}_{x} = \mc{Z}_{\tau}~  \lt[\dfrac{\mc{Z}_5}{\mc{Z}_4}\rt]^{1/2},\nn
& \mc{Z}_{\psi} = \dfrac{\mc{Z}_3}{\mc{Z}_x ~ \mc{Z}_{\tau}^{(d-1)}},
\quad 
\mc{Z}_{\phi} = \dfrac{\mc{Z}_4}{\mc{Z}_x ~ \mc{Z}_{\tau}^{(d+1)}},
\end{align}
with $\mc{Z}_i \equiv   1 + \mc{A}_i(v, c, g, u, \eps)$.
Here we made the choice $k_y = k_{B;y}$, which fixes the scaling dimension of $k_y$  to be $1$. 
This choice can be always made, even at the quantum level, because one can measure scaling dimensions of other quantities with respect to that of $k_y$. 
$\mc{Z}_\tau$ and $\mc{Z}_x$ encode the anisotropic quantum corrections, which lead to anomalous dimensions for $\mbf K$ and $k_x$.

The renormalization group equation is obtained by requiring that 
the bare Green's function is invariant under the change of the scale $\mu$ 
at which the renormalized vertex functions are defined. 
The renormalized Green's function, 
\begin{align}
& \langle \Phi(q_1) \ldots \Phi(q_b) \Psi(k_1) \ldots \Psi(k_f) \bar \Psi(k_{f+1}) \ldots \bar \Psi(k_{2f}) \rangle \nn
& \equiv G^{(2f,b)}(q_i, k_j; v, c, g, u; \mu) \nn
& \qquad \times \dl^{(d+1)}\lt( \sum_{i=1}^{b} q_i + \sum_{j=1}^{f} (k_j - k_{f+j}) \rt), ~ 
\end{align}
obeys the renormalization group equation,
\begin{widetext}
\begin{align}
& \Biggl[ z_{\tau} ~ \lt( \mbf{K}_j \cdot
\mbf{\nabla}_{\mbf{K}_j} + \mbf{Q}_i
\cdot \mbf{\nabla}_{\mbf{Q}_i} \rt) 
+ z_{x} ~ \lt( k_{j; x} \dow_{k_{j; x}} + q_{i; x} \dow_{q_{i; x}} \rt)
+ \lt(k_{j; y} \dow_{k_{j; y}} + q_{i; y} \dow_{q_{i; y}} \rt)  \nn
&
- \beta_v \frac{\dow}{\dow v} 
-  \beta_c \frac{\dow}{\dow c}
- \beta_g \frac{\dow}{\dow g} 
- \beta_{u_1}\frac{\dow}{\dow u_1}  
- \beta_{u_2}\frac{\dow}{\dow u_2}  \nn
& + 2f \lt( \frac{d + 2}{2} -  \eta_\psi \rt) + b
\lt( \frac{d + 3}{2} -  \eta_\phi \rt) 
 - \Bigl. \lt(z_{\tau} ~ (d-1) + z_{x} + 1 \rt) \Biggr] G^{(2f,b)}(q_i, k_j; v, c, g, u; \mu) = 0.
\label{eq: RGE}
\end{align}
\end{widetext}
Here $z_\tau$ and $z_x$ are the quantum scaling dimensions for $\mbf K$ and $k_x$ given by
\begin{align}
z_{\tau} = 1 + \frac{\dow \ln \mc{Z}_{\tau}}{\dow \ln{\mu}},
\qquad
z_{x} = 1 + \frac{\dow \ln \mc{Z}_{x}}{\dow \ln{\mu}},
\end{align}
and $\eta_\psi$ and $\eta_\phi$ are the anomalous dimensions of the fields, 
\begin{align}
\eta_\psi = \half \frac{\dow \ln \mc{Z}_{\psi}}{\dow \ln{\mu}},
\qquad
\eta_\phi = \half \frac{\dow \ln \mc{Z}_{\phi}}{\dow \ln{\mu}}.
\end{align}
The beta functions, which describe the change of couplings with an increasing energy scale, are defined as 
\begin{align}
& \beta_v = \frac{\dow v}{\dow \ln{\mu}},
\qquad
\beta_c = \frac{\dow c}{\dow \ln{\mu}},
\nn
& \beta_{g} = \frac{\dow g}{\dow \ln{\mu}},
\qquad
\beta_{u_i} = \frac{\dow u_i}{\dow \ln{\mu}}.
\end{align}

We use the relationship between the bare and renormalized quantities defined in \eq{eq: bare-ren} to obtain a set of coupled differential equations, 
\begin{align}
& \mc{Z}_1 \Bigl[ d(z_\tau - 1) + (z_x - 1) + 2\eta_\psi \Bigr] - \frac{\dow \mc{Z}_1}{\dow \ln{\mu}} = 0, \nn
& \mc{Z}_2 \Bigl[ \beta_v - v \lt\{(d-1)(z_\tau - 1) + 2(z_x - 1) + 2\eta_\psi \rt\} \Bigr] 
 \nn & \hspace{0.65\columnwidth} 
+ v ~\frac{\dow \mc{Z}_2}{\dow \ln{\mu}} = 0, \nn
& \mc{Z}_3 \Bigl[ (d-1)(z_\tau - 1) + (z_x - 1) + 2\eta_\psi \Bigr] - \frac{\dow \mc{Z}_3}{\dow \ln{\mu}} = 0, \nn
& \mc{Z}_4 \Bigl[ (d + 1)(z_\tau - 1) + (z_x - 1) + 2 \eta_\phi \Bigr] - \frac{\dow \mc{Z}_4}{\dow \ln{\mu}} = 0, \nn
& \mc{Z}_5 \Bigl[ (d - 1)(z_\tau - 1) + 3 (z_x - 1) + 2 \eta_\phi \Bigr] - \frac{\dow \mc{Z}_5}{\dow \ln{\mu}} = 0, \nn
& \mc{Z}_6 \Bigl[ 2 \beta_c - c \lt\{(d-1)(z_\tau - 1) + (z_x - 1) + 2\eta_\phi \rt\} \Bigr] 
\nn & \hspace{0.65\columnwidth} 
+ c ~\frac{\dow \mc{Z}_6}{\dow \ln{\mu}} = 0, \nn
& \mc{Z}_7 \Bigl[ \beta_{g} - g~ \Bigl\{ -\frac{3-d}{2} + 2 (d-1)(z_\tau - 1) + 2 (z_x - 1) 
\nn & \hspace{0.4\columnwidth} 
+ 2 \eta_\psi + \eta_\phi \Bigr\} \Bigr] + g ~\frac{\dow \mc{Z}_7}{\dow \ln{\mu}} = 0, \nn
& \mc{Z}_8 \Bigl[ \beta_{u_1} - u_1 ~ \Bigl\{ -(3-d) + 3 (d-1)(z_\tau - 1) + 3 (z_x - 1) 
\nn & \hspace{0.45\columnwidth} 
+ 4 \eta_\phi \Bigr\} \Bigr] + u_1 ~\frac{\dow \mc{Z}_8}{\dow \ln{\mu}} = 0, \nn
& \mc{Z}_9 \Bigl[ \beta_{u_2} - u_2 ~\Bigl\{ -(3-d) + 3 (d-1)(z_\tau - 1) + 3 (z_x - 1) 
\nn & \hspace{0.4\columnwidth} 
+ 4 \eta_\phi \Bigr\} \Bigr] + u_2 ~\frac{\dow \mc{Z}_9}{\dow \ln{\mu}} = 0,
\label{eq: coupled-eqn}
\end{align}
which are solved to obtain the expressions for the critical exponents and the beta functions,
\begin{widetext}
\begin{flalign}
z_\tau &= \lt[ 1 + \lt( \half g\dow_{g} + \sum_i u_i \dow_{u_i} \rt) (Z_{1,1} - Z_{3,1})\rt]^{-1},&
\label{eq: zTau} \\
z_x &= 1 - \half z_{\tau}\lt( \half g\dow_{g} + \sum_i u_i \dow_{u_i} \rt)(2 Z_{1,1} - 2 Z_{3,1} - Z_{4,1} + Z_{5,1}),& 
\label{eq: zX}\\
\eta_{\psi} &= \frac{\eps}{2}(z_{\tau} - 1) 
- \half \lt[ 2(z_\tau - 1) + (z_x - 1) + z_{\tau}\lt( \half g\dow_{g} + \sum_i u_i \dow_{u_i} \rt)Z_{3,1} \rt],& 
\label{eq: EtaPsi} \\
\eta_\phi &= \frac{\eps}{2} (z_{\tau} - 1) 
- \half \lt[ 4 (z_\tau - 1) + (z_x - 1) + z_{\tau}\lt( \half g\dow_{g} + \sum_i u_i \dow_{u_i} \rt)Z_{4,1} \rt],& 
\label{eq: EtaPhi} \\
\dow_{\ell} v &=  \half z_{\tau} v \lt( \half g\dow_{g} + \sum_i u_i \dow_{u_i} \rt)
(2 Z_{1,1} - 2 Z_{2,1} -  Z_{4,1} +  Z_{5,1}),& 
\label{eq: uvBetaV}\\
\dow_{\ell} c &=  - \half z_{\tau} c \lt( \half g\dow_{g} + \sum_i u_i \dow_{u_i} \rt) (2 Z_{1,1} - 2 Z_{3,1} - Z_{4,1} + Z_{6,1}),& 
\label{eq: uvBetaC}\\
\dow_{\ell} {g} &=  \frac{1}{4} z_\tau g \lt[ 2\eps + \lt( \half g\dow_{g} + \sum_i u_i \dow_{u_i} \rt)(2 Z_{1,1} + 2 Z_{3,1} + Z_{4,1} + Z_{5,1} - 4 Z_{7,1} ) \rt],&
 \label{eq: uvBetaG} \\
\dow_{\ell} {u_1} &=  \half z_{\tau} u_1  \lt[ 2\eps - \lt( \half g\dow_{g} + \sum_i u_i \dow_{u_i} \rt)(2 Z_{1,1} - 2 Z_{3,1} - 3 Z_{4,1} - Z_{5,1} + 2 Z_{8,1} ) \rt], &
\label{eq: uvBetaU1} \\
\dow_{\ell} {u_2} &=  \half z_{\tau} u_2  \lt[ 2\eps - \lt( \half g\dow_{g} + \sum_i u_i \dow_{u_i} \rt)(2 Z_{1,1} - 2 Z_{3,1} - 3 Z_{4,1} - Z_{5,1} + 2 Z_{9,1} ) \rt],&
\label{eq: uvBetaU2}
\end{flalign}
\end{widetext}
where we introduced the IR beta function, $\partial_{\ell} \lambda = - \beta_\lambda$ 
which describes the RG flow with an increase of the logarithmic {\it length} scale $\ell$.

In the absence of the Yukawa coupling, every quartic coupling $u_i$ is accompanied by $1/c$ in the perturbative series.
This reflects the IR singularity for the flat bosonic band in the $c \rtarw 0$ limit. 
Since the actual perturbative expansion is organized in terms of $u_i/c$,
 it is convenient to introduce $\chi_\sub{i} = u_i/c $.
 The beta functions for $\chi_\sub{i}$ can be readily obtained from those of $u_i$ and $c$,
\begin{widetext}
\begin{flalign}
\dow_{\ell} {\chi_\sub{1}} &=  \half z_{\tau} \chi_\sub{1} \lt[ 2\eps + \lt( \half g\dow_{g} + \sum_i u_i \dow_{u_i} \rt)(2 Z_{4,1} + Z_{5,1} + Z_{6,1} - 2 Z_{8,1} ) \rt],& \label{eq: uvBetaChi1} \\ 
\dow_{\ell} {\chi_\sub{2}} &=  \half z_{\tau} \chi_\sub{2} \lt[ 2\eps + \lt( \half g\dow_{g} + \sum_i u_i \dow_{u_i} \rt)(2 Z_{4,1} + Z_{5,1} + Z_{6,1} - 2 Z_{9,1} ) \rt].& 
\label{eq: uvBetaChi2}
\end{flalign}
\end{widetext}


\section{Spin Density Wave Criticality} 
\label{sec: SDW}

\begin{figure}[!h]
\centering
\begin{subfigure}[b]{0.35\columnwidth}
 \includegraphics[width=0.95\columnwidth]{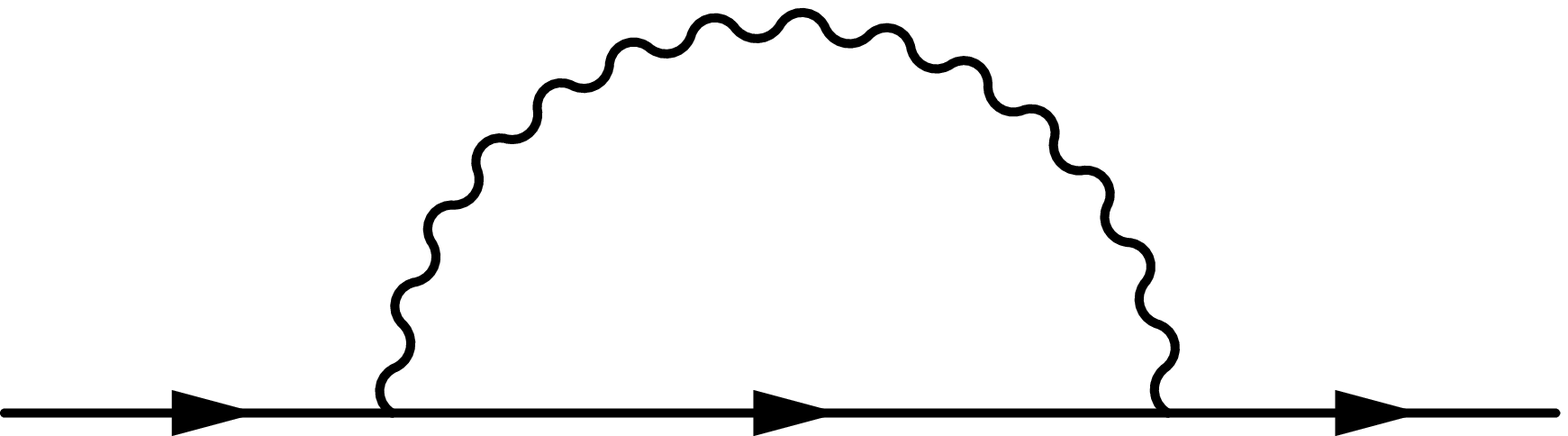}
 \caption{}
 \label{fig: CT-SEf}
\end{subfigure}
\hfill
\begin{subfigure}[b]{0.35\columnwidth}
 \includegraphics[width=0.95\columnwidth]{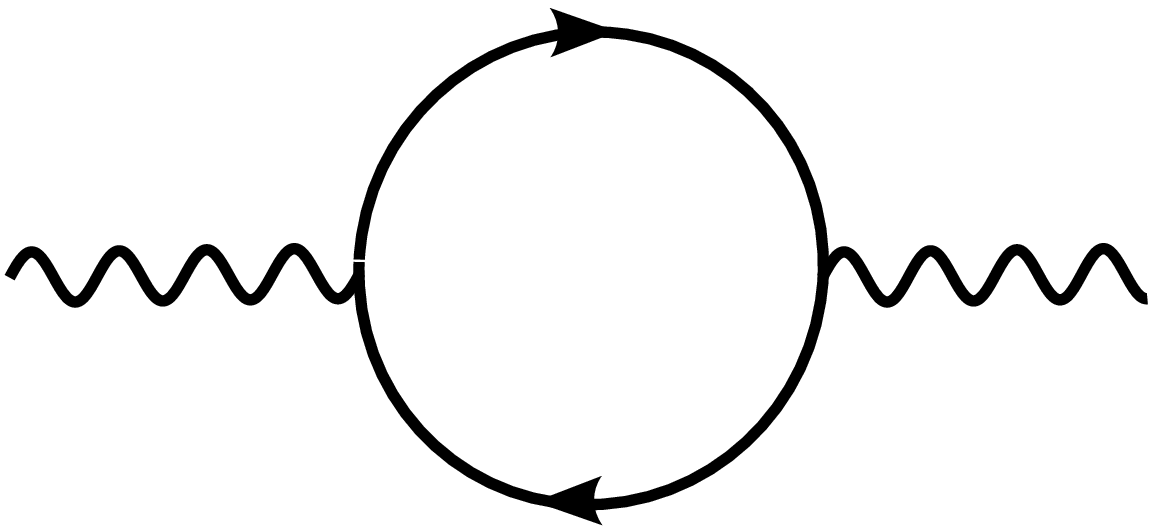}
 \caption{}
 \label{fig: CT-SEb}
\end{subfigure}
\hfill
\begin{subfigure}[b]{0.28\columnwidth}
 \includegraphics[width=0.9\columnwidth]{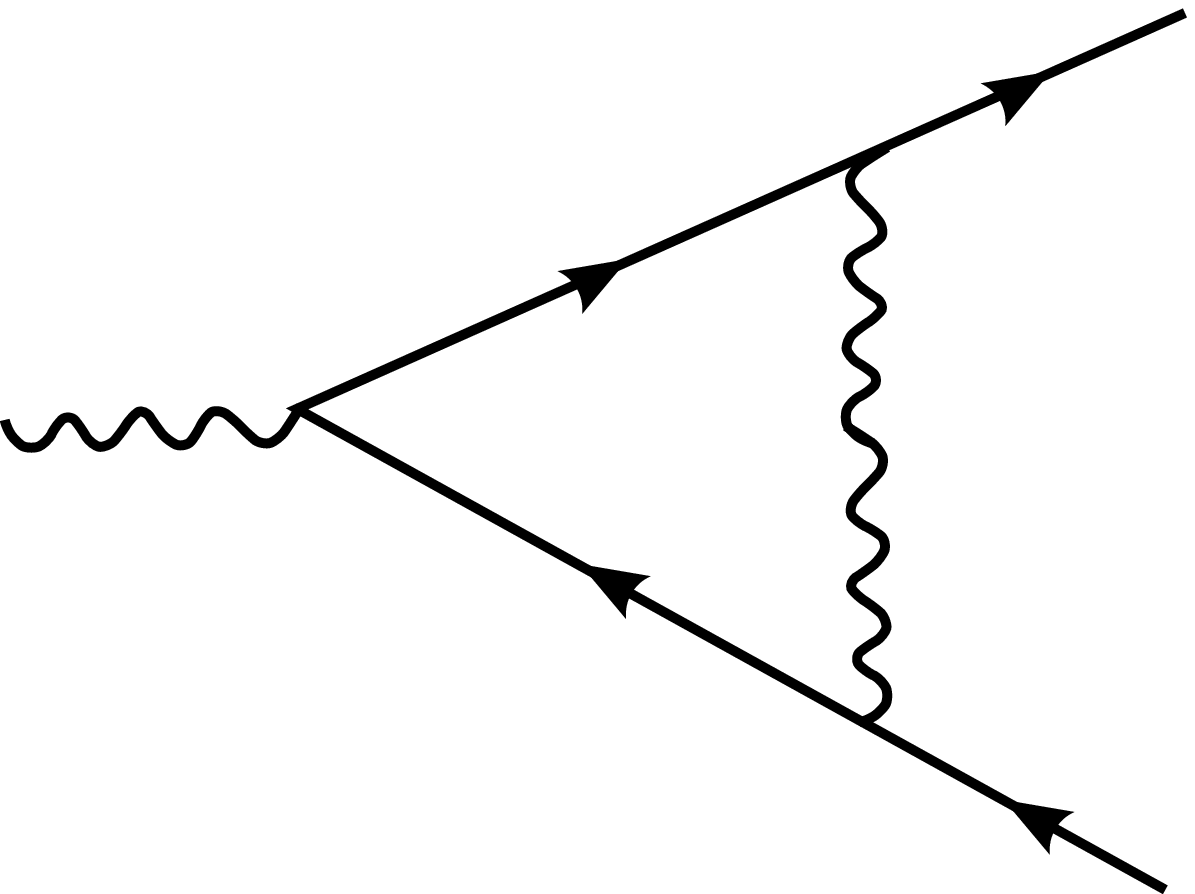}
 \caption{}
 \label{fig: CT-g}
\end{subfigure}
\hfill
\begin{subfigure}[b]{0.35\columnwidth}
 \includegraphics[width=0.95\columnwidth]{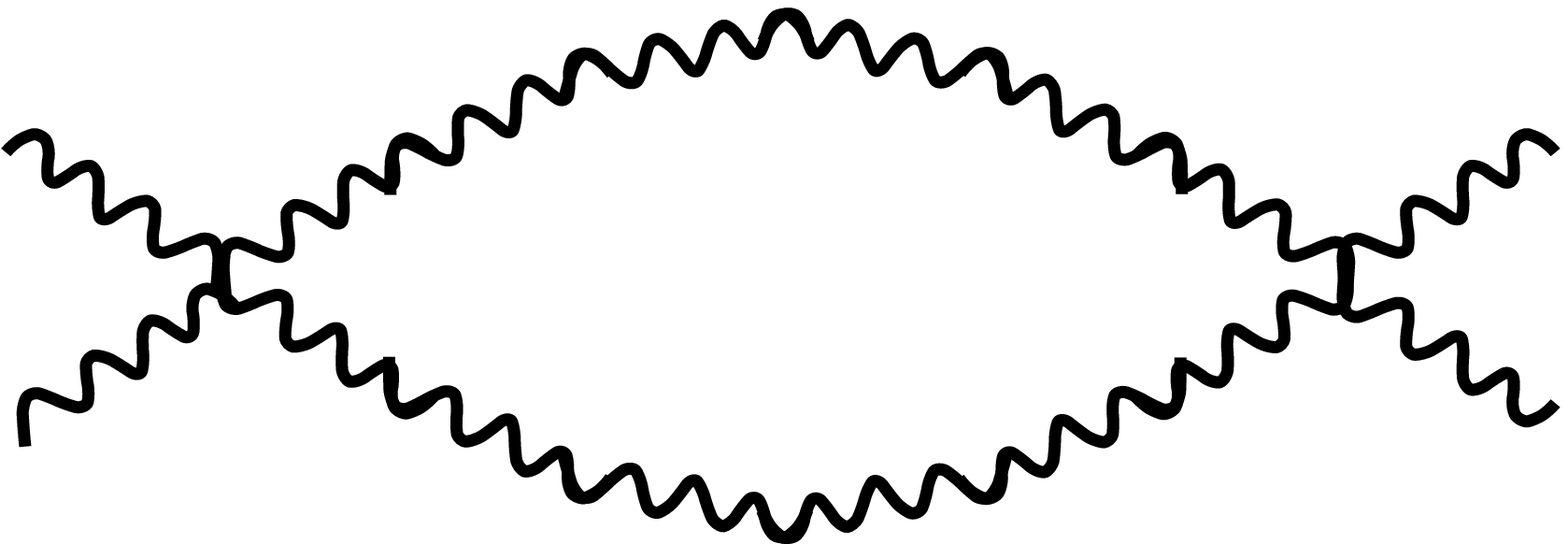}
 \caption{}
 \label{fig: CT-phi4-u}
\end{subfigure}
\hfill
\begin{subfigure}[b]{0.35\columnwidth}
 \includegraphics[width=0.6\columnwidth]{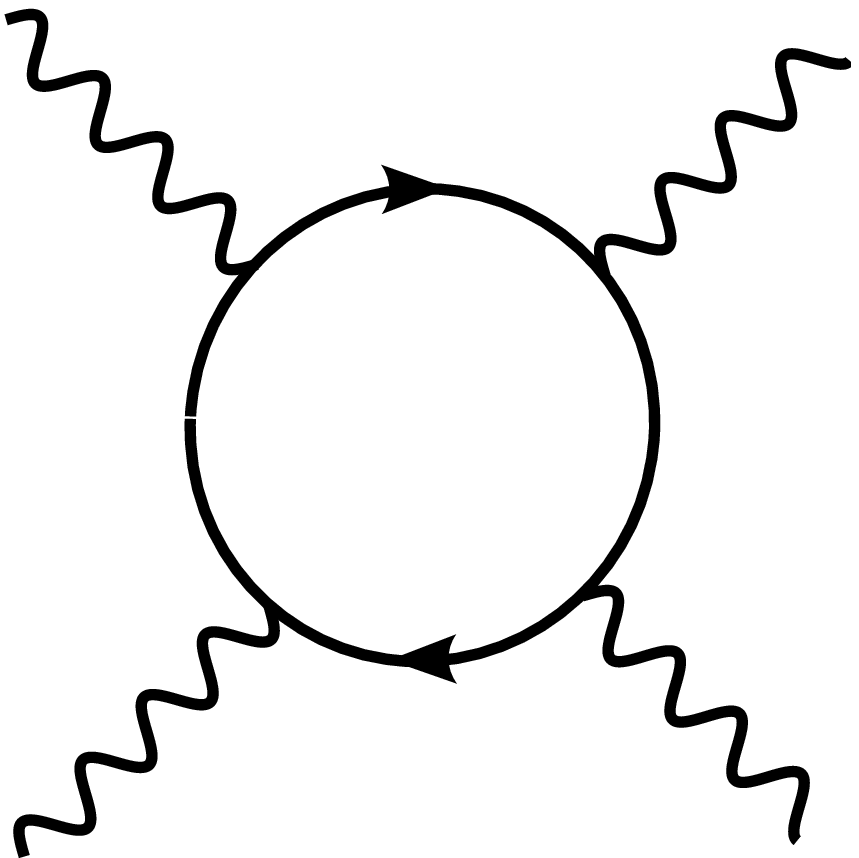}
 \caption{}
 \label{fig: CT-phi4-g1}
\end{subfigure}
\caption{
The one-loop Feynman diagrams. 
The solid (wiggly) line represents the electron (boson) propagator. 
}
\label{fig: 1L-CT}
\end{figure}

We have introduced the minimal theories for the SDW and CDW critical points. 
Despite the similarities between the two theories,
the behaviors of the two are quite different.
The difference originates from the non-abelian and abelian
nature of the Yukawa vertex  in  \eq{eq: Phi}
for the SDW and CDW theories, respectively.
In this section, we will focus on the SDW case, and 
return to the CDW case in section \ref{sec: CDW}.

\subsection{One Loop}
In this subsection we present the one-loop analysis for the SDW critical point.
From the one-loop diagrams shown in \fig{fig: 1L-CT},
we obtain the following counter terms (see Appendix \ref{app: diagrams} for details of the calculation),
\begin{align}
& Z_{1,1} = 
- \dfrac{(N_c^2 - 1)}{4 \pi^2 N_c N_f} ~ g^2  ~ h_1(v, c), \nn
& Z_{2,1} =  \dfrac{(N_c^2 - 1)}{4 \pi^2 N_c N_f} ~ g^2  ~ h_2(v, c) \nn
& Z_{3,1} =
- \dfrac{(N_c^2 - 1)}{4 \pi^2 N_c N_f} ~ g^2  ~ h_2(v, c)   \nn
& Z_{4,1} = 
- \dfrac{1}{8\pi} ~ \dfrac{g^2}{v}, \nn
&  Z_{5,1},~~ Z_{6,1} = 0,  \nn
& Z_{7,1} = 
- \dfrac{1}{8 \pi^3 N_c N_f} ~ g^2 ~ v ~ h_3(v, c),  \nn
&  Z_{8,1} = \frac{N_c^2 + 7}{2\pi^2} \chi_\sub{1} + \frac{2N_c^2 - 3}{\pi^2 N_c} \chi_\sub{2} + \frac{3(N_c^2 + 3)}{2\pi^2 N_c^2} \frac{\chi_\sub{2}^2}{\chi_\sub{1}} , \nn
& Z_{9,1} = \frac{6}{\pi^2} \chi_\sub{1} + \frac{2(N_c^2 - 9)}{2\pi^2 N_c} \chi_\sub{2} ,
\label{eq: Z-factors-SDW}
\end{align}
where
\begin{align}
& h_1(v, c) = \int_0^1 dx ~ \sqrt{\frac{1-x}{c^2 + x(1 - (1 - v^2) c^2)}}, 
\nn
& h_2(v, c) = c^2 \int_0^1 dx ~ \sqrt{\frac{1-x}{\lt[c^2 + x(1 - (1 - v^2) c^2) \rt]^3}}. \nn
& h_3(v,c) = \int_0^{2\pi} d\theta \int_0^1 dx_1 \int_0^{1-x_1} dx_2 ~ \lt[ \frac{1}{\zeta(\theta, x_1, x_2, v, c)} \rt. 
\nn & \hspace{0.5\columnwidth} 
- \lt. \frac{v^2 ~ \sin(2\theta)}{\zeta^2(\theta, x_1, x_2, v, c)}\rt]
\label{eq: h-functions}
\end{align}
with 
\begin{align}
& \zeta(\theta, x_1, x_2, v, c) = 2 v^2 [x_1 \sin^2(\theta) + x_2 \cos^2(\theta)] \nn
& \quad + (1 - x_1 - x_2)\lt[\sin^2\lt(\theta + \frac{\pi}{4}\rt) + c^2 v^2 \cos^2\lt(\theta + \frac{\pi}{4}\rt)\rt].
\label{eq: zeta}
\end{align}
From Eqs. \eqref{eq: zTau} - \eqref{eq: uvBetaChi2}, and \eq{eq: Z-factors-SDW},
we obtain the one-loop beta functions for the SDW critical point,
\begin{widetext}
\begin{align}
& \dow_\ell v = \frac{z_\tau ~ g^2}{16 \pi}   \lt[1  - \frac{4(N_c^2 - 1)}{\pi N_c N_f} ~ v~ (h_1(v,c) + h_2(v,c)) \rt],
\label{eq: 1L-Beta-v} \\
& \dow_\ell c = - \frac{z_\tau g^2 c}{16 \pi v} \lt[ 1 - \frac{4(N_c^2 - 1)}{\pi N_c N_f} ~ v~ (h_1(v,c) - h_2(v,c)) \rt],
\label{eq: 1L-Beta-c} \\
& \dow_\ell g = z_\tau ~ g \lt[ \frac{\eps}{2} - \frac{g^2}{32\pi v} 
\lt( 1 + \frac{4(N_c^2 - 1)}{\pi N_c N_f} ~ v~ (h_1(v,c)  
+ h_2(v,c)) \rt) + \frac{ g^2 v ~ h_3(v,c)}{8 \pi^3 N_c N_f}  \rt],
\label{eq: 1L-Beta-g} \\
& \dow_\ell \chi_\sub{1} = z_\tau ~ \chi_\sub{1} \lt[ \lt( \eps - \frac{g^2}{8\pi v} \rt) 
- \lt( \frac{N_c^2 + 7}{2\pi^2} \chi_\sub{1} + \frac{2N_c^2 - 3}{\pi^2 N_c} \chi_\sub{2} \rt. \rt. 
+ \lt. \lt. \frac{3(N_c^2 + 3)}{2\pi^2 N_c^2} \frac{\chi_\sub{2}^2}{\chi_\sub{1}} \rt) \rt],
\label{eq: 1L-Beta-chi1} \\
& \dow_\ell \chi_\sub{2} = z_\tau ~ \chi_\sub{2} \lt[ \lt( \eps - \frac{g^2}{8\pi v} \rt) 
- \lt(\frac{6}{\pi^2} \chi_\sub{1} + \frac{N_c^2 - 9}{\pi^2 N_c} \chi_\sub{2} \rt) \rt].
\label{eq: 1L-Beta-chi2}
\end{align}
\end{widetext}

\begin{figure}[!]
\centering
\includegraphics[width=0.5\columnwidth]{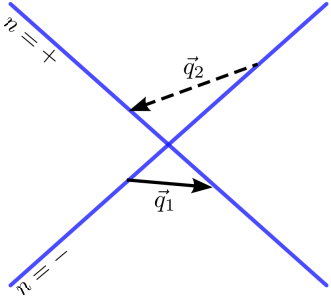}
\caption{
For any momentum $\vec q=(q_x,q_y)$ on the xy-plane, 
one can always find particle-hole pairs with zero energy 
across two patches of Fermi surface near hot spots. 
Since the fermionic dispersion is linear, 
the spectrum of particle-hole pair is independent of $\vec q$.
}
\label{fig: p-h}
\end{figure}
Now we explain the physical origin of each term in the beta functions based on the results obtained in Appendix \ref{app: diagrams}.
The boson self energy in Fig. \ref{fig: CT-SEb} is proportional to $|\mbf Q|^2$ and independent of $\vec q = (q_x, q_y)$.
This is because a boson with any $\vec q$ can be absorbed by a particle-hole pair on the Fermi surface (see \fig{fig: p-h}), 
and the energy spectrum of particle-hole excitations is independent of $\vec q$. 
Vanishing $Z_{5,1}$ and $Z_{6,1}$ at the one-loop order, along with the negative sign of $Z_{4,1}$ (the counter term and 
the quantum correction generated by integrating out high energy modes in the Wilsonian RG have opposite signs), leads to a weakened dependence of the dressed boson propagator on $q_x, q_y$ relative to that of $\mbf Q$.
As a result, $c_x$ and $c_y$ are renormalized to smaller values. 
Because $v = v_x / c_x$ and $c = c_y / v_y$, as defined in Eqs. \eqref{eq: original-action} and \eqref{eq: dw-action-2d}, the suppression of $c_x$ and $c_y$ enhances $v$ and suppresses $c$.
This is shown in the first terms on the right hand side of Eqs. \eqref{eq: 1L-Beta-v} and \eqref{eq: 1L-Beta-c}.
The $\mbf K$ dependent term ($Z_{1,1}$) in the fermion self energy in \fig{fig: CT-SEf}  
similarly reduces $v_x$ and $v_y$.
This reduces $v$ and enhance $c$ as is shown in the second terms ($\propto h_1$) on the right hand side of Eqs. \eqref{eq: 1L-Beta-v} and \eqref{eq: 1L-Beta-c}.
\fig{fig: CT-SEf} also directly renormalizes the Fermi velocity through $Z_{2,1}$, $Z_{3,1}$.
The spin fluctuations mix electrons from different hot spots. 
This reduces the angle between the Fermi velocities at the hot spots connected by $\vec Q_{ord}$,
thereby improving the nesting between the hot spots.
As $v_x$ and $v_y$ are renormalized to smaller and larger values respectively, 
$v$ and $c$ are suppressed. 
This is embodied in the third terms ($\propto h_2$) in the expressions for $\dow_\ell v$ and $\dow_\ell c$.

\begin{figure}[!]
\centering
\includegraphics[width=0.99\columnwidth]
{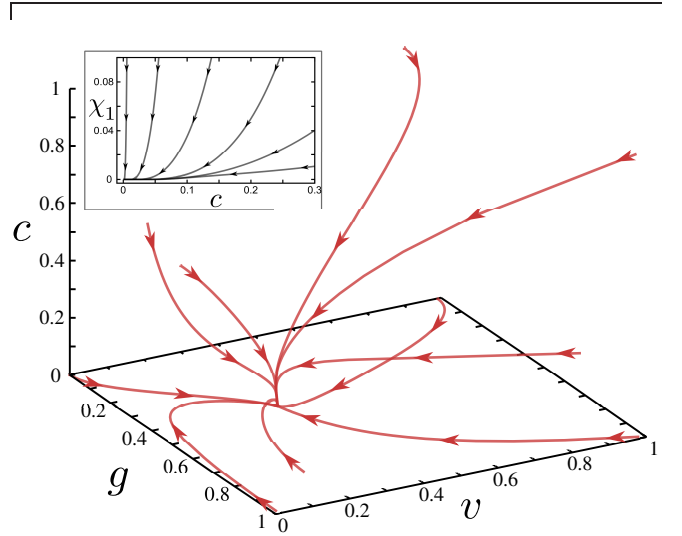}
\caption{
Projection of the RG flow in the space of $(g, v, c)$ for $N_c = 2$, $N_f = 1$ and $\eps = 0.01$. 
The arrows point towards decreasing energy. 
An IR fixed point given by \eq{eq: P*} exists on the $(g,v)$ plane with $c=0$.
(Inset) Projection of the RG flow in the $(\chi_\sub{1}, c)$ plane. 
$c$ ($\chi_\sub{1}$) flows to zero logarithmically (algebraically). 
}
\label{fig: flow-1L}
\end{figure}

The beta function for the Yukawa vertex includes two different contributions.
The second ($\propto 1$), third ($\propto h_1$) and fourth ($\propto h_2$) terms on the right hand side of \eq{eq: 1L-Beta-g} 
are the contributions from the boson and fermion self energies which alter the scaling dimensions of spacetime and the fields.
The contributions from the self-energies weaken the interaction at low energies 
because the virtual excitations in Figs. \ref{fig: CT-SEf}, \ref{fig: CT-SEb} screens the interaction.
This is reflected in the negative contributions to the beta function.
The last term ($\propto h_3$) in \eq{eq: 1L-Beta-g} is the vertex correction ($Z_{7,1}$) shown in \fig{fig: CT-g}.
Unlike the contributions from the self-energies, the vertex correction anti-screens the interaction,
which tends to make the interaction stronger.
The anti-screening is attributed to the fact that the SDW vertices anti-commute on average in the sense,
\begin{align}
\sum_{a = 1}^{N_c^2 - 1} \tau^{a} \tau^{b} \tau^{a} = - \dfrac{2}{N_c} ~ \tau^{b}.
\label{eq:ttt}
\end{align}
This is analogous to the anti-screening effect which results in the asymptotic freedom in non-abelian gauge theories.
The anti-screening effect also has a significant impact at the two-loop order as will be discussed in Sec. \ref{sec: beyond1L}.

The beta functions for $\chi_i$ can be understood similarly.
The second terms in the Eqs. \eqref{eq: 1L-Beta-chi1} and \eqref{eq: 1L-Beta-chi2} are the contributions from the boson self-energy.
Rest of the terms in these equations are the standard vertex corrections ($Z_{8,1}$, $Z_{9,1}$) from \fig{fig: CT-phi4-u}.
We note that \fig{fig: CT-phi4-g1} does not contribute to the beta functions, 
because it is UV finite at $d=3$ \cite{Sur2}.

In \fig{fig: flow-1L} we plot the one-loop RG flow of the four parameters, $(g, v, c, \chi_\sub{1} = u_1/c)$ for $N_c = 2$ and $N_f = 1$. 
Here we set $u_2=0$.
The RG flow shows the presence of a stable IR fixed point with vanishing $c$ and $\chi_\sub{1}$.
In order to find the analytic expression of the couplings at the fixed point for general $N_c$ and $N_f$, 
we expand $h_i(v,c)$ to the linear order in $c$ with $v \sim 1$,
\begin{align}
& h_1(v,c) = \frac{\pi}{2} - 2c + \ordr{c^2}, ~~~
h_2(v,c) = 2c + \ordr{c^2}, \nn
& h_3(v,c) = \frac{2\pi^2}{v(1+v)} - \frac{4\pi}{v} c + \ordr{c^2}.
\label{eq: SmallC}  
\end{align}
In the small $c$ limit, the beta functions become 
\begin{widetext}~~~
\begin{align}
\dow_\ell v &= \frac{z_\tau}{16 \pi}~ g^2  \lt[1 -  \frac{2(N_c^2 - 1)}{N_c N_f} ~ v \rt],
\label{eq: 1L-Beta-v-SmallC} \\
\dow_\ell c &= - \frac{z_\tau}{16 \pi}~ 
\frac{g^2 c}{v} \lt[\lt(1 
- \frac{2(N_c^2 - 1)}{N_c N_f} ~ v \rt) 
+ \frac{16 (N_c^2 - 1)}{\pi N_c N_f}~ v ~ c \rt],
\label{eq: 1L-Beta-c--SmallC} \\
\dow_\ell g &= \half z_\tau ~ g \lt[ \eps - \frac{g^2}{16\pi v} 
\lt\{ 1 + \frac{2(N_c^2 - 1)}{N_c N_f} ~ v - \frac{8 v}{N_c N_f (1+v)} \rt\} \rt],
\label{eq: 1L-Beta-g-SmallC} \\
\dow_\ell \chi_\sub{1} &= z_\tau ~ \chi_\sub{1} \lt[ \lt( \eps - \frac{g^2}{8\pi v} \rt) 
- \lt( \frac{N_c^2 + 7}{2\pi^2} \chi_\sub{1} + \frac{2N_c^2 - 3}{\pi^2 N_c} \chi_\sub{2} + \frac{3(N_c^2 + 3)}{2\pi^2 N_c^2} \frac{\chi_\sub{2}^2}{\chi_\sub{1}} \rt) \rt],
\label{eq: 1L-Beta-chi1-SmallC} \\
\dow_\ell \chi_\sub{2} &= z_\tau ~ \chi_\sub{2} \lt[ 
\lt( \eps - \frac{g^2}{8\pi v} \rt) 
- \lt(\frac{6}{\pi^2} \chi_\sub{1} + \frac{N_c^2 - 9}{\pi^2 N_c} \chi_\sub{2} \rt) \rt].
\label{eq: 1L-Beta-chi2-SmallC}
\end{align}
\end{widetext}
Although the anti-screening vertex correction 
(the last term in \eq{eq: 1L-Beta-g-SmallC})
tends to enhance the coupling, 
the screening from the self-energies is dominant for any $N_c \geq 2$.
As a result, $g$ is stabilized at a finite value below three dimensions.
The stable one-loop fixed point is given by
\begin{align}
& v_* = \dfrac{N_c N_f}{2(N_c^2 - 1)}, \nn
&  g_*^2 = \dfrac{4\pi  N_c N_f}{(N_c^2 - 1)} ~ \aleph(N_c, N_f)~  \eps,  \nn
& c_*  = 0, \nn
& \chi_{\,\dsty{\!_{1*}}} =  \chi_{\,\dsty{\!_{2*}}}  = 0,
\label{eq: P*}
\end{align}
where
\begin{align}
\aleph(N_c, N_f) = \dfrac{2(N_c^2 - 1) + N_c N_f}{2(N_c^2 - 3) + N_c N_f}.
\label{eq: N}
\end{align}
At the one-loop order,
the dynamical critical exponent 
$z_\tau = 1 + \frac{\aleph(N_c, N_f)}{2} \eps$
becomes greater than one, 
while $z_x$ retains its classical value.
However, $z_x$ deviates from one at the two-loop order as will be shown later.
It is remarkable that 
the quantum scaling dimensions of the quartic vertices 
$\lt( \eps - {g^2}/{8\pi v} \rt)$ become negative at the fixed point,
resulting in their irrelevance even below three dimensions. 
This is due to the fact that 
the effective spacetime dimension, $d_{eff} = (2-\eps)z_\tau + z_x + 1$, at the one-loop fixed point is greater than $d = 4 - \eps$ of the classical theory.
In this sense the upper critical dimension  for the quartic vertices 
is pushed down below $3-\epsilon$  at the interacting fixed point \cite{Hertz,Millis}.

\begin{figure}[!]
\centering
\begin{subfigure}{0.8\columnwidth}
\includegraphics[width=.98\columnwidth]
{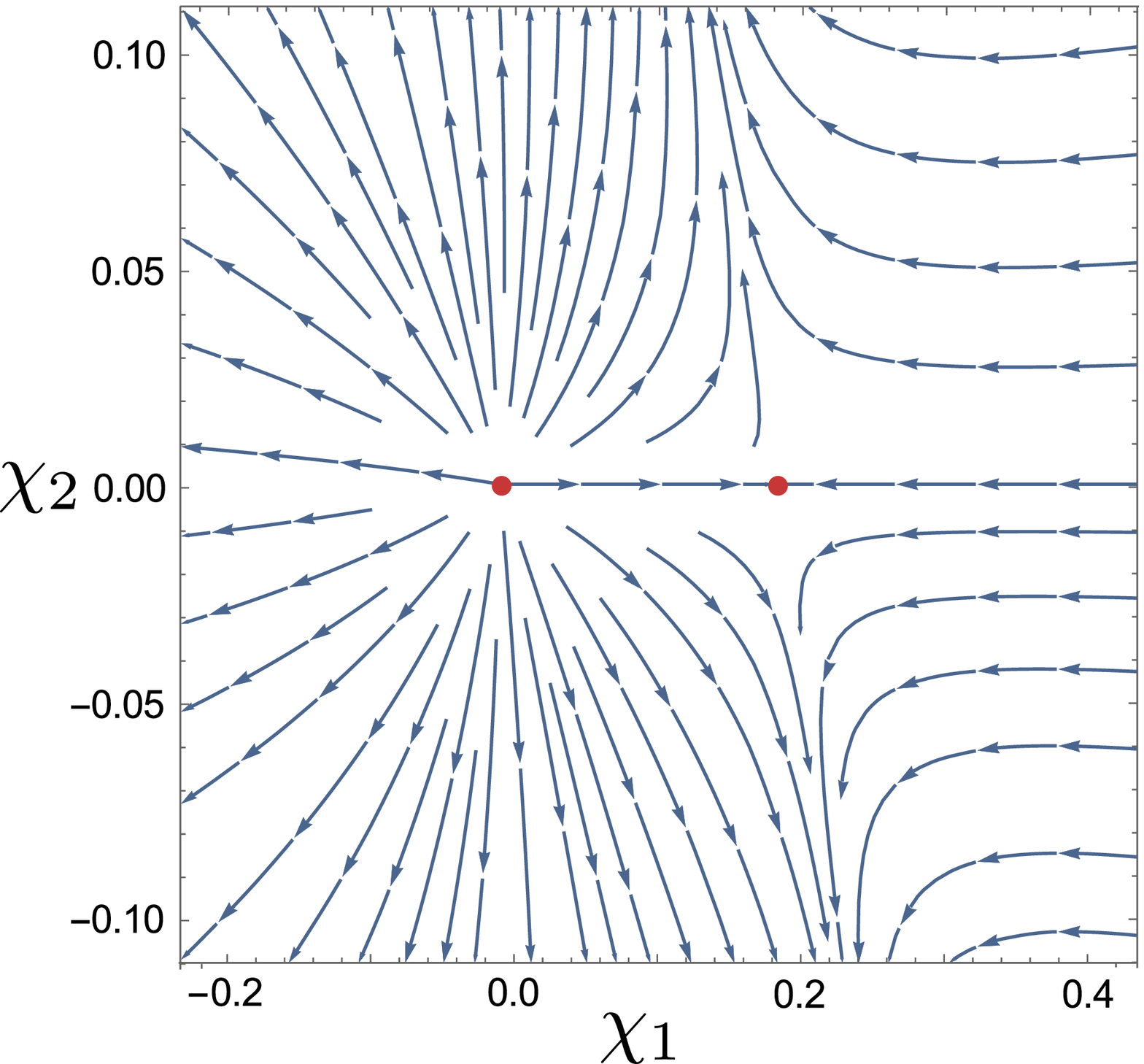}
\caption{}
\label{fig: chi1-chi2-g0}
\end{subfigure}
\hfill
\begin{subfigure}{0.8\columnwidth}
\includegraphics[width=.98\columnwidth]
{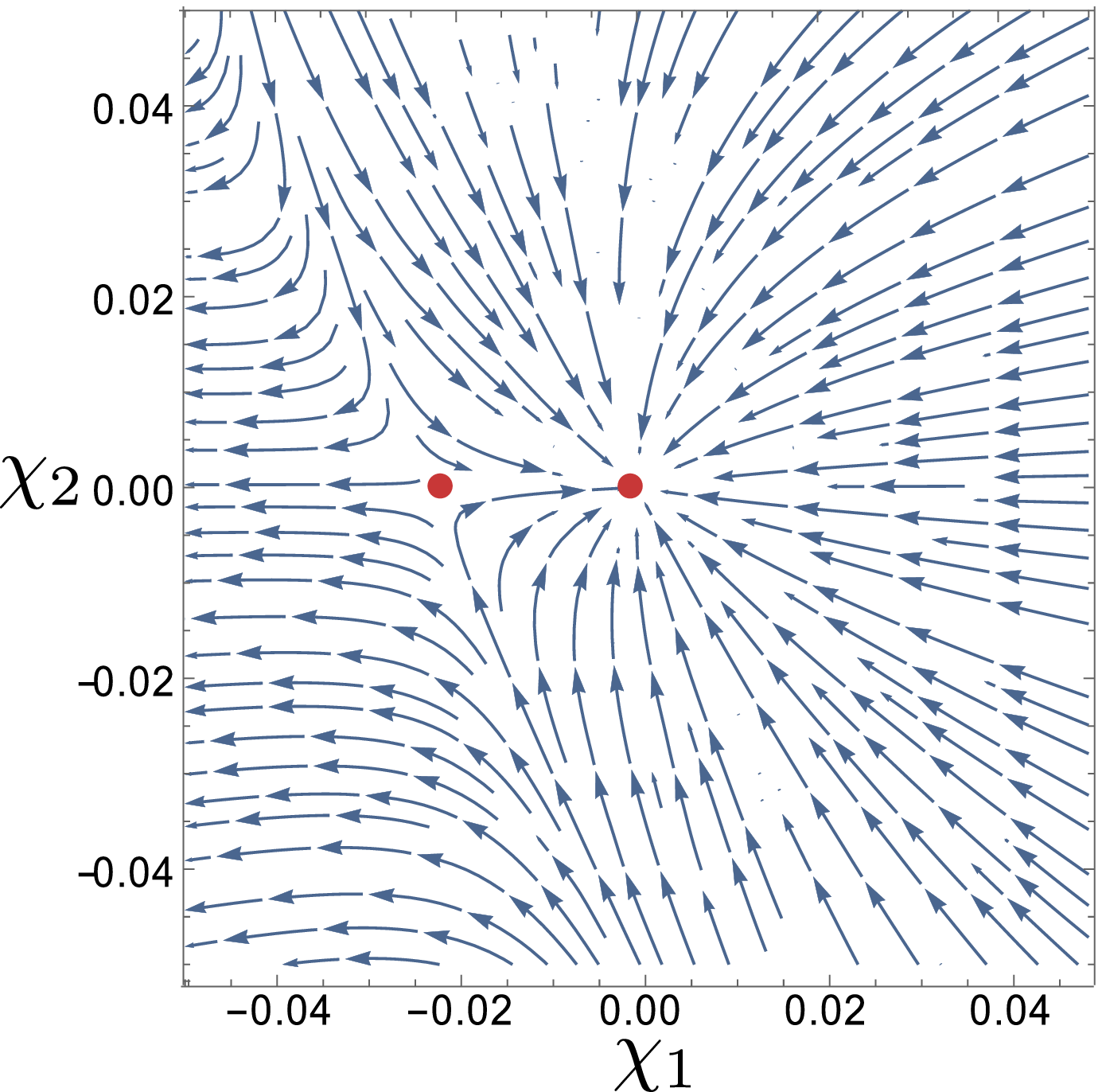}
\caption{}
\label{fig: chi1-chi2-g}
\end{subfigure}
\caption{
RG Flow in the $\chi_\sub{1} - \chi_\sub{2}$ plane with $N_c = 4, N_f = 1$, and $\eps = 0.2$. 
(a) In the subspace of $g = 0$, 
there are two unstable fixed points at $\chi_\sub{1} = 0$ and $\chi_\sub{1} \sim \eps$ 
with $\chi_\sub{2}=0$. 
The former is the Gaussian fixed point, while the latter is the Wilson-Fisher fixed point.
(b) In the subspace of $g = g_*$ and $v = v_*$, 
there exist an unstable non-Fermi liquid at $\chi_\sub{1} \sim - \eps$, 
and a stable non-Fermi liquid at $\chi_\sub{1} = 0$ with $\chi_\sub{2}=0$. 
The arrows in both plots point towards increasing length scale, and the (red) dots indicate the fixed points.
}
\label{fig: chi1-chi2}
\end{figure}

\begin{table*}[!]
\centering
\begin{tabular}{c || c | c | c | c | c}
\# & $\chi_\sub{1}$ & ~~~$\chi_\sub{2}$ ~~~& ~~~$g$~~~ & Relevant deformation & State \\
\hline &&&&& \\[-2ex]
\rom{1} & 0 & 0 & 0 & $g$, $\chi_\sub{1}$, $\chi_\sub{2}$ & Free fermion and boson\\
\hline &&&&& \\[-2ex]
\rom{2} & $\dfrac{2\pi^2 \eps}{N_c^2 + 7}$ & $0$ & $0$ & $g$, $\chi_\sub{2}$ & Free fermion +  Wilson-Fisher  \\[2ex]
\hline &&&&& \\[-2ex]
\rom{3} & $ -\dfrac{2\pi^2 (\aleph(N_c, N_f)-1) \eps}{N_c^2 + 7}$ & 0 & $g_*$ & $\chi_\sub{1}$ & Unstable non-Fermi liquid \\[2ex]
\hline &&&&& \\[-2ex]
\rom{4} & $0$ & $0$ & $g_*$ & None & Stable non-Fermi liquid \\[2ex]
\hline
\end{tabular}
\caption{
The four fixed points in the three dimensional space of couplings. 
$g_*$ is defined in \eq{eq: P*}.
The penultimate column lists the couplings that need to be tuned to reach the fixed point
(besides the mass that has been tuned to reach the critical point).
}
\label{tab: FP}
\end{table*}
We note that \eq{eq: P*} is the fixed point of the full beta functions in Eqs. \eqref{eq: 1L-Beta-v} - \eqref{eq: 1L-Beta-chi2},  
because the truncation of higher order terms in \eq{eq: SmallC} becomes exact in the small $c$ limit.
Besides the Gaussian and stable non-Fermi liquid fixed points,  
there exist two unstable interacting fixed points as listed in Table \ref{tab: FP}.
At the fixed point \rom{2}, the fermions are decoupled from the bosons, 
and the dynamics of the boson is controlled by the Wilson-Fisher fixed point.
Here $g$ and $\chi_\sub{2}$ are relevant perturbations.
The other fixed point (\rom{3}) is realized at $(\chi_\sub{1}, \chi_\sub{2}) = \lt( -\frac{2\pi^2 (\aleph(N_c, N_f)-1)}{N_c^2 + 7}  ~ \eps,0 \rt)$  
with the same values of $g, c, v$ as in \eq{eq: P*}.
A deviation of $\chi_1$ from \rom{3} is the relevant perturbation,  
which takes the flow either towards the stable fixed point (\rom{4}) at the origin of $(\chi_1, \chi_2)$-plane, 
or towards strong coupling where $\chi_\sub{1}$ becomes large and negative.
The full RG flow in the $(\chi_1, \chi_2)$-plane at fixed $g$ and $v$ is shown in Fig. \ref{fig: chi1-chi2}.

We now focus on the stable fixed point (\rom{4}), which is realized at the critical point without further fine tuning.
At the fixed point the electron and boson propagators satisfy the scaling forms, 
\begin{align}
& G_{l,m}(k) = \frac{1}{|k_y|^{1 - 2 \wtil{\eta}_\psi}} ~ 
\mc{G}_{l,m}\lt(\frac{k_x}{|k_y|^{z_x}}, \frac{\mbf{K}}{|k_y|^{z_\tau}} \rt), \\
& D(q) = \frac{1}{|q_y|^{2 - 2\wtil{\eta}_\phi}} ~  \mc{D}\lt(\frac{q_x}{|q_y|^{z_x}}, \frac{\mbf{Q}}{|q_y|^{z_\tau}} \rt).
\end{align}
The anomalous dimensions that dictate the scaling of the two-point functions are deduced from \eq{eq: RGE}, and they 
are given by combinations of $\eta_\psi$, $\eta_\phi$, $z_\tau$ and $z_x$.
\begin{align}
\wtil{\eta}_\psi &= \frac{2 z_\tau + z_x - 3}{2} + \eta_\psi, \nn
\wtil{\eta}_\phi &= \frac{2 z_\tau + z_x - 3}{2} + \eta_\phi.
\end{align}
At the one-loop order, $\wtil{\eta}_\psi = 0$ and $\wtil{\eta}_\phi = 0$.
$\mc{G}_{l,m}(x,y)$ and $\mc{D}(x,y)$ are universal functions of the dimensionless ratios of momentum and frequency.
Due to the non-trivial dynamical critical exponent, 
the single-particle excitations are not well-defined, and the electrons near the hot spots become non-Fermi liquid below three dimensions.
Since at one-loop order the critical exponents are solely determined by the Yukawa coupling and the velocities, the unstable fixed point \rom{3} is also a non-Fermi liquid.

The velocity $c$ which measures the boson velocity along the direction of the ordering vector with respect to the Fermi velocity flows to zero logarithmically.
The vanishing velocity leads to enhanced fluctuations of the collective mode at low energies,
which can make higher-loop corrections bigger than naively expected.
This can, in principle, pose a serious threat to a controlled expansion.
In order to see whether the perturbative expansion is controlled beyond one-loop,
one first needs to understand how higher-loop corrections change the flow of  $c$.
In the following two sub-sections, 
we show that $c$ flows to a non-zero value 
which is order of $\eps^{1/3}$ due to a two-loop correction,
and the perturbative expansion is controlled.


\subsection{Beyond One Loop} \label{sec: beyond1L}

\subsubsection{Estimation of general diagrams }
\label{sec:estimation}

The vanishing boson velocity at the one-loop fixed point can enhance higher order diagrams 
which are nominally suppressed by the small coupling $g^2 \sim \epsilon$.
A $L$-loop diagram with $V_g$ Yukawa vertices and $V_u$ quartic vertices takes the form of
\begin{align}
& F(p_i; v, c, g, \chi; \eps, V_g, V_u, L) \propto   g^{V_g} \chi^{V_u} c^{V_u} \nn
& ~~\times \int \left[\prod_{i=1}^L d p'_i \right]
\prod_{l=1}^{I_f} \left( 
\frac{1}{ \mathbf{\Gamma} \cdot \mathbf{K}_l  
+  \gamma_{d-1} \left[ v k_{l,x} + n_l k_{l,y} \right]   }
\right) \nn 
& \qquad \times \prod_{m=1}^{I_b} \left(
\frac{1}{| \mathbf{Q}_m |^2 + q_{m,x}^2 + c^2  q_{m,y}^2 } \right).
\end{align}
Here $p_i$ ($p'_i$) are external (internal) momenta, 
and $k_i$ and $q_i$ are linear combinations of $p_i$ and $p'_i$.
$\chi$ represents either $\chi_\sub{1}$ or $\chi_\sub{2}$, whose difference is not important for the current purpose.
$I_f$ and $I_b$ are the numbers of the internal electron and boson propagators, respectively.
$n_l$ is either $+$ or $-$ depending on the hot spot index carried by the $l$-th electron propagator. 

When $c$ is zero, some loop integrations can diverge 
as the dependence on $q_y$ drops out in the boson propagator.
This happens in the \textit{bosonic loops}, which are solely made of boson propagators.
For example, $y$-component of the internal momentum in \fig{fig: CT-phi4-u} is unbounded at $c=0$.
For a small but nonzero $c$, the UV divergence is cut-off at a scale proportional to $1/c$.
As a result, the diagram is enhanced by $1/c$.
This is why $Z_{8,1}$ and $Z_{9,1}$ in \eq{eq: Z-factors-SDW} is order of $\chi$ not $\chi c$.

\begin{figure}[!]
\centering
\includegraphics[width=0.8\columnwidth]{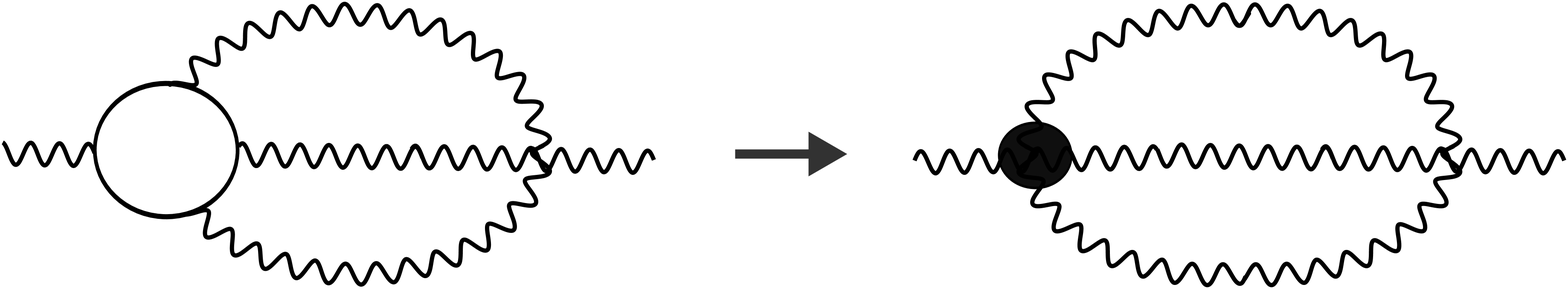}
\caption{
A three-loop diagram which can be enhanced by upto $1/c^2$.
This diagram does not include any loop that is solely made of boson propagators.
Nonetheless, this can exhibit  an enhancement in powers of $1/c$ 
as the fermionic loop plays the role of a bosonic quartic vertex (represented by shaded circles),
which is not suppressed at large momentum.
}
\label{fig: 2L-u-b}
\end{figure}

Such enhancement can also arise if bosonic loops are formed out of dressed vertices and dressed propagators.
Let us first consider the case with dressed vertices.
Superficially, the diagram in \fig{fig: 2L-u-b} does not have any boson loop.
However, the fermion loop can be regarded as a quartic boson vertex 
which is a part of a bosonic loop. 
Since the quartic vertex is 
dimensionless at the tree-level in $3$ dimensions, 
it is not suppressed at large momentum.
Therefore, the diagram can exhibit an enhancement of $1/c^2$
as the boson propagators lose dispersion in the small $c$ limit.

\begin{figure}[!]
\centering
\begin{subfigure}[b]{0.3\columnwidth}
\includegraphics[width=0.9\columnwidth]{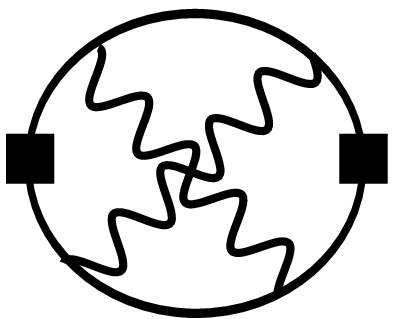}
\caption{}
\label{fig: bSE-C}
\end{subfigure}
\hfill
\begin{subfigure}[b]{0.35\columnwidth}
\includegraphics[width=0.9\columnwidth]{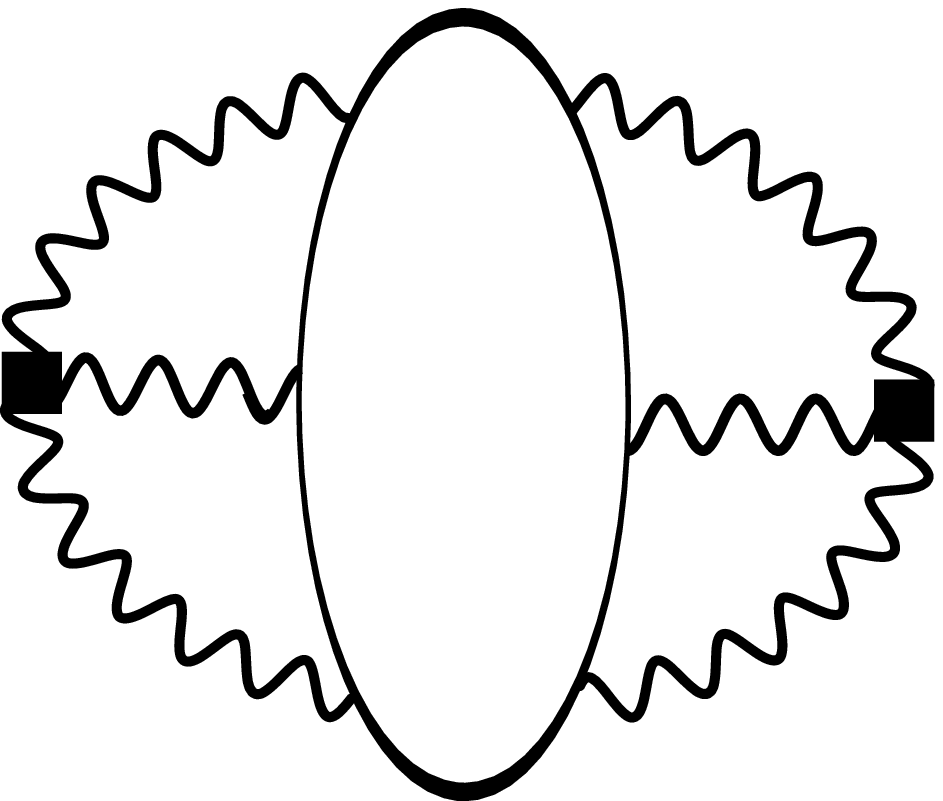}
\caption{}
\label{fig: bSE-D}
\end{subfigure}
\caption{
Examples of (amputated) boson self-energy which can potentially
diverge with order of $1$ coefficient 
in the large $q_y$ limit. 
}
\label{fig: N-Pi1}
\end{figure}

\begin{figure}[!]
\centering
\begin{subfigure}[b]{0.35\columnwidth}
\includegraphics[width=0.9\columnwidth]{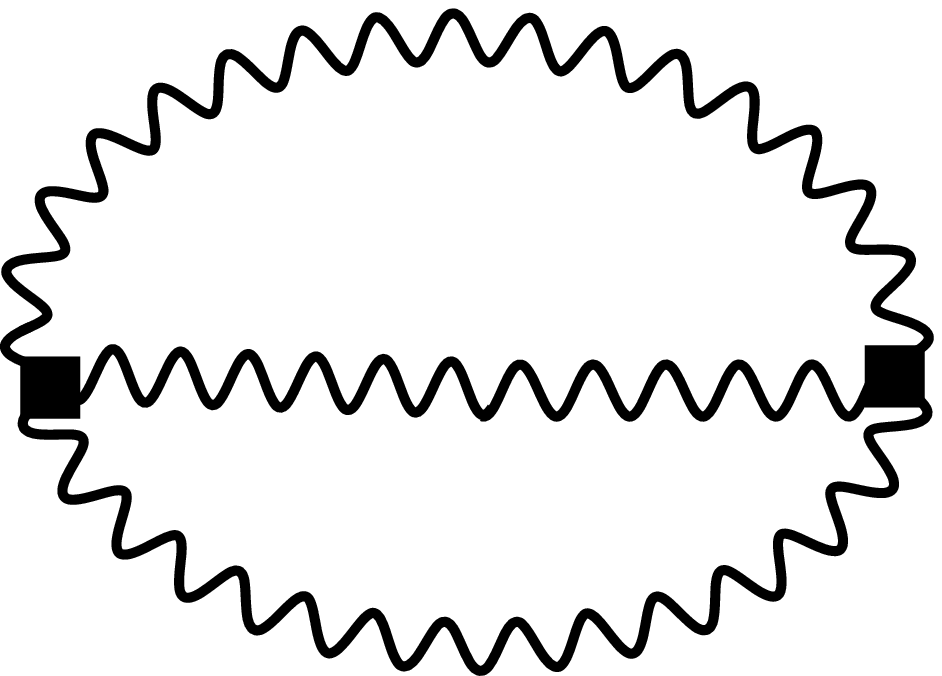}
\caption{}
\label{fig: bSE-A}
\end{subfigure}
\hfill
\begin{subfigure}[b]{0.55\columnwidth}
\includegraphics[width=0.9\columnwidth]{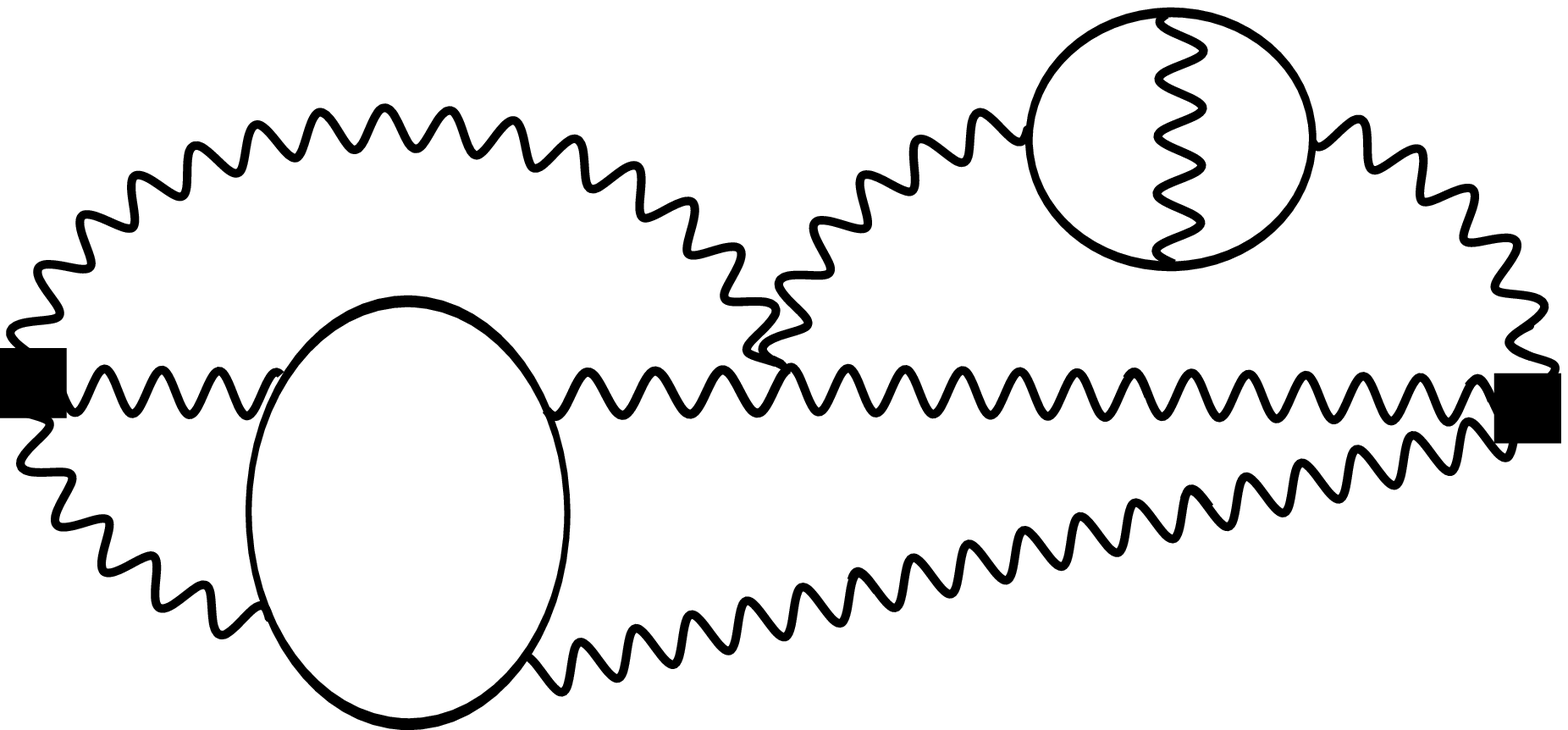}
\caption{}
\label{fig: bSE-B}
\end{subfigure}
\caption{
Examples of (amputated) boson self-energy diagrams whose dependences on $q_y$ are suppressed by $c$. 
This can be seen from that fact that the external momentum can be directed to go through only boson propagators  
which are independent of $y$-momentum in the small $c$ limit.
Consequently, the self-energy depends on $q_y$ only through $c q_y$.
}
\label{fig: N-Pi2}
\end{figure}

Similarly, boson loops made of dressed boson propagators can exhibit enhancements.
However, the situation is a bit more complicated in this case.
Since boson self-energy has scaling dimension $2$,
it can diverge quadratically in the momentum that flows through the self-energy.
Therefore, there can be an additional enhancement of $1/c^2$
in the small $c$ limit 
because the typical $y$-component of internal momentum is order of $1/c$ in the boson loops. 
In order to account for the additional enhancements 
from the boson self-energy more precisely, 
it is convenient to divide diagrams for boson self-energy into two groups.
The first group includes those diagrams  which  diverge  
in the large $q_y$ limit with order of $1$ coefficient
as $c$ goes to zero.
Potentially, the diagrams in \fig{fig: N-Pi1} have un-suppressed dependence on $q_y$  
because the external momentum must go through at least one fermion propagator
whose dispersion is not suppressed  in the small $c$ limit.
Each boson self-energy of the first kind in bosonic loops contributes a factor of atmost $1/c^2$. 
The second group includes those diagrams that are either independent of $q_y$ for any $c$, 
or  become independent of $q_y$ as $c$ goes to zero.
For example, the one-loop self-energy in \fig{fig: CT-SEb} is independent of $q_y$.
The diagrams in \fig{fig: N-Pi2} depend on $q_y$ through the combination  $c q_y$ 
because the external momentum can be directed to go through only boson propagators 
which are independent of $y$-momentum in the small $c$ limit.  
Therefore, the self-energies in the second group do not contribute an additional enhancement of $1/c$.

Owing to the aforementioned reasons a general diagram can be enhanced \textit{at most} by a factor of $c^{-L_b - 2 N_\Pi}$,
where $L_b$ is the number of loops solely made of bosonic propagators once 
fermion loops are replaced by the corresponding effective $\phi^{2n}$ vertices, 
and $N_\Pi$ is the total number of boson self-energy of the first kind
in bosonic loops.
Therefore, we estimate the upper bound for the magnitude of general higher-loop diagrams to be
\begin{align}
&F(p_i; v, c, g, \chi; \eps, V_g, V_u, L, L_b, N_\Pi) \nn 
& =  \lt(\frac{g^2}{c}\rt)^{V_g/2} \chi^{V_u} ~c^{(E-2)/2 + (L - L_b - 2 N_\Pi)} ~ f(p_i; v, c; \eps, L), 
\label{eq: 1PI}
\end{align} 
where we have used the relation $L = (V_g + 2V_u + 2 - E)/2$ with $E$ being the number of external legs. 
The function $f(p_i; v, c; \eps, L)$ is regular in the small $c$ limit.
We emphasize that \eq{eq: 1PI} is an upper bound in the small $c$ limit.
The actual magnitudes may well be smaller by positive powers of $c$.
For example, \fig{fig: BosonSE2La} is nominally order of $g^4/c$ in the small $c$ limit according to \eq{eq: 1PI}.
However, an explicit computation shows that it is order of $g^4$.
Currently, we do not have a full expression for the actual magnitudes of general diagrams in the small $c$ limit.
Our strategy here is to use the upper bound, 
which is sufficient to show that the perturbative expansion is controlled.

\subsubsection{Two-loop correction }

\begin{figure}[!]
\centering
\begin{subfigure}[b]{0.45\columnwidth}
\includegraphics[width=0.9\columnwidth]{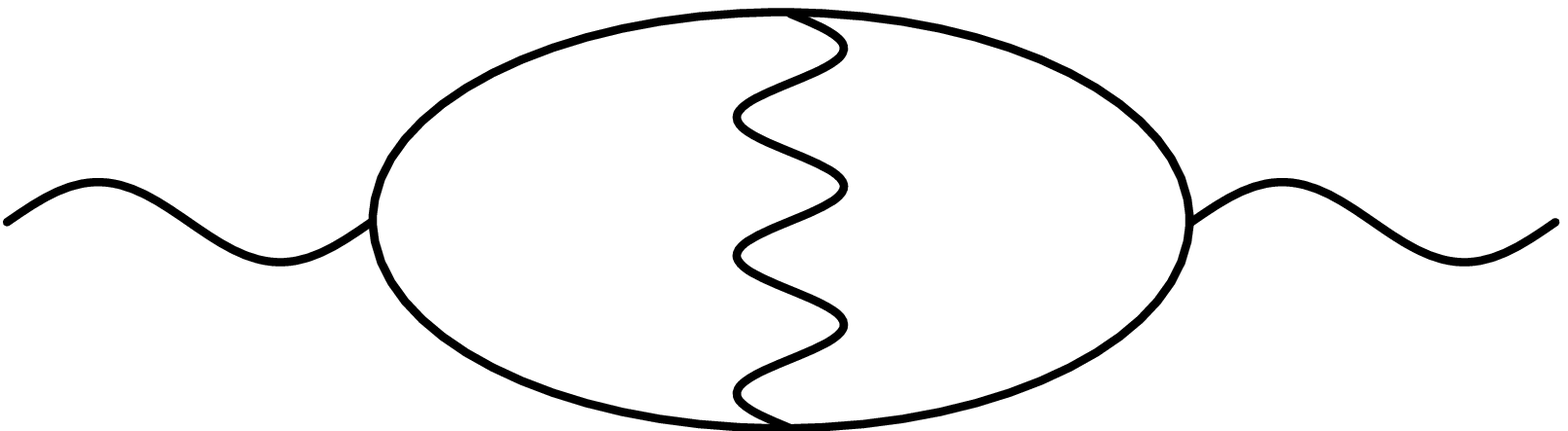}
\caption{}
\label{fig: BosonSE2La}
\end{subfigure}
\hfill
\begin{subfigure}[b]{0.45\columnwidth}
\includegraphics[width=0.9\columnwidth]{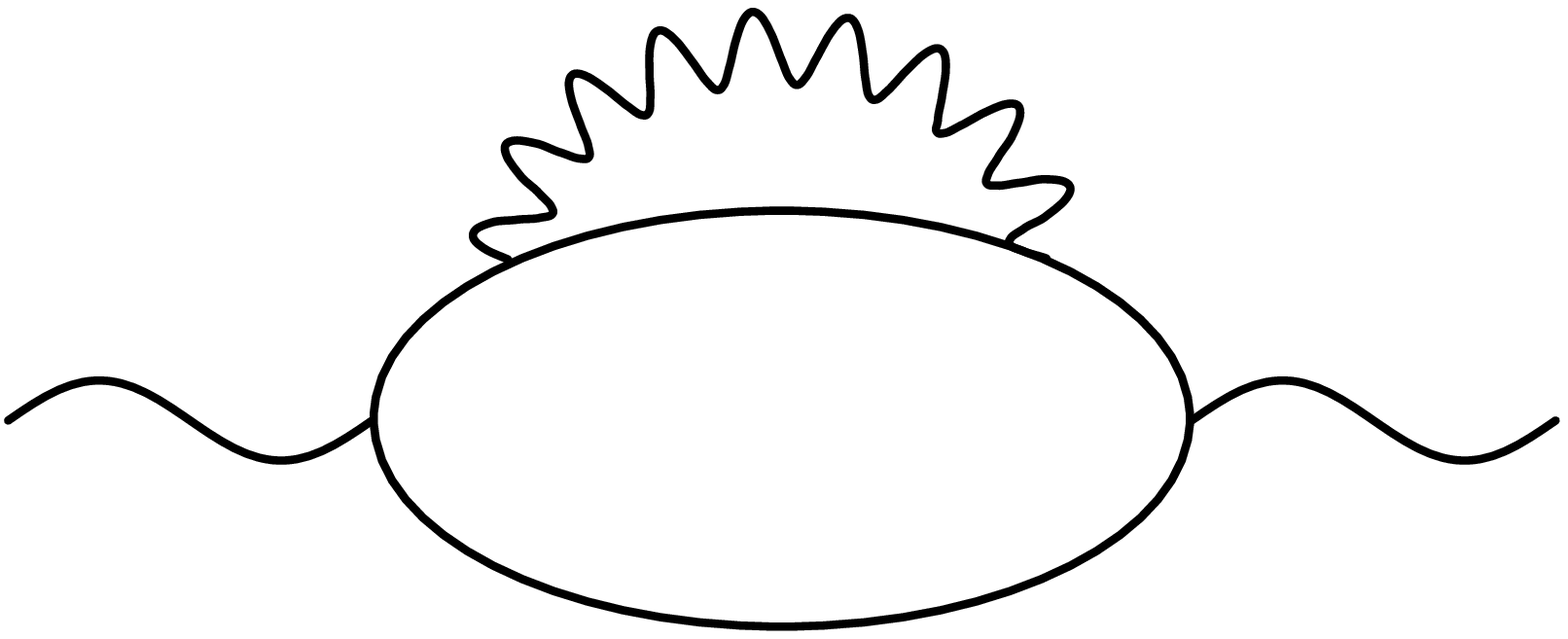}
\caption{}
\label{fig: BosonSE2Lb}
\end{subfigure}
\hfill
\begin{subfigure}[b]{0.45\columnwidth}
\includegraphics[width=0.9\columnwidth]{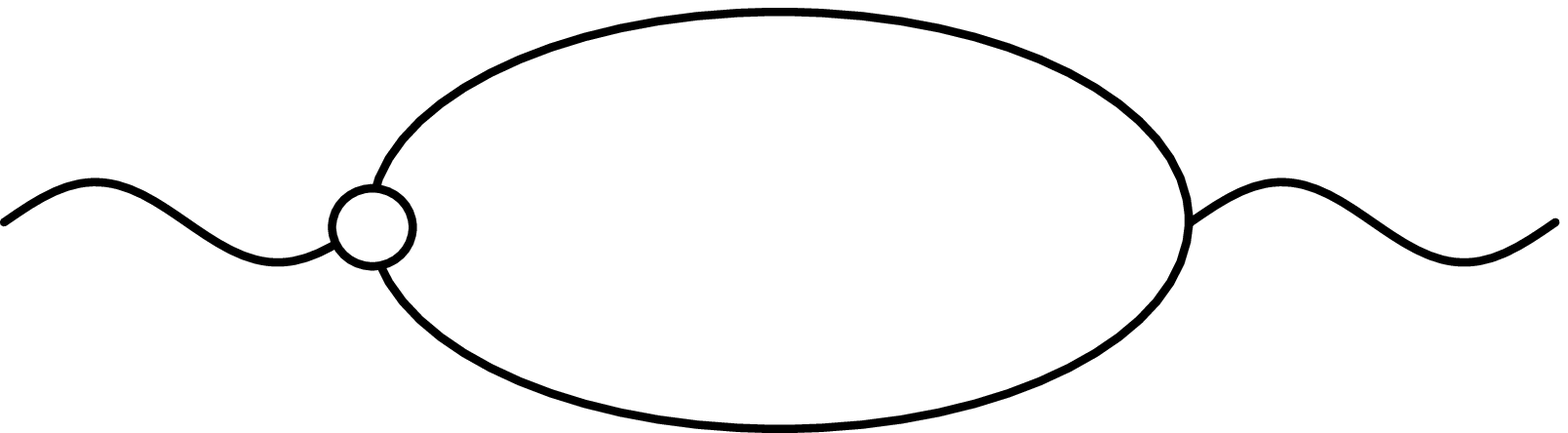}
\caption{}
\label{fig: BosonSE2Lc}
\end{subfigure}
\hfill
\begin{subfigure}[b]{0.45\columnwidth}
\includegraphics[width=0.9\columnwidth]{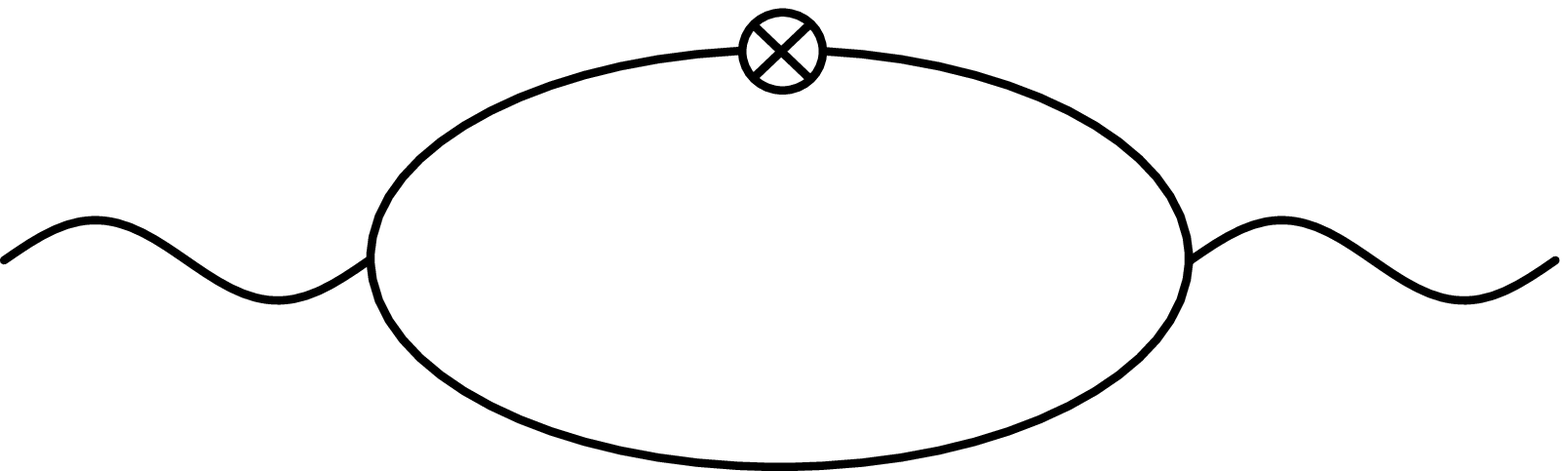}
\caption{}
\label{fig: BosonSE2Ld}
\end{subfigure}
\hfill
\begin{subfigure}[b]{0.4\columnwidth}
\includegraphics[width=0.8\columnwidth]{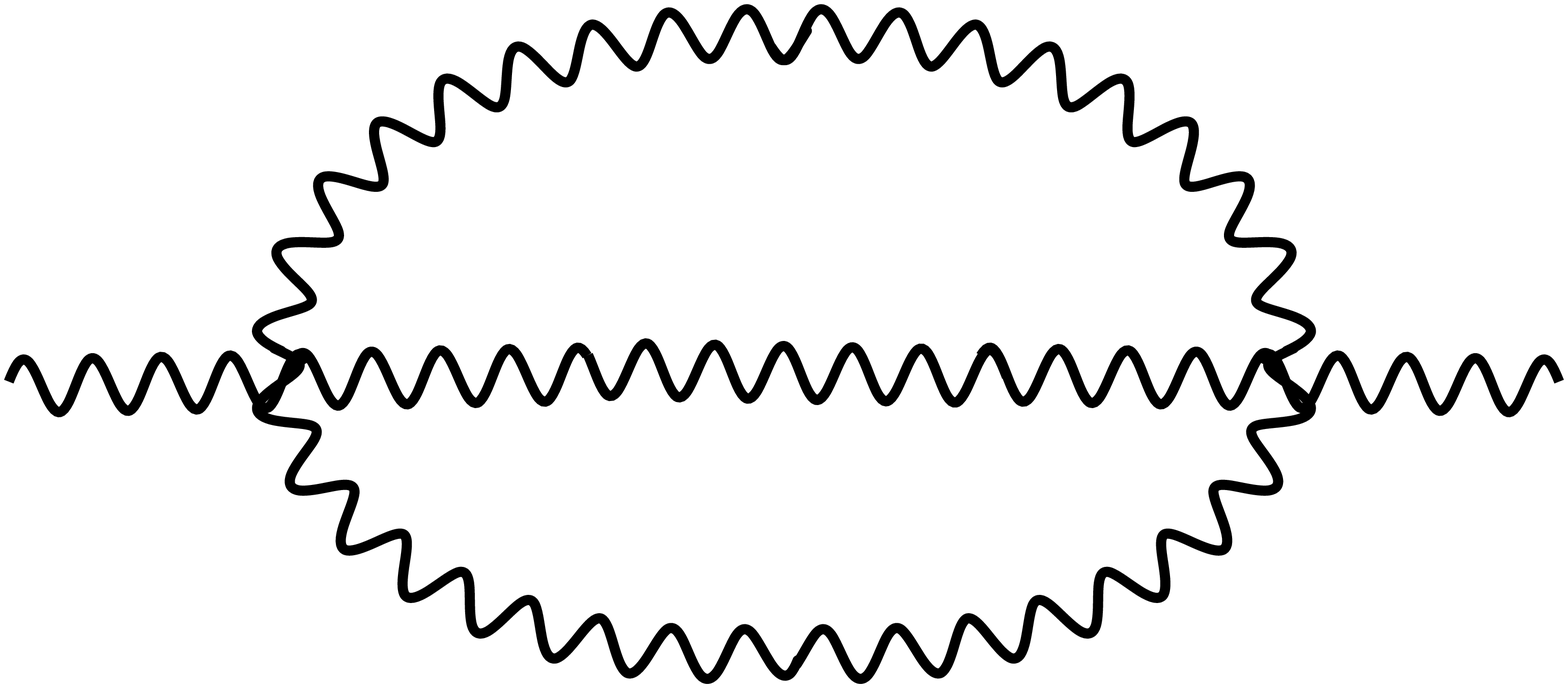}
\caption{}
\label{fig: BosonSE2Le}
\end{subfigure}
\caption{
The diagrams for the two-loop boson self energy.
The small circles in (c) and (d) denote the one-loop counter terms.
}
\label{fig: BosonSE2L}
\end{figure}

The ratio $g^2/c$ which diverges at the one-loop fixed point may spoil the control of the perturbative expansion. 
However, such a conclusion is premature 
because higher-loop diagrams that are divergent at the one-loop fixed point 
can feed back to the flow of $c$ and stabilize it at a nonzero value. 
As long as $c$ is not too small, higher-loop diagrams can be still suppressed. 
In order to include the leading quantum correction to $c$, 
we first focus on the two-loop diagrams for the boson self-energy shown in \fig{fig: BosonSE2L}.

An explicit calculation in Appendix \ref{app: BosonSE2L} shows that only 
\fig{fig: BosonSE2La} 
renormalizes $c$ in the limit $c \rtarw 0$.
Other two-loop diagrams are suppressed by additional factors of $c$, $g$, or $\chi_i$ compared to Figs. \ref{fig: BosonSE2La}. 
Because stabilization of $c$ at a non-zero value can occur only through two or higher loop effect, 
the non-zero value of $c$ must be order of $\eps^b$ with $b>0$.
The two-loop diagram in \fig{fig: BosonSE2La} is proportional to $g^4 q_y^2$, 
which is strictly smaller than the upper bound in \eq{eq: 1PI} by a factor of $c$.
Its contribution to $Z_{6,1}$ is given by
\begin{align}
Z_{6,1} = - \frac{8}{N_c N_f} ~ \dfrac{g^4}{v^2 c^2} ~ \lt( h_6(v) + \ordr{c} \rt),
\label{eq:Z6}
\end{align}
where $h_6(v)$ is defined in \eq{eq:h6}.
The extra factor of $1/c^2$ in \eq{eq:Z6} originates from the fact that 
$Z_{6,1}$ is the multiplicative renormalization to the boson kinetic term, $c^2 q_y^2$.
Since the quantum correction from the two-loop diagram does not vanish in the small $c$ limit, 
it is relatively large compared to the vanishingly small classical action $c^2 q_y^2$. 
Because \fig{fig: BosonSE2La} generates a positive kinetic term at low energy, 
it stabilizes $c$ at a nonzero value.
The non-commuting nature of the SDW vertex in \eq{eq:ttt} is crucial for the stabilization of $c$.
Without the anti-screening effect, \eq{eq:Z6} would come with the opposite sign, 
and $c$ would flow to zero even faster by the two-loop effect.
As we will see, the two-loop diagram indeed suppresses $c$ in the CDW case,
where there is no anti-screening effect.

\begin{figure}[!]
\centering
\includegraphics[width=0.99\columnwidth]
{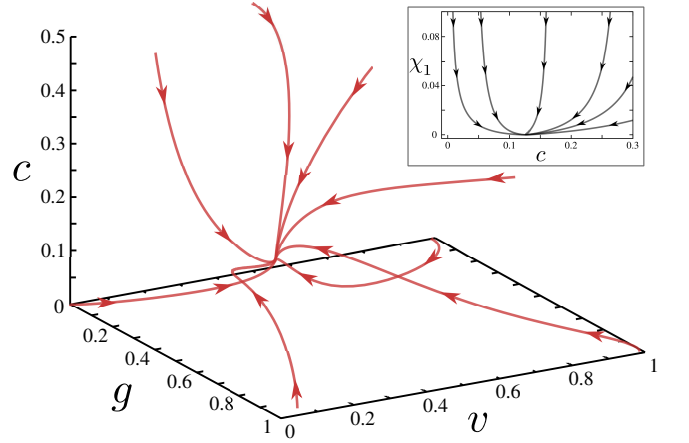}
\caption{
Projection of the RG flow in the $(g, v, c)$ space for $N_c = 2$, $N_f = 1$ and $\eps = 0.01$. The fixed point in \fig{fig: flow-1L} is modified by the two-loop correction (\fig{fig: BosonSE2La}) 
such that $c$ flows to a non-zero value as shown in  \eq{eq: barP*}.
(Inset) Projection of the RG flow in the $(\chi_\sub{1}, c)$ plane. Although $\chi_\sub{1}$ still flows to zero, $c$ does not. }
\label{fig: flow-mod1L}
\end{figure}
\begin{figure}[!]
\centering
\includegraphics[width=0.8\columnwidth]{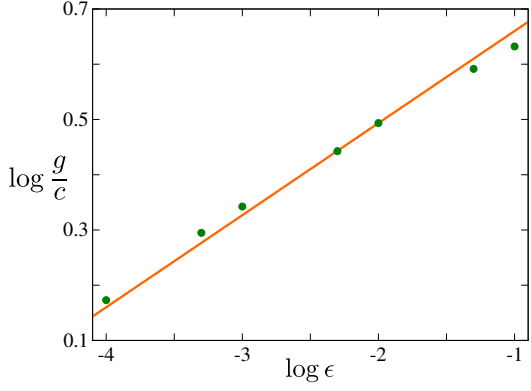}
\caption{
The ratio $g/c$ 
as a function of $\eps$ 
obtained from the numerical solution 
of the full beta functions
in the low energy limit
for $N_f = 1$ and $N_c = 2$. 
The filled circles are from the numerical solution of the beta functions, 
and the straight line is a fit, $g/c = 6.7~ \eps^{1/6}$. 
}
\label{fig: log-log}
\end{figure}

The RG flow which includes the two-loop effect is shown in \fig{fig: flow-mod1L} for $N_c=2$ and $N_f=1$.
$c$ flows to a small but non-zero value in the low energy limit, 
while the other three parameters flow to  values that are similar to those obtained at the one-loop order.
$g/c$ becomes order of $\epsilon^{1/6}$ at the fixed point as is shown in \fig{fig: log-log}.
To find the fixed point for general $N_c$ and $N_f$ analytically, 
we analyze the beta functions in the region where $v \sim 1$ and $0 < c \ll 1$. 
The beta functions can be written 
as an expansion in $g/c$, $\chi_i$ and $c$,
\begin{align}
\dow_{\ell}{\lambda} = \lambda \sum_{l,m,n=0}^{\infty} J_{l,m,n}^{(\lambda)}(v, \eps) ~ \lt(\frac{g}{c}\rt)^{2l} ~  \chi_\sub{i}^{m-1} c^{n-1},
\label{eq: Dl}
\end{align}
where $\lambda$ represents a velocity or a coupling, 
and $J_{l,m,n}^{(\lambda)}(v, \eps)$ are functions of $v$ and $\eps$.
From general considerations some $J_{l,m,n}^{(\lambda)}(v, \eps)$ can be shown to be zero
\cite{Note1}.
To $ \ordr{(g/c)^4~ c^2, \chi}$  
in the small $g/c, \chi, c$ limit, 
the beta functions are given by
\begin{widetext}
\begin{align}
& \dow_\ell v = \frac{z_\tau }{16 \pi}~ v \lt[\frac{g^2}{v} - \frac{2(N_c^2 - 1)}{N_c N_f} ~ g^2 \rt] ,
\label{eq: 1L-mod-Beta-v} \\
& \dow_\ell c = - \frac{z_\tau}{16 \pi}~ 
\frac{g^2 c}{v} \lt[\lt(1  
- \frac{2(N_c^2 - 1)}{N_c N_f} ~ v \rt) 
+ \frac{16(N_c^2 - 1) v}{\pi N_c N_f}~  
\lt( c - \frac{8 \pi^2}{(N_c^2 - 1)} \frac{g^2 ~ h_6(v)}{v^2 c^2} \rt)\rt] ,
\label{eq: 1L-mod-Beta-c} \\
& \dow_\ell g = \half z_\tau ~ g \lt[ \eps - \frac{g^2}{16 \pi v} 
\lt\{ 1 + \frac{2(N_c^2 - 1)}{N_c N_f} ~ v - \frac{8 v}{N_c N_f (1+v)}
\lt( 1  - \frac{2}{\pi} (1+v) c \rt)  \rt\} \rt],
\label{eq: 1L-mod-Beta-g} \\
& \dow_\ell \chi_\sub{1} = z_\tau ~ \chi_\sub{1} \lt[ \lt( \eps - \frac{g^2}{8\pi v} - \frac{8}{N_c N_f} \frac{g^4 ~ h_6(v)}{v^2 c^2} \rt) 
- \lt( \frac{N_c^2 + 7}{2\pi^2} \chi_\sub{1} + \frac{2N_c^2 - 3}{\pi^2 N_c} \chi_\sub{2} + \frac{3(N_c^2 + 3)}{2\pi^2 N_c^2} \frac{\chi_\sub{2}^2}{\chi_\sub{1}} \rt) \rt],
\label{eq: 1L-mod-Beta-chi} \\
& \dow_\ell \chi_\sub{2} = z_\tau ~ \chi_\sub{2} \lt[ \lt( \eps - \frac{g^2}{8\pi v} - \frac{8}{N_c N_f} \frac{g^4 ~ h_6(v)}{v^2 c^2}  \rt) 
- \lt(\frac{6}{\pi^2} \chi_\sub{1} + \frac{N_c^2 - 9}{\pi^2 N_c} \chi_\sub{2} \rt) \rt].
\label{eq: 1L-mod-Beta-chi2}
\end{align}
\end{widetext}
We note that $ J^{(c)}_{1,1,1}(v,\eps)$, $J^{(c)}_{1,1,2}(v,\eps)$, $J^{(c)}_{2,1,0}(v,\eps)$, $J^{(c)}_{2,1,1}(v,\eps)$, $J^{(c)}_{2,1,2}(v,\eps) = 0$ in \eq{eq: 1L-mod-Beta-c}.
This underscores the fact that the actual magnitudes of the diagrams can be smaller than \eq{eq: 1PI} which is only the upper bound.

In the small $c$ limit, only the flows of $c$ and $\chi_\sub{i}$ are affected by the two-loop diagram through the fourth term in \eq{eq: 1L-mod-Beta-c} and the third terms in Eqs. \eqref{eq: 1L-mod-Beta-chi} and \eqref{eq: 1L-mod-Beta-chi2}, respectively.
While $c$ tends to decrease under the one-loop effect,
the two-loop correction enhances $c$ due to the anti-screening produced by the vertex correction.
These opposite tendencies eventually balance each other to yield a stable fixed point for $c$.
At the fixed point of $v_* \sim \ordr{1}$ and $g^{2}_* \sim \ordr{\epsilon}$, the beta function for $c$ is proportional to $-c + r \epsilon/c^2$ with a constant $r > 0$,
such that $c$ flows to $\ordr{\epsilon^{1/3}}$ in the low energy limit.
This confirms that $g/c$ is $\ordr{\epsilon^{1/6}}$ at the fixed point.

\begin{figure}[!]
\centering
\begin{subfigure}[b]{0.4\columnwidth}
 \includegraphics[width=0.9\columnwidth]{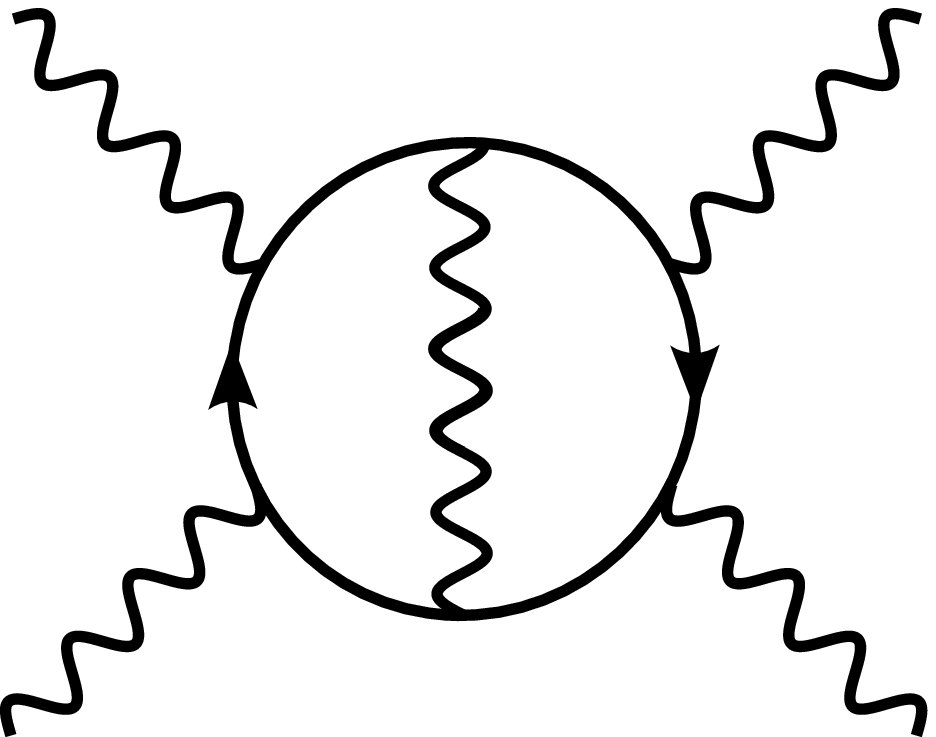}
 \caption{}
 \label{fig: CT-phi4-g2}
\end{subfigure}
\hfill
\begin{subfigure}[b]{0.4\columnwidth}
 \includegraphics[width=0.8\columnwidth]{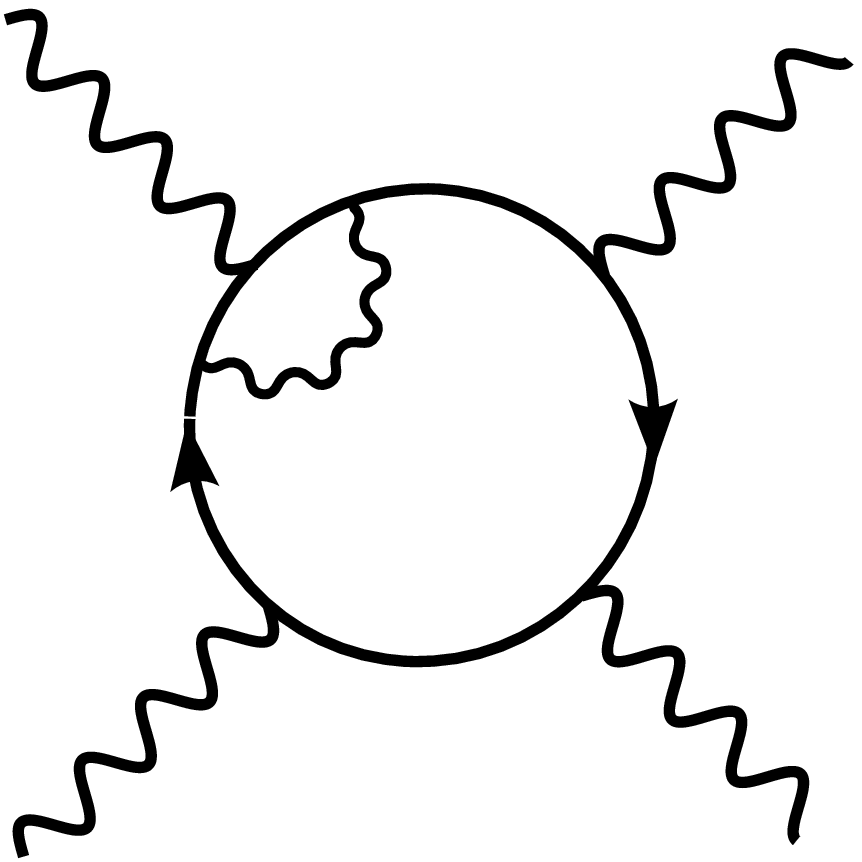}
 \caption{}
 \label{fig: CT-phi4-g3}
\end{subfigure}
\caption{
The leading two-loop diagrams that generate bosonic quartic interaction.
}
\label{fig: phi4-g}
\end{figure}

The RG flow in the $\chi_\sub{1} - \chi_\sub{2}$ plane resembles \fig{fig: chi1-chi2} for small $\eps$, 
and the  $\chi_\sub{i}$'s remain irrelevant at the fixed point.
The two-loop diagrams in \fig{fig: phi4-g}
will generate non-zero quartic couplings which are at most order of 
$\ordr{g^6/c^2}$ 
in the beta function for $\chi_i$.
This is no longer singular because $g^2 \sim \epsilon$ and $c \sim \epsilon^{1/3}$ at the fixed point.
If the leading order term of $\ordr{g^6/c^2}$ survives,
the beta functions for $\chi_i$ has the form of $-\epsilon\chi_i + r' g^6/c^2$ with a constant $r'$.
This suggests that $\chi_i$ is at most $\ordr{\eps^{4/3}}$ at the fixed point.
Other two-loop diagrams and higher-loop diagrams are suppressed by
additional powers of $\epsilon^{1/3}$ compared to 
the one-loop diagrams
and the two-loop diagram in \fig{fig: BosonSE2La},
which are already included.
Therefore, the two-loop effect modifies the fixed point as 
\begin{widetext}
\begin{align}
& v_* = \dfrac{N_c N_f}{2(N_c^2 - 1)} + \ordr{\eps^{1/3}},  \nn
\quad 
& g_*^2  = 
\dfrac{4\pi ~N_c N_f}{(N_c^2 - 1)} ~ \aleph(N_c, N_f) 
\lt[\eps - \frac{16}{N_c^2 - 1} \lt(\frac{2~ \aleph^{4}(N_c, N_f) ~ h_6(v_*)}{N_c N_f}\rt)^{1/3} ~ \eps^{4/3} \rt] + \ordr{\eps^{5/3}}, \nn
& c_* = 4\pi \lt[\frac{2 \aleph(N_c, N_f)}{N_c N_f}~ h_6(v_*) \rt]^{1/3}~ \eps^{1/3} + \ordr{\eps^{2/3}},  \nn
& \chi_{\,\dsty{\!_{i;*}}}  =  \ordr{\eps^{4/3}}.
\label{eq: barP*}
\end{align}
\end{widetext}
It is noted that $\ordr{\epsilon^\alpha}$ in the above equations represent the upper bounds of the sub-leading terms.
The actual sub-leading terms may be smaller than the upper bound.
For example, the actual sub-leading correction to $v_*$ is $\ordr{\eps^{2/3}}$ because
the $\ordr{c}$ term in $h_1(v,c)  + h_2(v,c)$ is zero (see \eq{eq: SmallC}).

Eqs. (\ref{eq: zTau})-(\ref{eq: EtaPhi}) along with \eq{eq: 1PI} 
implies that the anomalous dimensions at the fixed point can be expressed as
\begin{align}
\Dl = \sum_{l,m,n = 0} A_{l,m,n} ~ \lt(\frac{g_*}{c_*} \rt)^{2l} ~ \chi_*^{m} ~ c_*^n,
\label{eq: C-E-gen}
\end{align}
where 
$\Delta$ represents either $z_\tau-1, z_x-1,\eta_\psi$, or $\eta_\phi$, 
and  $A_{l,m,n}$ are constants with $A_{0,0,n} = 0$. 
Using the expressions of the parameters at the fixed point (\eq{eq: barP*}), we compute the critical exponents up to order $\eps^{4/3}$,
\begin{align}
& z_\tau = 1 + \frac{\aleph(N_c, N_f)}{2}~ \eps 
\nn & \quad 
-  8 \lt( 2 + \frac{\aleph(N_c, N_f)}{N_c^2 - 1}\rt) \lt(\frac{2 \aleph^4(N_c, N_f)}{N_c N_f}~ h_6(v_*) \rt)^{1/3} \eps^{4/3}, \nn
& z_x = 1 - 16 \lt(\frac{2 \aleph^4(N_c, N_f)}{N_c N_f}~ h_6(v_*) \rt)^{1/3}~ \eps^{4/3}, \nn
& \wtil{\eta}_\psi =  4 \lt(\frac{2 \aleph^4(N_c, N_f)}{N_c N_f}~ h_6(v_*) \rt)^{1/3}~ \eps^{4/3},  \nn
& \wtil{\eta}_\phi = 16 \lt(\frac{2 \aleph^4(N_c, N_f)}{N_c N_f}~ h_6(v_*) \rt)^{1/3}~ \eps^{4/3}. 
\label{eq: barP*-CE}
\end{align}
The fixed point value of $\chi_{\sub{i};*}$ does not affect the critical exponents up to $\ordr{\eps^{4/3}}$ because a single $\phi^4$ vertex does not contribute to any of the nine counter terms. 
Because $z_x$ differs from one at the fixed point, the system develops an anisotropy in the $(k_x, k_y)$ plane.


\subsection{Control of the perturbative expansion}

In the previous subsection, we incorporated one particular two-loop diagram
which stabilizes the boson velocity at a non-zero value
in order to compute the critical exponents to the order of $\epsilon^{4/3}$.
A natural question is whether it is safe to ignore other two-loop diagrams,
and more generally whether the perturbative expansion is under control for small $\epsilon$.
From \eq{eq: barP*} we note that  $g^2 \sim c^3$ at the fixed point, 
and  \eq{eq: 1PI} can be expressed in terms of $c$ and $\chi$,
\begin{align}
& F(p_i; v, c, g, \chi; \eps, V_g, V_u, L, L_b, N_\Pi) \nn
& =  \chi^{V_u} ~c^{(E-2)/2 + (L - L_b) + (V_g - 2 N_\Pi)} ~ f'(p_i; v, c; \eps, L).
\label{eq: 1PI-1}
\end{align}
Here $f'(p_i; v, c; \eps, L)$ is finite in the small $c$ limit.
The exponents of $\chi$, $c$ are non-negative
because $L \geq L_b$ and $V_g > 2 N_\Pi$.
The latter inequality follows from the fact that any diagram for boson self-energy of the first kind 
must contain at least four Yukawa vertices.
For a fixed $E$, new vertices cannot be added without increasing either $V_u$, $(L-L_b)$, or $(V_g - 2 N_\Pi)$.
Therefore, \eq{eq: 1PI-1} implies that quantum corrections 
are systematically suppressed by powers of 
$c \sim \epsilon^{1/3}$ and $\chi \lesssim \epsilon^{4/3}$ 
as the number of loops increases,
and there exist only a finite number of diagrams at each order. 
This shows that other two-loop diagrams and higher-loop diagrams are indeed sub-leading,
and they do not modify the critical exponents in \eq{eq: barP*-CE} up to the order of $\eps^{4/3}$.

Sub-leading terms come in two ways.
The first is from the $c$-expansions of $h_i(v,c)$ defined in \eq{eq: SmallC}.  
The second is from higher-loop diagrams.
For $Z_{n,1}$ with $n=1, 2, 3, 7$, 
two-loop diagrams are suppressed at least by $\eps^{5/3}$ according to \eq{eq: 1PI-1} 
because $(L - L_b)  \geq 1$  and $V_g - 2 N_\Pi \geq 4$
for the fermion self energy
and the Yukawa vertex correction.
$\ordr{c^2}$ terms in the expansion of $h_i(v,c)$ from the one-loop diagrams
are also at most order of $\eps^{5/3}$.
For $Z_{4,1}$, the one-loop diagram does not contain any sub-leading term in $c$.
According to \eq{eq: 1PI-1}, higher-loop contributions to $Z_{4,1}$ are at most order of $g^4/c \sim \eps^{5/3}$.
As noted earlier, the first non-vanishing contributions to $Z_{5,1}$ and $Z_{6,1}$ arise at the two-loop order. 
In Appendix \ref{app: BosonSE2L}, we show that the leading order term in $Z_{5,1}$ is at most order of $g^4 c \sim \eps^{7/3}$.
In contrast, \eq{eq:Z6} shows that the leading order term in $Z_{6,1}$ is  $g^4/c^2 \sim \eps^{4/3}$ which is already included.
The sub-leading terms are suppressed by $c, g^2/c^2 \sim \epsilon^{1/3}$.
As discussed below \eq{eq: 1L-mod-Beta-chi2}, 
$\chi_i$ are at most $\ordr{\eps^{4/3}}$ from two-loop contributions.
However, $\chi_{\sub{i}} \sim \eps^{4/3}$ can not affect the critical exponents up to $\ordr{\eps^{4/3}}$ 
because a single quartic vertex only renormalizes boson mass.
Thus the critical exponents in  \eq{eq: barP*-CE} are accurate up to $\ordr{\eps^{4/3}}$.

It is interesting to note that the perturbative expansion is not simply organized by the number of loops.
Instead, one has to perform an expansion in terms of the couplings and the boson velocity together.
There are notable consequence of this unconventional expansion.
First, the perturbative expansion is in power series of $\epsilon^{1/3}$.
Second, not all diagrams at a given loop play the same role; 
only one two-loop diagram (\fig{fig: BosonSE2La}) is important 
for the critical exponents to the order of $\epsilon^{4/3}$.


\subsection{Physical Properties} 
\label{sec: PhysicalProperties}

\begin{figure}[!]
\centering
\begin{subfigure}[b]{0.9\columnwidth}
\includegraphics[width=0.95\columnwidth]{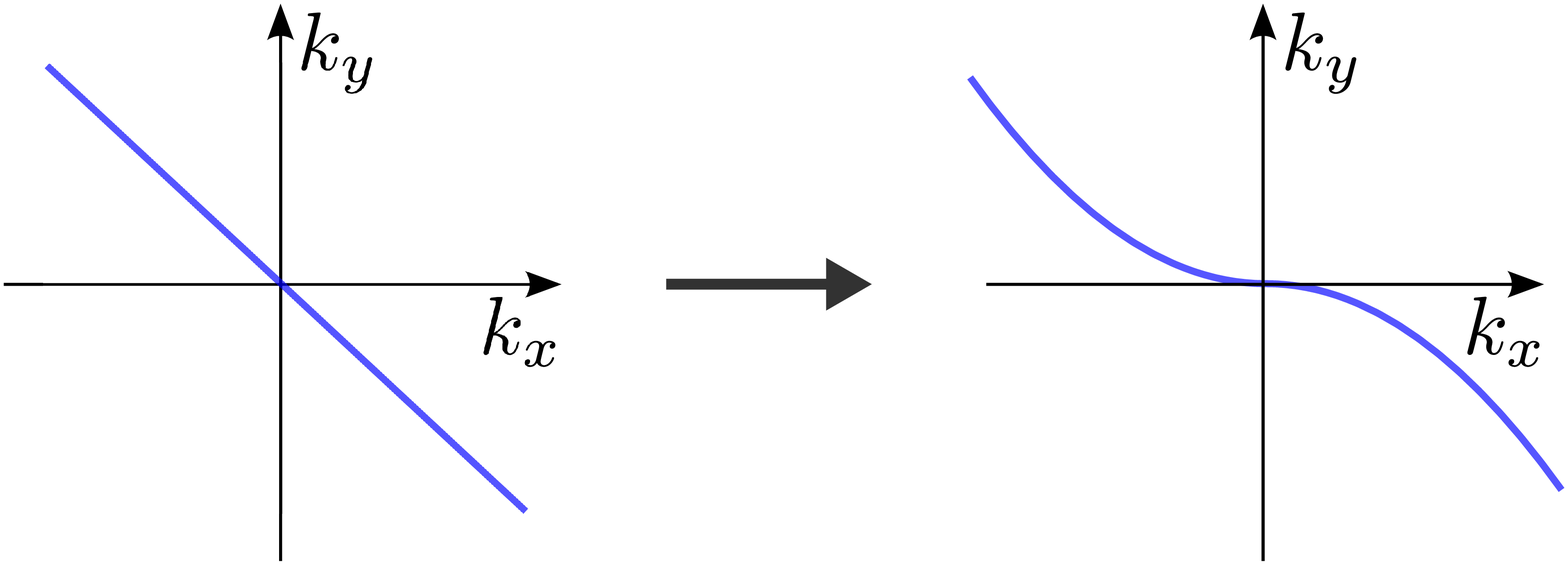}
\caption{}
\label{fig: patch-evolve}
\end{subfigure}
\begin{subfigure}[b]{0.8\columnwidth}
\includegraphics[width=0.9\columnwidth]{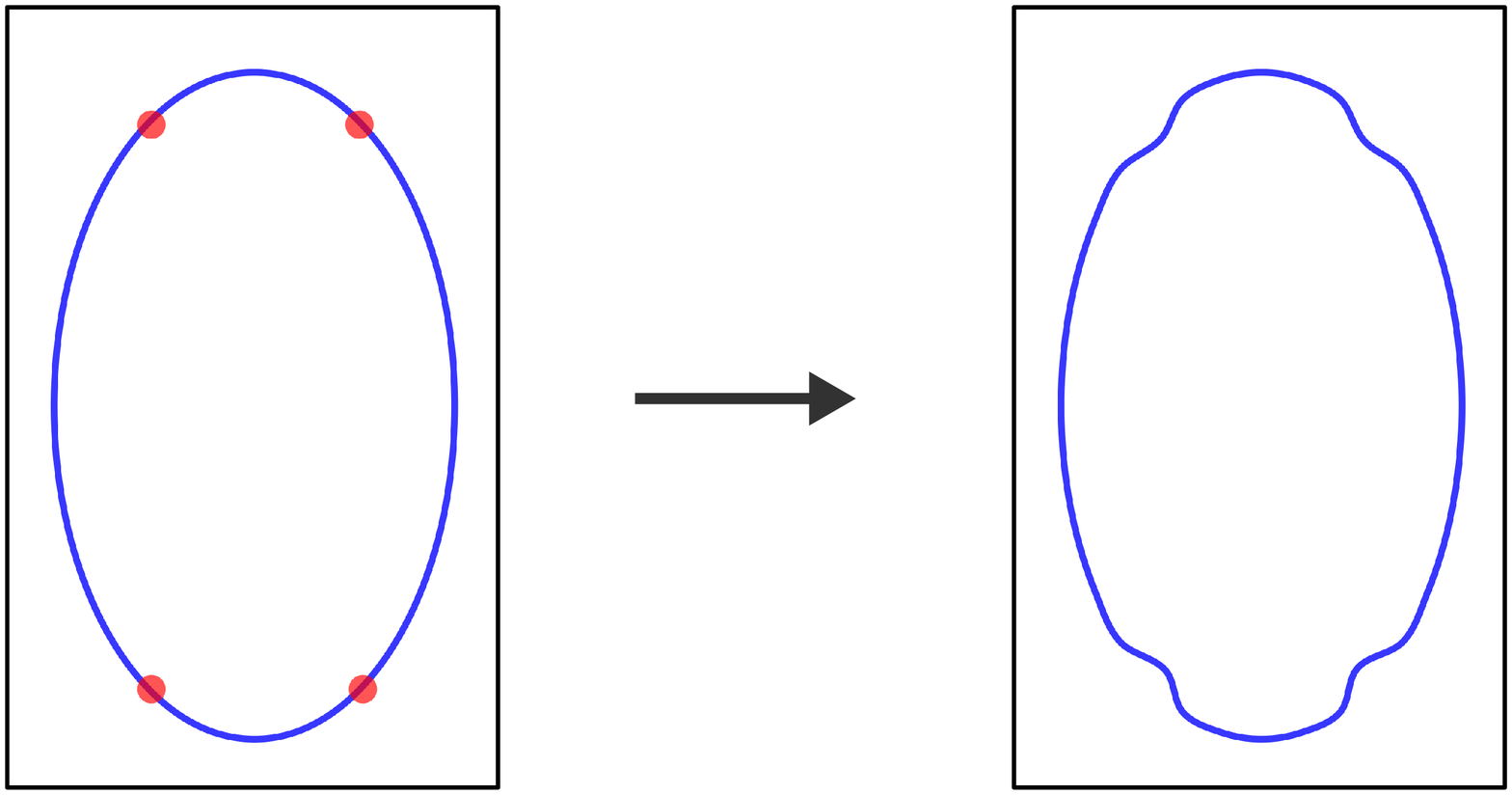}
\caption{}
\label{fig: FS-evolve}
\end{subfigure}
\caption{
(a) The patches of Fermi surface near the hot spots are deformed 
into a universal non-analytic curve. 
(b) The hot spots become algebraically nested near the hot spots.
}
\label{fig: UnivCurv}
\end{figure}

In this section we discuss the physical  properties of the non-Fermi liquid state that is realized at the SDW critical point.
The anomalous dimension of $k_x$ implies that the Fermi surface near the hot spots are deformed into a universal curve,
\begin{align}
k_y \sim \sgn{k_x} |k_x|^{1/z_x}
\label{eq: FS-patch}
\end{align}
as is illustrated in \fig{fig: UnivCurv}.
The algebraic nesting of the Fermi surface  near the hot spots
is in contrast to the  $C_4$-symmetric case, 
where the emergent nesting is only logarithmic such that $k_y \sim k_x/(\ln{k_x})$   \cite{Abanov1,Metlitski2,Sur2}.
The electronic spectral function at the hot spot 
scales with frequency as 
\begin{align}
\mc{A}_{l,m}(\om) \sim \frac{1}{\om^{(1 - 2\wtil \eta_{\psi})/z_{\tau}}},
\end{align}
and the dynamical spin structure factor at momentum $\vec Q_{ord}$, 
\begin{align}
\mc{S}(\om) \sim \frac{1}{\om^{(2-2\wtil \eta_\phi)/z_\tau}}.
\end{align}
As one moves away from the hot spots or the ordering vector 
in the $x$ ($y$) directions, 
the electron spectral function 
and the spin structure factor 
will exhibit incoherent peak at frequency 
$\om \sim |k_x|^{z_\tau/z_x}$ 
and
$\om \sim |k_y|^{z_\tau}$
depending on the direction of momentum.
In principle, all exponents $z_\tau, z_x, \wtil \eta_\psi, \wtil \eta_\phi$ 
can be determined from the angle resolved photoemission 
and inelastic neutron scattering experiments.
In Table \ref{tab: values}, we list the exponents at the fixed point for $(N_c, N_f) = (2, 1)$ and $(3,1)$.
It is noted that $z_x$ becomes smaller than one below three dimensions, 
which is consistent with the intuition that the interaction enhances nesting.  
However, the fact that $z_x$ becomes negative for $\epsilon=1$ and $N_c=2$ should not be taken  seriously, 
since higher order contributions need to be taken into account in two dimensions.

\begin{table*}[t]
\centering
\begin{tabular}{c | c  c  c  c}
\hline \hline \\[-2ex]
$N_c$ ~&~  $z_\tau$ ~&~  $z_x$ ~&~  $\wtil{\eta}_\psi$ ~&~  $\wtil{\eta}_\phi$\\
\hline \\[-1ex]
$2$ ~&~  $1 + \eps - 1.9 ~\eps^{4/3}$ ~~&~~  $1 - 1.4~ \eps^{4/3}$ ~~&~~  $0.36~ \eps^{4/3}$ ~~&~~  $1.4~ \eps^{4/3}$\\[2ex]
$3$ ~&~ $1 + 0.63~\eps - 0.68 ~\eps^{4/3}$ ~~&~~  $1 - 0.63~ \eps^{4/3}$ ~~&~~  $0.16~ \eps^{4/3}$ ~~&~~  $0.63~ \eps^{4/3}$\\[2ex]
\hline \hline
\end{tabular}
\caption{
Fixed point values of the critical exponents  for $N_c = 2,3$ with $N_f=1$.}
\label{tab: values}
\end{table*}

We also estimate the contribution of electrons near the hot spots 
to the specific heat and the optical conductivity following the work 
by Patel \textit{et al.} for the $C_4$-symmetric model \cite{Patel}.  
The scaling dimension of the free energy density $\mc{F}$ is 
\begin{align}
[\mc F] = z_\tau (d-1) + 1 + z_x.
\end{align}
The current density $\mc{J}_\mu$ has the dimension of 
$[\mc J_\mu] =  [\mc F] - [k_\mu]$. 
Because $k_x$ and $k_y$ have different scaling dimensions,
the two diagonal elements of the optical conductivity 
have different scaling dimensions,
\begin{align}
& [\sig_{x x}] = z_\tau (1 - \eps) + 1 - z_x, \nn
& [\sig_{y y}] = z_\tau (1 - \eps) + z_x - 1,
\label{eq: sig-dim}
\end{align}
at $d = 3 -\eps$.
The contribution of the hot spot electrons obeys the hyperscaling
because temperature or frequency provides a cut-off for the momentum along the Fermi surface.
As a result, the size of Fermi surface does not enter 
in the scaling of the contributions from the hot spots.
This is analogous to the phenomenon where the thermodynamic responses 
from inflection points obey the hyperscaling relation in non-Fermi liquids where Fermi surface is coupled with
a $\vec Q_{ord}=0$ critical boson \cite{Sur1}. 
As a result, the hot spot contribution to the specific heat scales with temperature as 
\bqa
c^{hot} \sim T^{ 1-\epsilon + \frac{1+z_x}{z_\tau} },
\eqa
and the hot spot contributions to the optical conductivity scales with frequency as
\begin{align}
& \sig_{x x}^{hot}(\om) \sim \om^{1 - \eps + \frac{1 - z_x}{z_\tau}}, \nn
& \sig_{y y}^{hot}(\om) \sim \om^{1 - \eps - \frac{1 - z_x}{z_\tau} }.
\label{eq: sig-om}
\end{align}
Because $z_x < 1$, the optical conductivity 
is  greater along the ordering vector than the perpendicular direction at low frequency  \cite{Nakajima, Dusza}.
We emphasize that the anisotropy in \eq{eq: sig-om} arises from anisotropic spatial scaling, rather than anisotropic carrier  velocity   \cite{Tanatar,Valenzuela}.
Electrons away from the hot spots are expected to violate the hyperscaling,
and contribute to the specific heat and the optical conductivity as
$c^{cold} \sim k_F T^{2-\epsilon}$ and
$\sigma^{cold} \sim k_F \omega^{-\epsilon}$,
where $k_F$ is the size of Fermi surface.
For small $\epsilon$, the contributions from cold electrons dominate the hot spot contributions.

Near three dimensions, there is no perturbative instability,
and the anisotropic non-Fermi is stable.
However, near two dimensions the non-Fermi liquid state can become
unstable against other ordered phases.
As far as the hot spot electrons are concerned,
a charge density wave is the leading instability,
followed by the $d$-wave pairing and pair density wave \cite{Sur2}. 
However, the order of leading instability can change due to cold electrons
away from the hot spots,  which favour zero-momentum pairing due to the lack of nesting.

\section{Charge Density Wave Criticality}
\label{sec: CDW}

In this section, we discuss the low energy properties of the CDW critical point.
Since many aspects are similar to the SDW case, we will highlight the differences between the two critical points.
The main differences arise from the commuting versus non-commuting nature of the respective interaction vertices 
as is shown in Eqs. \eqref{eq: original-action} and \eqref{eq: Phi}.
It is analogous to the difference between the nematic and ferromagnetic critical points    \cite{Rech}.

Since the CDW order parameter couples to the global $U(1)$ charge, the interaction vertex is diagonal in both the spin and flavor space.
One can also set $\chi_\sub{2} = 0$ for any $N_f$ and $N_c$ since the two quartic vertices are equivalent.
As derived in Appendix \ref{app: diagrams}, 
the counter terms resulting from the one-loop diagrams in \fig{fig: 1L-CT} are
\begin{alignat}{2}
& Z_{1,1} = 
- \dfrac{1}{4 \pi^2 \wtil{N}_f} ~ g^2  ~ h_1(v, c), 
&& \qquad 
Z_{2,1} =  \dfrac{1}{4 \pi^2 \wtil{N}_f} ~ g^2  ~ h_2(v, c) \nn
& Z_{3,1} =
- \dfrac{1}{4 \pi^2 \wtil{N}_f} ~ g^2  ~ h_2(v, c)   
&& \qquad Z_{4,1} = 
- \dfrac{1}{8\pi} ~ \dfrac{g^2}{v}, \nn
&  Z_{5,1} = 0,  
&& \qquad  
Z_{6,1} = 0,  \nn
& Z_{7,1} = 
\dfrac{1}{8 \pi^3 \wtil{N}_f} ~ g^2 ~ v ~ h_3(v, c),  
&&\qquad 
Z_{8,1} = 
\dfrac{9}{2 \pi^2} ~ \chi_\sub{1},
\label{eq: Z-factors-CDW}
\end{alignat}
where $\wtil{N}_f = N_c N_f$.
Using the general expressions of the beta functions in section \ref{sec: RG},
 we obtain the one loop beta-functions for the CDW critical point.
As in the SDW case, the boson velocity flows to zero in the low energy limit at the one-loop order.
Therefore, we focus on the regime with small $c$
where the one-loop beta functions take the form,
\begin{align}
\dow_\ell v &= \frac{z_\tau}{16 \pi}~ g^2  \lt[1 -  \frac{2 v}{\wtil{N}_f} \rt],
\label{eq: CDW1LBetV} \\
\dow_\ell c &= - \frac{z_\tau}{16 \pi}~ 
\frac{c}{v} ~ g^2 \lt[\lt(1 - \frac{2 v}{\wtil{N}_f} \rt) 
+ \frac{16~ v ~ c}{\pi \wtil{N}_f} \rt],
\label{eq: CDW1LBetC} \\
\dow_\ell g &= \frac{z_\tau}{2} ~ g 
\lt[ \eps - \frac{g^2}{16\pi v} 
\lt\{ \lt( 1 + \frac{2v}{\wtil{N}_f} \rt) \rt. \rt. 
\nn & \qquad 
+ \lt. \lt. \frac{8 v}{ \wtil{N}_f (1+v)} \lt( 1 - \frac{2}{\pi} (1+v) c \rt) \rt\} \rt],
\label{eq: CDW1LBetG} \\
\dow_\ell \chi_\sub{1} &= z_\tau ~ \chi_\sub{1} \lt[ \lt( \eps - \frac{g^2}{8\pi v} \rt) - \frac{9}{2 \pi^2} ~ \chi_\sub{1} \rt]
\label{eq: CDW1LBetChi}.
\end{align}
We note that the sign of $Z_{7,1}$ for the CDW critical point is opposite to that of the SDW critical point.
This is due to the fact that the three CDW vertices (identity matrix) that appear in \fig{fig: CT-g}
are mutually commuting, while the SDW vertices ($SU(N_c)$ generators) are mutually anti-commuting as shown in \eq{eq:ttt}.
Consequently, the vertex correction screens the interaction at the CDW critical point in contrast to the SDW case.
A stable one-loop fixed point arises at
\begin{align}
v_* &= \dfrac{\wtil{N}_f}{2},  \nn
g_*^2 & = 4\pi \wtil{N}_f \dfrac{\wtil{N}_f + 2}{\wtil{N}_f + 6}  ~ \eps, \nn
c_* & = 0, \nn
\chi_{\,\dsty{\!_{1;*}}} &= \frac{8 \pi^2 ~ \eps}{9(\wtil{N}_f + 6)}.
\label{eq: CDW-P*}
\end{align}
To the leading order in $\eps$, the critical exponents become
\begin{align}
z_\tau &= 1 + \frac{\wtil{N}_f + 2}{2(\wtil{N}_f + 6)} ~ \eps,
\quad z_x = 1, \quad
\wtil\eta_\psi = 0 \quad
 \mbox{and} 
\quad \wtil\eta_\phi = 0.
\label{eq: CDW-exponents}
\end{align}
Since the one-loop vertex correction screens the Yukawa interaction in \eq{eq: CDW1LBetG}, 
the Yukawa coupling is not strong enough to push the 
upper critical dimension for $\chi_\sub{1}$ below $3-\epsilon$,  
in contrast to the SDW case. 
As a result, $\chi_\sub{1}$ remains non-zero at the one-loop fixed point below three dimensions.
It is interesting to note that the weaker (better screened) Yukawa coupling 
makes it possible for the quartic coupling to be stronger at the CDW critical point
as compared to the SDW case.

We now investigate how the two-loop correction  modifies the flow of $c$.
The two-loop diagram in \fig{fig: BosonSE2L} leads to 
\begin{align}
Z_{5,1} = 0, \qquad
Z_{6,1} = \frac{8}{\wtil{N}_f} ~ \dfrac{g^4}{v^2 c^2} ~h_6(v)
\label{eq: 2lCT-CDW}
\end{align}
to the leading order in $c$.
The modified beta functions for $c$ and $\chi_\sub{1}$ are given by
\begin{align}
& \dow_{\ell} c = - \frac{z_\tau}{16 \pi}~ 
\frac{c}{v} ~ g^2 \lt[\lt(1 - \frac{2 v}{\wtil{N}_f} \rt) \rt. 
\nn & \qquad \quad 
+ \lt.\frac{16~ v}{\pi \wtil{N}_f} \lt( c + \frac{8 \pi^2 g^2 h_6(v)}{v^2 c^2} \rt)\rt] ,
\label{eq: CDWModBetC} 
\\
&\dow_\ell \chi_\sub{1} = z_\tau ~ \chi_\sub{1} \lt[ \lt( \eps - \frac{g^2}{8\pi v} + \frac{8}{\wtil{N}_f} ~ \dfrac{g^4}{v^2 c^2} ~h_6(v) \rt) - \frac{9}{2 \pi^2} ~ \chi_\sub{1} \rt]. 
\label{eq: CDWModBetChi}
\end{align}
We note that the sign of $Z_{6,1}$ in \eq{eq: 2lCT-CDW}, which contributes the fourth term in \eq{eq: CDWModBetC} and the third term in \eq{eq: CDWModBetChi}, is opposite
to that of $Z_{6,1}$ for the SDW case in \eq{eq:Z6}.
This is again due to the commuting CDW vertices in contrast to the anti-commuting SDW vertices in \eq{eq:ttt}.
Therefore the two-loop diagram further reduces $c$ for the CDW case, 
while it stops $c$ from flowing to zero for the SDW case.
Since \eq{eq: CDW1LBetV} is not modified by the two-loop diagram, $v$ flows to $v_* = \wtil{N}_f/2$, irrespective of how the other parameters flow, as long as $c$ remains small.
To understand the fate of the system in the low energy limit,
it is useful to examine the flow of $g,c,\chi_\sub{1}$ with fixed $ v = v_*$,
\begin{align}
& \dow_\ell g = \frac{z_\tau}{2} ~ g 
\lt[ \eps - \frac{g^2}{4\pi \wtil{N}_f} 
\lt\{ 1 + \frac{4}{ \wtil{N}_f + 2} \lt( 1 - \frac{\wtil{N}_f + 2}{\pi} ~c \rt) \rt\} \rt],
\label{eq: CDW1LModBetG} \\
& \dow_{\ell} c = 
- \frac{32 z_\tau }{\wtil{N}_f^3}~ h_6(\wtil{N}_f/2)~ g^2 ~ c^2 \lt( \frac{\wtil{N}_f^2}{32 \pi^2  h_6(\wtil{N}_f/2)} + \frac{g^2}{c^3} \rt) , \label{eq: CDW1LModBetC} \\
& \dow_\ell \chi_\sub{1} = z_\tau ~ \chi_\sub{1} \lt[ \lt( \eps - \frac{g^2}{4\pi \wtil{N}_f} + \frac{32}{\wtil{N}_f^3} ~ \dfrac{g^4}{c^2} ~h_6(\wtil{N}_f/2) \rt) \rt. 
\nn & \qquad \qquad 
- \lt. \frac{9}{2 \pi^2} ~ \chi_\sub{1} \rt]. 
\label{eq: CDW1LModBetChi}
\end{align}

\begin{figure}[!]
\centering
\begin{subfigure}{0.75\columnwidth}
\includegraphics[width=.9\columnwidth]{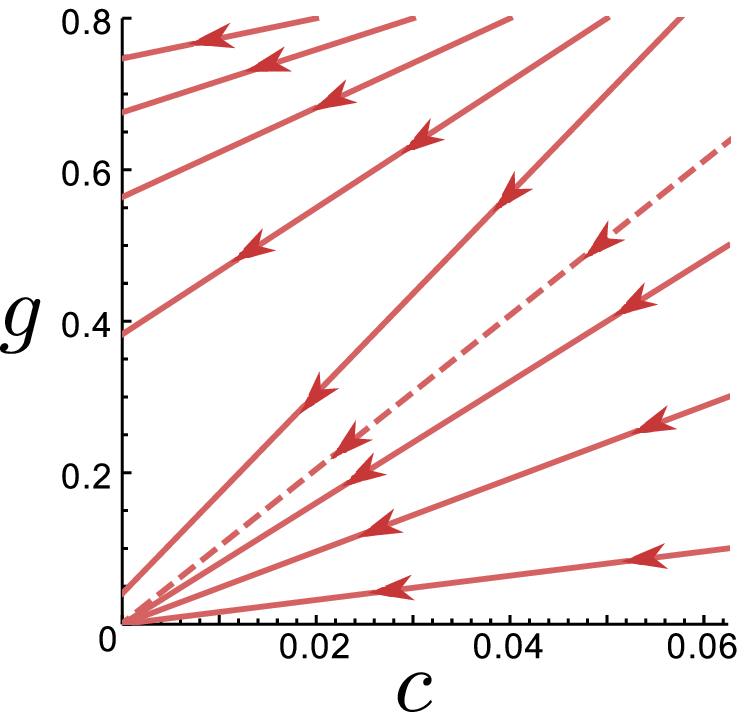}
\caption{}
\label{fig: flow-CDW-3d}
\end{subfigure}
\hfill
\begin{subfigure}{0.75\columnwidth}
\includegraphics[width=.9\columnwidth]{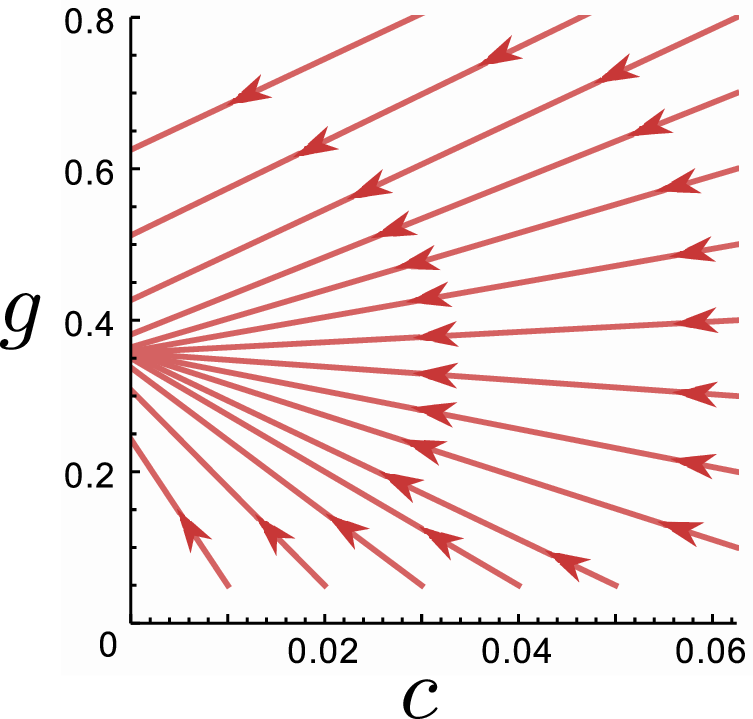}
\caption{}
\label{fig: flow-CDW-3-eps}
\end{subfigure}
\caption{
RG flow of $g$ and $c$ at the CDW critical point with $\wtil N = 2$ and $v = v_* = 1$. 
(a) In $d = 3$ there is a fixed point at the origin with a finite basin of attraction; for flows originating outside the basin, $c$ flows to zero after a finite RG time. 
The dashed line in (a) denotes the  separatrix 
which divides the flows to the stable quasilocal marginal Fermi liquid (below the separatrix)
from the flow toward the non-perturbative regime (above the separatrix).
(b) In $d = 3 - \eps$,
 $c$ flows to zero after a finite RG time with a finite $g$. For the plot we have chosen $\eps = 0.01$.}
\label{fig: flow-CDW}
\end{figure}

The analysis of the beta functions is rather involved,
and the details are in Appendix B.
Key elements of the final result are summarized in \fig{fig: flow-CDW}.
In three dimensions, the system flows to a weakly coupled quasilocal marginal Fermi liquid
if the initial Yukawa coupling is smaller than the boson velocity (below the dashed separatrix in \fig{fig: flow-CDW-3d}).
On the other hand, the system flows to strong coupling regime as the boson velocity is renormalized to zero 
when initial Yukawa coupling is large (above the dashed separatrix in \fig{fig: flow-CDW-3d}).
Below three dimensions, the boson velocity always flows to zero within a finite RG time (\fig{fig: flow-CDW-3-eps}),
and the theory becomes non-perturbative. 
When the system flows to the strong coupling regime, there are several  possibilities.
First, the system may still flow to a strongly interacting non-Fermi liquid fixed point.
Second, $c^2$ may become negative at low energies, which results in a shift of the ordering vector,
possibly towards an incommensurate CDW ordering \cite{Castellani}.
In this case, the commensurate CDW can not occur without further fine tuning.
Third, the system may develop an instability toward other competing order, 
such as superconductivity \cite{Wang}.
Finally, a first order transition is a possibility.

\begin{table*}[t]
\centering
\begin{tabular}{  c | l   l   }
\hline \hline \\[-2ex]
 ~ & \multicolumn{1}{c}{SDW} & \multicolumn{1}{c}{CDW} \\
 \hline \\[-1ex]
$d = 3$ ~&  ~Quasilocal MFL &  ~~\tabitem Non-perturbative $(~g_0^2 >  c_0^2 ~ )$ \\[1em]
& ~~& ~~\tabitem ~Quasilocal MFL $(~g_0^2 <   c_0^2~)$ \\[1em]
\hline \\[-2ex]
$d = 3-\epsilon$~ & ~Anisotropic NFL & ~~~ Non-perturbative \\[0.1em]
\hline \hline
\end{tabular}
\caption{
Fate of the SDW and CDW critical points in metals with $C_2$ symmetry.
Here NFL and MFL represents non-Fermi liquid  and marginal Fermi liquid, respectively.
}
\label{tab: SDW-CDW}
\end{table*}
The difference between the SDW and CDW critical points is summarized in Table \ref{tab: SDW-CDW}.
The differences arise from 
the fact that the vertex correction screens  (anti-screens) the interaction
for the CDW (SDW) critical point.
Within the present framework, it is also possible to consider a SDW critical point where the $SU(N_c)$ spin rotational symmetry is explicitly broken down to a subgroup.
In the Ising case where only one mode becomes critical at the critical point, the Yukawa vertex is commuting as in the CDW case.
Therefore, we expect that the Ising SDW critical point will be similar to the CDW critical point.
The easy-plane SDW criticality with $N_c=2$  \cite{Varma15} 
is special in that the one-loop Yukawa vertex correction vanishes 
due to $\sum_{a=1}^2 \tau^a \tau^b \tau^a = 0$.
In this case, the two-loop diagram fails to prevent $c$ from flowing to zero.
Thus, one has to consider higher order self-energy and vertex corrections
to determine the fate of the critical point.

\section{Summary and Discussion}
\label{sec: conclusion}


In this work,
we studied the spin and charge density wave critical point in metals with the $C_2$ symmetry, 
where a one dimensional Fermi surface is embedded in space dimensions three and below. 
Within one-loop RG analysis augmented by a two-loop diagram, 
we obtained an anisotropic non-Fermi liquid below three dimensions at the SDW critical point.
The Green's function near the hot spots and the spin-spin correlation function obey the anisotropic scaling, 
where not only frequency but also different components of momentum acquire non-trivial anomalous dimensions.
Consequently, the Fermi surface develops an algebraic nesting near the hot spots with a universal shape.
The stable non-Fermi liquid fixed point turns into a quasilocal marginal Fermi liquid in three dimensions, 
where the boson velocity along the ordering vector flows to zero compared to the Fermi velocity.
In contrast to the SDW criticality, the CDW critical point flows to a  non-perturbative regime below three dimensions,
while there is a finite parameter regime where the marginal Fermi liquid is still stable in three dimensions.

At the SDW critical point, 
it is expected that superconducting, pair density wave and charge density wave fluctuations 
are enhanced \cite{Metlitski2,Sur2,Berg_MonteCarlo,Berg15}. 
At the one-loop order, the pattern of enhancement is expected to be similar to the case with the $C_4$ symmetry.
However, it will be of interest to examine the effects of anisotropic scaling through a comparative study.
In particular, the stronger nesting in the $C_2$ case will increase the phase space for the zero-energy particle-particle excitations with momentum $2k_F$.
This will help enhance the pair density wave fluctuations,
which was found to be as strong as the $d$-wave superconducting fluctuations at the one-loop order in the $C_4$ case  \cite{Sur2}.

\section{Acknowledgment}
We thank Luis Balicas, Peter Lunts and  Subir Sachdev for helpful discussions, and Philipp Strack for drawing our attention to several recent works.
The research was supported in part
by the Natural Sciences and Engineering Research Council of Canada (Canada) 
and the Early Research Award from the Ontario Ministry of Research and Innovation. 
Research at the Perimeter Institute is supported in part by the Government of Canada (Canada) through Industry Canada, and by the Province of Ontario through the Ministry of Research and Information.





\newpage

\onecolumngrid
\singlespacing
\appendix



\section{Computation of Feynman diagrams} \label{app: diagrams}

In this appendix we show the key steps for  computing the Feynman 
diagrams.

\subsection{One loop diagrams}


\subsubsection{Electron self energy}

The quantum correction to the electron self-energy
from the diagram in \fig{fig: CT-SEf} is
\begin{align}
\dl S^{(2,0)} &=   \mu^{3-d} ~ 2 \mc{B}^{(1L)}_{(2,0)} ~ g^2 \sum_{n=\pm} \sum_{s=1}^{N_c} \sum_{j=1}^{N_f} 
\int dk ~ 
\bar{\Psi}_{n,s,j}(k) 
~ \Upsilon_{(2,0)}^{(n)}(k) ~ \Psi_{n,s,j}(k), 
\label{eq: dS-20-defn}
\end{align}
where
\begin{align}
\mc{B}^{(1L)}_{(2,0)} =
\begin{cases}
\dfrac{N_c^2 - 1}{N_f N_c} & \mbox{for ~ SDW} \\[1em] 
\dfrac{1}{N_f N_c} & \mbox{for ~ CDW}
\end{cases}
\end{align}
and
\begin{align}
\Upsilon_{(2,0)}^{(n)}(k)
&= \int \frac{d^{d-1} \mbf{Q}}{(2 \pi)^{d-1}} \frac{d^2 
\vec q}{(2 \pi)^2}  ~ \gamma_{d-1} G_{\bar{n}}(k + q) 
\gamma_{d-1} ~ D(q).
\label{eq: upsilon-20-defn}
\end{align}
The bare Green's functions are given by
\begin{align}
& G_n(k) = -i ~ \frac{\mathbf{\Gamma} \cdot \mbf{K} +
\gamma_{d-1} \veps_n(\vec k)}{\abs{\mbf{K}}^2 +
\veps_n^2(\vec k)}, \\
& D(q) = \frac{1}{\abs{\mbf{Q}}^2 + q_x^2 + c^2 q_y^2}. 
\label{eq: propagators}
\end{align}
After the integrations over $\vec q$ and ${\bf Q}$,
\eq{eq: upsilon-20-defn} can be expressed in terms of
a Feynman parameter, 
\begin{align}
  \Upsilon_{(2,0)}^{(n)}(k)
&=  \frac{i}{ (4 \pi)^{(d+1)/2}} 
\Gamma\lt(\frac{3-d}{2} \rt) 
\int_0^1 dx ~ \sqrt{\frac{1-x}{c^2 + x(1 - (1 - v^2) c^2)}} \nn 
& \times \lt[ x(1-x) \lt\{ |\mbf{K}|^2 + \frac{c^2 ~
\veps_{\bar{n}}^2(\vec k)}{c^2 + x(1 - (1 - v^2) c^2) } \rt\} 
\rt]^{-\frac{3-d}{2}} \lt[ \mbf{K} \cdot \mbf{\Gamma} - \frac{c^2 ~ \veps_{\bar{n}}(\vec k) ~ \gamma_{d-1}}{c^2 + x(1 - (1 - v^2) c^2) }   \rt].
\end{align}
The UV divergent part in the $d \rightarrow 3$ limit is given by
\begin{align}
\Upsilon_{(2,0)}^{(n)}(k)
=  \frac{i}{8 \pi^2~  \eps} \lt[ h_{1}(v, c) ~
\mbf{K} \cdot  \mbf{\Gamma} - h_2(v, c) ~ 
\veps_{\bar{n}}(\vec k)\gamma_{d-1} \rt],
\end{align}
where
\begin{align}
h_1(v, c) &= \int_0^1 dx ~ \sqrt{\frac{1-x}{c^2 + x(1 - (1 - v^2) c^2)}}, 
\quad
h_2(v, c) = c^2 \int_0^1 dx ~ \sqrt{\frac{1-x}{\lt[c^2 + x(1 - (1 - v^2) c^2) \rt]^3}}.
\end{align}
This leads to the one-loop counter term
for the electron self-energy,
\begin{align}
 S^{(2,0)}_{CT} 
 &= - \frac{ \mc{B}^{(1L)}_{(2,0)}}{4\pi^2 \eps} ~g^2
\sum_{n=\pm}
\sum_{s=1}^{N_c} 
\sum_{j=1}^{N_f} 
\int dk ~ \nn
& \qquad \qquad \times
\bar{\Psi}_{n,s,j}(k) 
\lt[ i h_{1}(v, c) ~
\mbf{K} \cdot  \mbf{\Gamma} - i h_2(v, c) ~ 
\veps_{\bar{n}}(\vec k)\gamma_{d-1} \rt] 
\Psi_{n,s,j}(k).
\label{eq: CT-20}
\end{align}


\subsubsection{Boson self energy}

The boson self energy in \fig{fig: CT-SEb} is given by
\begin{align}
\dl S^{(0,2)} &= - \mu^{3-d} ~ \frac{g^2}{2}  
 \int dq 
~ \Upsilon_{(0,2)}(q) ~ \tr{{\Phi}(-q) {\Phi}(q)}, \label{eq: dS-02-defn}
\end{align}
where
\begin{align}
\Upsilon_{(0,2)}(q) &= \sum_{n = \pm}
\int dk ~
\mbox{Tr}\lt[ \gamma_{d-1} G_{n}(k + q) 
\gamma_{d-1} G_{\bar  n}(k) \rt].
\label{eq: upsilon-02-defn}
\end{align}
We first integrate over $\vec k$.
Because $\vec q$ can be absorbed into the internal momentum $\vec k$, 
$\Upsilon_{(0,2)}(q) $ is independent of $\vec q$.
Using the Feynman parameterization, we write the resulting
expression as
\begin{align}
 \Upsilon_{(0,2)}(q) &=  \frac{1}{2 \pi v} 
 \int_0^1 dx \int \frac{d^{d-1} \mbf{K}}{(2 \pi)^{d-1}}
 ~ \frac{[x(1-x)]^{-\half} ~~ \mbf{K}\cdot (\mbf{K} + 
\mbf{Q})}
{x \abs{\mbf{K} + \mbf{Q}}^2 + (1 - x) \abs{\mbf{K}}^2}.
\end{align}
The quadratically divergent term is the mass renormalization,
which is automatically tuned away at the critical point in the present scheme.
The remaining correction to the kinetic energy of the boson becomes
\begin{align}
\Upsilon_{(0,2)}(q)  =  - \frac{\abs{\mbf{Q}}^2}{16 \pi v \eps}
\end{align} 
up to finite terms.
Accordingly we add the following counter term,
\begin{align}
 S^{(0,2)}_{CT} &= - \frac{1}{4} ~ \frac{1}{8 \pi ~ \eps} ~ \frac{g^2}{v} \int dq 
 ~ \abs{\mbf{Q}}^2 ~ \tr{{\Phi}(-q) {\Phi}(q)}.
\label{eq: CT-02}
\end{align}


\subsubsection{Yukawa vertex correction} \label{app: 
yukawa-1L}

The diagram in Fig. \ref{fig: CT-g} gives rise to the vertex
correction in the quantum effective action,
\begin{align}
 \dl S^{(2,1)} &=  i \frac{g}{\sqrt{N_f}} ~ \mu^{\frac{3(3-d)}{2}} ~
 2\mc{B}^{(1L)}_{(2,1)}~ g^2  
\sum_{j=1}^{N_f}  \sum_{s,s' = 1}^{N_c}   \nn
& \qquad \times 
\int dk ~ dq ~ \lt[ \bar \Psi_{+,j,s}(k+q)~ \Upsilon^{(+, -)}_{(2,1)}(k,q) ~ \Phi_{s,s'}(q) ~\Psi_{-,j,s'}(k) - \mbox{h.c.} \rt],
\label{eq: dS-21-defn}
\end{align}
where
\begin{align}
 \Upsilon^{(n, \bar n)}_{(2,1)}(k,q) 
 &=  \int dp ~ \gamma_{d-1} 
G_{\bar{n}}(p+q+k) \gamma_{d-1} G_{n}(p + k) 
\gamma_{d-1} ~ D(p)
\label{eq: upsilon-21-defn}
\end{align}
and
\begin{align}
\mc{B}^{(1L)}_{(2,1)} =
\begin{cases}
 \dfrac{1}{N_c N_f} & \mbox{for ~ SDW} \\[1.0em] 
-\dfrac{1}{N_c N_f} & \mbox{for ~ CDW}
\end{cases}.
\label{B1L}
\end{align}
The minus sign in $\mc{B}^{(1L)}_{(2,1)} $ for the SDW case is due to the 
anti-commuting nature of the $SU(N_c)$ generators, 
$\sum_{a = 1}^{N_c^2 - 1} \tau^{a} \tau^{b} \tau^{a} = - \dfrac{2}{N_c} ~ \tau^{b}$.
The UV divergent part in the $\epsilon \rightarrow 0$ limit can be extracted by setting all external frequency and momenta to zero except $\mbf{Q}$,
\begin{align}
 \Upsilon^{(n,\bar n)}_{(2,1)}(\mbf{Q}) 
 &=  \gamma_{d-1} \int dp ~
 \frac{ | \mbf{P} |^2 - \veps_{\bar{n}}(\vec p) 
\veps_{n}(\vec p) }
{\Bigl[|\mbf{P}|^2 + p_x^2 + c^2 p_y^2 \Bigr] ~ 
\Bigl[ | \mbf{Q} + \mbf{P} |^2  +  \veps_{\bar n}^2(\vec p) \Bigr] ~ 
\Bigl[ |\mbf{P} |^2  + \veps_{n}^2(\vec p) \Bigr]}.
\label{eq: upsilon-21-1}
\end{align}
%
\eq{eq: upsilon-21-1} is evaluated following the computation in Ref. \cite{Sur2} to obtain
\begin{align}
 \Upsilon^{(n, \bar n)}_{(2,1)}(\mbf{K}) 
 =  \frac{\gamma_{d-1} ~ v}{16 \pi^3 ~ \eps}  ~ 
h_3(v, c) + \ordr{\eps^0} ,
\end{align}
where
\begin{align}
h_3(v,c) &= \int_0^{2\pi} d\theta \int_0^1 dx_1 \int_0^{1-x_1} dx_2 ~ \lt[ \frac{1}{\zeta(\theta, x_1, x_2, v, c)} - \frac{v^2 ~ \sin(2\theta)}{\zeta^2(\theta, x_1, x_2, v, c)}\rt], 
\end{align}
with 
\begin{align}
\zeta(\theta, x_1, x_2, v, c) &= 2 v^2 [x_1 \sin^2(\theta) + x_2 \cos^2(\theta)] \nn
& \qquad + (1 - x_1 - x_2)\lt[\sin^2\lt(\theta + \frac{\pi}{4}\rt) + c^2 v^2 \cos^2\lt(\theta + \frac{\pi}{4}\rt)\rt].
\end{align}
Note that the UV divergent part 
of $\Upsilon^{(n, \bar n)}_{(2,1)}$ is independent of $(n, \bar n)$.
From this, we identify the counter term for the Yukawa vertex,
\begin{align}
 S^{(2,1)}_{CT} &=  -i \frac{g}{\sqrt{N_f}}  ~
 \frac{\mc{B}^{(1L)}_{(2,1)}}{8\pi^3 \eps}~ g^2 v ~ h_3(v, c)
\sum_{j=1}^{N_f} \sum_{s,s' = 1}^{N_c}   \nn
& \qquad \times 
\int dk ~ dq ~ \lt[ \bar \Psi_{+,j,s}(k+q)~ \gamma_{d-1} ~ \Phi_{s,s'}(q) ~\Psi_{-,j,s'}(k) - \mbox{h.c.} \rt].
\label{eq: CT-21}
\end{align}


\subsubsection{$\phi^4$ vertex corrections}
\label{app: phi4}

There are two types of one-loop diagrams 
that can potentially contribute to the renormalization of the quartic vertex 
as is shown in \fig{fig: CT-phi4-u} and \fig{fig: CT-phi4-g1}. 
The diagram in \fig{fig: CT-phi4-g1} is UV finite at $d=3$   \cite{Sur2}, 
which implies that it  does not contain an $\eps^{-1}$ pole in $d=3 - \eps$. 
The second type of diagrams are produced by the boson vertices only. They lead to non-zero counter terms,
\begin{align}
 S^{(0,4)}_{CT} 
&= \frac{1}{8 \pi^2 c ~ \eps} 
 \int dq_1 dq_2 dq_3  \nn
 &  \times \Bigg\{
 \lt[ 
 \mc{B}^{(1L;1a)}_{(0,4)} u_1^2 
 + \mc{B}^{(1L;1b)}_{(0,4)} u_1 u_2
 + \mc{B}^{(1L;1c)}_{(0,4)} u_2^2
  \rt]
 \tr{ \Phi(-q_1+q_2) \Phi(q_1) } \tr{ \Phi(-q_3-q_2) \Phi(q_3)} \nn
 & \qquad +  \lt[ 
 \mc{B}^{(1L;2a)}_{(0,4)} u_1 u_2  
 + \mc{B}^{(1L;2b)}_{(0,4)} u_2^2
  \rt]
 \tr{ \Phi(-q_1+q_2) \Phi(q_1)  \Phi(-q_3-q_2) \Phi(q_3)} 
 \Bigg\}.
\label{eq: CT-04}
\end{align}
Here
\begin{align}
&& \mc{B}^{(1L;1a)}_{(0,4)} = N_c^2 + 7, ~~
\mc{B}^{(1L;1b)}_{(0,4)} = \frac{2(2N_c^2 - 3)}{N_c}, ~~ 
\mc{B}^{(1L;1c)}_{(0,4)} = \frac{3(N_c^2 + 3)}{N_c^2}, \nn
&& \mc{B}^{(1L;2a)}_{(0,4)} = 12, ~~
\mc{B}^{(1L;2b)}_{(0,4)} = \frac{2(N_c^2 - 9)}{N_c}
\end{align}
for the SDW case.
For the CDW case, one can set $u_2=0$ 
and $\mc{B}^{(1L;1a)}_{(0,4)} =9$.

\subsection{Two loops boson self-energy}
\label{app: BosonSE2L}

There are five diagrams, shown in \fig{fig: BosonSE2L}, that contribute to the boson self energy at the two-loop order.
We will first show that only \fig{fig: BosonSE2La}
contributes to the renormalization of $c$ to the leading order in $c$.
We will also outline the key steps for an explicit computation of \fig{fig: BosonSE2La}.

Let us denote the loop integrations in figures 
\ref{fig: BosonSE2Lb} - \ref{fig: BosonSE2Ld} 
for fixed electron flavor $n$  as $\Upsilon_{(0,2)}^{2L;b}(q)$, $\Upsilon_{(0,2)}^{2L;c}(q)$ and $\Upsilon_{(0,2)}^{2L;d}(q)$, respectively,  with $q$ being the external frequency-momentum. 
At $c = 0$ the loop integrations in \fig{fig: BosonSE2Lb} is given by
\begin{align}
\Upsilon_{(0,2)}^{2L;b}(q) = 
\int \frac{d^4 p}{(2\pi)^4} \frac{d^4 k}{(2\pi)^4} 
\frac{\tr{\gamma_{d-1} G_{\bar n}(k) \gamma_{d-1} G_{n}(k+p) \gamma_{d-1} G_{\bar n}(k) \gamma_{d-1} G_{n}(k + q)}}{|\mbf P|^2 + p_x^2}.
\end{align}
The integrand depends on $p_y$ only through 
\begin{align}
G_{n}(k+p) = \lt[ i (\mbf{K} + \mbf{P}) \cdot \mbf{\Gamma} + (\veps_{n}(k) + vp_x + np_y) \gamma_{d-1} \rt]^{-1}.
\end{align}
Changing coordinate as $p_y \mapsto p_y - n((\veps_{n}(k) + vp_x)$, $G_{n}(k+p)$ becomes independent of $\vec k$.
Since $\veps_n(\vec k)$ and $\veps_{\bar n}(\vec k)$ are linearly independent, 
we can change coordinates as $(k_x, k_y) \mapsto (\veps_n(\vec k), \veps_{\bar n}(\vec k))$ and shift $\veps_n(\vec k) \mapsto \veps_n(\vec k) - \veps_n(\vec q)$ to make $\Upsilon_{(0,2)}^{2L;b}(q)$ independent of $\vec q$.
This shows that \fig{fig: BosonSE2Lb} does not depend on $\vec q$ in the small $c$ limit. 
Note that such dependence may arise at order $g^4 c$ or higher, 
but these contributions are sub-dominant to that of \fig{fig: BosonSE2La}.

$\Upsilon_{(0,2)}^{2L;c}(q)$ and $\Upsilon_{(0,2)}^{2L;d}(q)$ closely resemble $\Upsilon_{(0,2)}(q)$. 
Because the one-loop counter terms are independent of the $x$ and $y$ components of momentum, 
it is straightforward to shift the internal integration variable 
to show that $\Gamma_{(0,2)}^{2L;c}(q)$ and $\Gamma_{(0,2)}^{2L;d}(q)$ 
are independent of $\vec q$, irrespective of the value of $c$.
Therefore, diagrams in Figs. \ref{fig: BosonSE2Lc} and \ref{fig: BosonSE2Ld} do not contribute to $Z_{5,1}$ and $Z_{6,1}$.
\ref{fig: BosonSE2Le} is also sub-leading because $\chi_i=0$ at the one-loop fixed point.


The quantum correction due to \fig{fig: BosonSE2La} is
\begin{align}
\dl S_{(0,2)}^{2L;a} = \mu^{2(3-d)}~ \frac{8 \mc{B}^{(2L)}_{(0,2)}}{4} ~ g^4 
\int dq ~ \Upsilon_{(0,2)}^{2L;a}(q) ~ \tr{\Phi(-q)\Phi(q)} 
\end{align}
where
\begin{align}
\Upsilon_{(0,2)}^{2L;a}(q) = -\int dp ~dk~
\frac{\tr{\gamma_{d-1} G_{+}(k+q) \gamma_{d-1} G_{-}(p+q) \gamma_{d-1} G_{+}(p) \gamma_{d-1} G_{-}(k)}}{|\mbf P - \mbf{K}|^2 + (p_x - k_x)^2 + c^2(p_y - k_y)^2}
\end{align}
and $\mc{B}^{(2L)}_{(0,2)} =
\mc{B}^{(1L)}_{(2,1)}$
as defined in \eq{B1L}.
In order to extract the leading order term that depends on $\vec k$ in the small $c$ limit, 
we set $c=0$ and $\mbf{Q} = 0$ in  $\Upsilon_{(0,2)}^{2L;a}(q)$ to write
\begin{align}
\Upsilon_{(0,2)}^{2L;a}(\vec q) = 
-\int dp ~dk~
\frac{\tr{\gamma_{d-1} G_{+}(\mbf{K}, \vec k + \vec q) \gamma_{d-1} G_{-}(\mbf{P}, \vec p + \vec q) \gamma_{d-1} G_{+}(p) \gamma_{d-1} G_{-}(k)}}{|\mbf P - \mbf{K}|^2 + (p_x - k_x)^2}.
\end{align}
Using $\tr{\gamma_\mu  \gamma_\nu} = 2 \dl_{\mu,\nu} \mc{I}_2$, we evaluate the trace in the numerator to obtain
\begin{align}
& \Upsilon_{(0,2)}^{2L;a}(\vec q) = 
-2 \int dp ~dk~ \nn
& \times \frac{ 
[ |\mbf{K}|^2 - \veps_{+}(\vec k + \vec q) \veps_{-}(\vec k) ] 
[ |\mbf{P}|^2 - \veps_{-}(\vec p + \vec q) \veps_{+}(\vec p) ]
-  \mbf{K}\cdot \mbf{P} 
[ \veps_{+}(\vec k + \vec q) + \veps_{-}(\vec k) ]
[\veps_{-}(\vec p + \vec q) + \veps_{+}(\vec p) ]
}
{
[|\mbf{K}|^2 + \veps_{+}^2(\vec k + \vec q)]
[|\mbf{K}|^2 + \veps_{-}^2(\vec k)]
[|\mbf{P}|^2 + \veps_{-}^2(\vec p + \vec q)]
[|\mbf{P}|^2 + \veps_{+}^2(\vec p)]
[|\mbf P - \mbf{K}|^2 + (p_x - k_x)^2]
}.
\end{align}
We change coordinates for both $\vec p$ and $\vec k$ as $(k_x, k_y) \mapsto (k_+, k_-)$ with $k_\pm = \veps_\pm(\vec k)$, 
and shift $k_+ \mapsto k_+ - \veps_+(\vec q)$ and $p_- \mapsto p_- - \veps_-(\vec q)$ to rewrite the expression as
\begin{align}
& \Upsilon_{(0,2)}^{2L;a}(\vec q) = 
-\frac{1}{2 v^2} \int dp' ~dk'~ \nn
& \times \frac{ 
[ |\mbf{K}|^2 - k_+ k_- ] 
[ |\mbf{P}|^2 - p_+ p_- ]
- \mbf{K}\cdot \mbf{P} 
[ k_+ + k_- ] [ p_+ + p_- ]
}
{
[|\mbf{K}|^2 + k_{+}^2]
[|\mbf{K}|^2 + k_{-}^2]
[|\mbf{P}|^2 + p_{-}^2]
[|\mbf{P}|^2 + p_{+}^2]
[|\mbf P - \mbf{K}|^2 + \frac{1}{4v^2}(p_+ + p_- - k_+ - k_- + 2q_y )^2]
},
\end{align}
where $dk' \equiv \dfrac{d^{d-1}\mbf{K} dk_+ dk_-}{(2\pi)^{d+1}}$. 
It is noted that $\Upsilon_{(0,2)}^{2L;a}(\vec q)$ has become \textit{independent} of $q_x$
in the small $c$ limit.
This implies that $Z_{5,1}$ is at most order of $g^4 c$ which is negligible.
From now on, we will focus on $Z_{6,1}$.

We integrate over $\mbf P$ and $\mbf{K}$ after introducing Feynman parameters, $x, y$ and $u, w$.
Employing a Schwinger parameter, $\alpha$, we have
\begin{align}
& \Upsilon_{(0,2)}^{2L;a}(\vec q) = 
- \frac{1}{2\pi^2 (4\pi)^{d+1} v^2} 
\int_0^1 dx ~ du \int_{0}^{1-x} dy \int_0^{1-u} dw ~ 
\frac{(1-u-w)^{(3-d)/2}}{A^{(d-1)/2}} ~ \int_0^{\infty}d \alpha ~ e^{-\alpha M^2} \nn
& \quad \times  \int_{-\infty}^{\infty} dp_+ dp_- dk_+ dk_- 
\Bigl[ \Bigl\{ \frac{(d-1)^2}{4 A} + \frac{d^2 - 1}{4 A^2} ~(1-u-w)(1-x-y)^2  \Bigr\} \alpha^{3-d}  \Bigr. \nn
&\qquad - \frac{(d-1)(1-u-w)}{2A} \Bigl\{ p_+ p_- + \lt((1-x-y)^2 + \frac{A}{1-u-w} \rt) k_+ k_- \Bigr. \nn 
& \qquad  + \Bigl. \Bigl. (1-x-y)(k_+ + k_-)(p_+ + p_-)  \Bigr \} \alpha^{4-d}  
+ (1-u-w) ~ p_+ p_- k_+ k_- ~ \alpha^{5 - d} \Bigr], 
\end{align}
where 
\begin{align}
& A \equiv A(x,y,u,w) = (u+w) + (x+y)(1-x-y)(1-u-w), \quad \mbox{and} \nn
& M^2 \equiv M^2(k_\pm, p_\pm; x,y,u,w; v, q_y) = u k_+^2 + w k_-^2 + x(1-u-w)p_+^2 + y(1-u-w)p_-^2 \nn
&\qquad \qquad + \frac{(1-u-w)(1-x-y)}{4v^2} (p_+ + p_- - k_+ - k_- + 2q_y )^2.
\end{align}
At this stage, we subtract the mass renormalization from $\Upsilon_{(0,2)}^{2L;a}(\vec q)$ and proceed with the computation of $\Dl \Upsilon_{(0,2)}^{2L;a}(\vec q) = \Upsilon_{(0,2)}^{2L;a}(\vec q) - \Upsilon_{(0,2)}^{2L;a}(0)$.
After integrating over $p_\pm$, $k_\pm$ and $\alpha$, 
we extract the pole in $\epsilon$ as
\begin{align}
\Dl \Upsilon_{(0,2)}^{2L;a}(\vec q) = 
\frac{q_y^2~ h_6(v)}{\eps ~ v^2}.
\end{align}
Here the function $h_6(v)$ is defined as
\begin{align}
& h_6(v) = \frac{2}{(4\pi)^4} 
\int_0^1 dx ~ du \int_{0}^{1-x} dy \int_0^{1-u} dw ~ \frac{1}{A} 
\Bigl[ \Bigl\{ 1 + \frac{2}{A} (1-u-w)(1-x-y)^2 \Bigr\} ~ \frac{J_1}{A} \Bigr. \nn
& \quad - \Bigl. \Bigl\{ 1 + \frac{1}{A} (1-u-w)(1-x-y)^2 \Bigr\} ~ J_2 
- \frac{1-u-w}{A} \Bigl\{ J_3 + (1-x-y) J_4 \Bigr\} 
+ (1-u-w) J_5 \Bigr],
\label{eq:h6}
\end{align}
where
\begin{align}
& J_1 \equiv J_1(\eta_i) = 
\frac{\eta_5}{\sqrt{\eta_1 \eta_2 \eta_3 \eta_4}}, \nn
& J_2 \equiv J_2(\eta_i, f_i) = 
\frac{f_3}{\sqrt{\eta_1 \eta_2 \eta_3 \eta_4}} \Bigl[ \frac{\eta_5}{2\eta_4} - f_4(1+ f_4)\Bigr], \nn
& J_3 \equiv J_3(\eta_i, f_i) = 
\frac{f_1}{\sqrt{\eta_1 \eta_2 \eta_3 \eta_4}} 
\Bigl[ \frac{\eta_5}{2}\Bigl\{ f_2(1-f_2) \Bigl( \frac{1}{\eta_3} + \frac{(1+f_3)^2}{\eta_4} \Bigr) - \frac{1}{\eta_2} \Bigr\} \Bigr. \nn
&  ~ \quad - \Bigl. f_2(1- f_2)(1+f_3)^2 (1+f_4)^2 \Bigr], \nn
& J_4 \equiv J_4(\eta_i, f_i) = 
\frac{f_1 + f_2(1-f_1)}{\sqrt{\eta_1 \eta_2 \eta_3 \eta_4}} 
\Bigl[ \frac{\eta_5}{2} \Bigl( \frac{1}{\eta_3} + \frac{(1+f_3)^2}{\eta_4} \Bigr) 
- (1+f_3)(1+f_4)(f_3 + f_4(1 + f_3)) \Bigr], \nn
& J_5 \equiv J_5(\eta_i, f_i) = 
\frac{f_1}{2\sqrt{\eta_1 \eta_2 \eta_3 \eta_4}} 
\lt[ \frac{\eta_5}{2 \eta_4} 
\lt( \frac{3 f_2 f_3 (1-f_2)(1+f_3)^2}{\eta_4} + \frac{f_2(1-f_2)(2 + 3 f_3)}{\eta_3} - \frac{f_3}{\eta_2} \rt) \rt. \nn
& ~ \quad - \lt. (1+f_4)\lt( \frac{3 f_2 f_3 (1-f_2)(1+f_3)^2(1+ 2f_4)}{\eta_4} + \frac{f_2 f_4 (1-f_2)(2 + 3 f_3)}{\eta_3} - \frac{f_3 f_4}{\eta_2} \rt) \rt],
\end{align}
with 
\begin{flalign}
& \eta_1 \equiv \eta_1(a_1, F) = 
a_1 + F, & \nn
& \eta_2 \equiv \eta_2(a_i, F) =
\frac{a_2 F + a_1(a_2 + F)}{a_1 + F}, & \nn
& \eta_3 \equiv \eta_3(a_i, u, F) =
\frac{a_2 u F + a_1(a_2(u + F) + u F)}{a_2 F + a_1(a_2 + F)}, & \nn
& \eta_4 \equiv \eta_4(a_i, u, w, F) =
\frac{a_2 u w F + a_1(a_2(w(u + F) + u F) + u w F)}{a_2 u F + a_1(a_2(u + F) + u F)}, & \nn
& \eta_5 \equiv \eta_5(a_i, u, w, F) =
\frac{a_1 a_2 u w F}{a_2 u w F + a_1(a_2(w(u + F) + u F) + u w F)}, &
\end{flalign}
\begin{flalign}
& f_1 \equiv f_1(a_1, F) = 
\frac{F}{a_1 + F}, & \nn
& f_2 \equiv f_2(a_i, F) = 
\frac{a_1 F}{a_2 F + a_1(a_2 + F)}, & \nn
& f_3 \equiv f_3(a_i, u, F) = 
- \frac{a_1 a_2 F}{a_2 u F + a_1(a_2(u + F) + u F)}, & \nn
& f_4 \equiv f_4(a_i, u, w, F) = 
- \frac{a_1 a_2 u F}{a_2 u w F + a_1(a_2(u F + w(u + F)) + u w F)}. &
\end{flalign}
The functions $a_i$ and $F$ are defined as,
\begin{flalign}
& a_1 \equiv a_1(x,u,w) = x(1-u-w), 
\qquad \qquad
 a_2 \equiv a_2(y,u,w) = y(1-u-w), & \nn
& F \equiv F(x,y, u,w; v) = \frac{(1-x-y)(1-u-w)}{4 v^2}. &
\end{flalign}

Therefore, the counter term to $\vec q$ dependent part of the bosonic kinetic energy is given by
\begin{align}
S^{(2L)}_{(0,2);CT} &= - \dl S_{(0,2)}^{2L;a} \nn 
&= - \frac{\mc{B}^{(2L)}_{(0,2)}}{4 \eps} ~\frac{8 g^4 h_6(v)}{v^2} 
\int dq ~ q_y^2 ~ \tr{\Phi(-q)\Phi(q)},
\label{eq: BosonSE2L}
\end{align}
which gives
\begin{align}
& Z_{5,1} = 0, \\
& Z_{6,1} = - 8 \mc{B}^{(2L)}_{(0,2)} ~ \frac{g^4 ~ h_6(v)}{v^2 c^2}
\end{align}
to the leading order in $c$.

\section{Analysis of RG flow at the CDW critical point}

In this appendix, we provide an analysis of the RG flow 
predicted by the beta functions in Eqs. 
(\ref{eq: CDW1LModBetG}) - (\ref{eq: CDW1LModBetChi}).
Let us first analyze the flow in $d=3$ (\fig{fig: flow-CDW-3d}).
According to \eq{eq: CDW1LModBetC}, $c(\ell)$ always decreases with increasing length scale.
If the initial value of $c$ is sufficiently small such that $c_0 \ll \pi/(\wtil{N}_f + 2)$, 
the inequality will be always satisfied at lower energies.
In this case, one can ignore the last term in \eq{eq: CDW1LModBetG} to obtain a logarithmically decreasing Yukawa coupling,
\begin{align}
g^2(\ell) = \frac{g^2_0}{1 + \alpha_g(\wtil{N}_f)~ g_0^2 ~ \ell},
\label{eq: soln-g}
\end{align}
where $g_0 \equiv g(\ell = 0)$ and
$ \alpha_g(\wtil{N}_f) = \dfrac{1}{4\pi \wtil{N}_f}\lt(1 + \dfrac{4}{\wtil{N}_f + 2} \rt)$.

The RG flow of $c$ is relatively more complicated due to the important role of the two-loop correction.
When $g^2 \ll \dfrac{c^3}{\pi^2 \wtil{N}_f \alpha_c(\wtil N_f)}$, 
where $ \alpha_c(\wtil{N}_f) = \dfrac{32}{\wtil{N}_f^3}~ h_6(\wtil{N}_f/2)$,
the second term on the right hand side of \eq{eq: CDW1LModBetC} is negligible.
In this case, the beta function for $c$ gives
\begin{align}
c(\ell) = c_0 \lt[ 1 + \dfrac{c_0}{\pi^2 \wtil{N}_f ~ \alpha_g(\wtil{N}_f)} ~ \ln\lt( 1 + \alpha_g(\wtil{N}_f) ~ g^2_0 ~\ell\rt)\rt]^{-1},
\label{eq: soln-c-1}
\end{align}
where we have utilized the expression for $g(\ell)$ in \eq{eq: soln-g}.
Since $g^2(\ell) \sim \ell^{-1}$ and $c(\ell) \sim (\ln{\ell})^{-1}$ in the $\ell \rtarw \infty$ limit,
the second term on the right hand side of \eq{eq: CDW1LModBetC} becomes even smaller compared to the first term as $\ell$ increases,
which justifies \eq{eq: soln-c-1} at all $l > 0$.
The quartic coupling flows to zero as $\chi_\sub{1}(\ell) \sim \ell^{-1}$.
Therefore, all the parameters, except for $v$, flow to zero in the low energy limit.
Although $c$ flows to zero, the critical point remains perturbatively controlled because $g$ flows to zero much faster, 
such that $g^2/c \ll 1$.
This is a stable quasilocal marginal Fermi liquid (MFL)  \cite{Varma}.

In contrast, the first term on the right hand side of \eq{eq: CDW1LModBetC} is negligible if 
$g^2 \gg \dfrac{c^3}{\pi^2 \wtil{N}_f \alpha_c(\wtil N_f)}$, in which case we obtain
\begin{align}
c(\ell) = \sqrt{
\dfrac{c_0^2 -  (2 \alpha_c(\wtil N_f) ~ g_0^2 - \alpha_g(\wtil N_f) ~  c_0^2)~g_0^2 ~\ell }
{1 + \alpha_g(\wtil N_f) ~ g_0^2~\ell}
}.
\label{eq: soln-c-2}
\end{align}
We note that the coefficient of $\ell$ in the numerator in \eq{eq: soln-c-2} depends on the initial values of $g$ and $c$.
If $g_0^2 > \dfrac{ \alpha_g(\wtil{N}_f)~ c_0^2}{2\alpha_c(\wtil{N}_f)}$, 
which automatically implies $g^2 \gg \dfrac{c^3}{\pi^2 \wtil{N}_f \alpha_c(\wtil N_f)}$ for small $c$,
the boson velocity becomes zero at a \textit{finite} RG time
\begin{align}
\ell_0 = \dfrac{c_0^2 / g_0^2}{2 \alpha_c(\wtil N_f) ~ g_0^2 - \alpha_g(\wtil N_f) ~ c_0^2}.
\end{align}
This is different from the first case where $c$ vanishes only asymptotically 
while the ratio $g^2/c$ remains small.
In the current case, the ratio $g^2/c$ blows up, resulting in a loss of control over the perturbative expansion. 
For example, as $\ell \rtarw \ell_0$, $\chi_\sub{1}$ diverges as $(\ell_0 - \ell)^{-a(c_0, g_0, \wtil N_f)}$ with $a(c_0, g_0, \wtil N_f) = \mbox{min}\lt\{1, \dfrac{\alpha_c(\wtil N_f)~g_0^4~ \ell_0}{c_0^2 (1 + \alpha_g(\wtil N_f) g_0^2 \ell_0)} \rt\}$, which results in the theory becoming non-perturbative.

Finally, let us consider the case where
$\dfrac{c_0^3}{\pi^2 \wtil{N}_f \alpha_c(\wtil N_f)} \ll g_0^2 < \dfrac{ \alpha_g(\wtil{N}_f)~ c_0^2}{2\alpha_c(\wtil{N}_f)}$.
In this case, $c$ initially approaches a nonzero constant dictated by \eq{eq: soln-c-2}.
However, the system eventually enters into the regime with $g^2 \ll \dfrac{c^3}{\pi^2 \wtil{N}_f \alpha_c(\wtil N_f)}$ at sufficiently large length scale.
This is because $g$ decreases much faster than $c$ in this regime.
Therefore the system again flows to the quasilocal marginal Fermi liquid.

Having understood the fate of the critical point in three dimensions, we consider the case below three dimensions (\fig{fig: flow-CDW-3-eps}).
For $\eps > 0$, $g^2 \sim \eps$ as long as $c$ is initially small.
On the other hand, $c$ flows to zero in a finite RG time. Consequently, the system becomes strongly coupled in the low energy limit.


\begin{thebibliography}{99}
%
\bibitem{Stewart} G. R. Stewart, Rev. Mod. Phys. {\bf 73}, 797 (2001).
\bibitem{Lohneysen} H. v. L\"{o}hneysen, A. Rosch, M. Vojta, and P. W\"{o}lfle, Rev. Mod. Phys. {\bf 79}, 1015 (2007).
\bibitem{Gegenwart} P. Gegenwart, Q. Si, and F. Steglich, Nature Physics {\bf 4}, 186 - 197 (2008).
\bibitem{Armitage} N. P. Armitage, P. Fournier, and R. L. Greene, Rev. Mod. Phys. {\bf 82}, 2421 (2010).
%
\bibitem{Helm} T. Helm, M. V. Kartsovnik, I. Sheikin, M. Bartkowiak, F. Wolff-Fabris, N. Bittner,W. Biberacher,M. Lambacher, A. Erb,
J.Wosnitza, and R. Gross, Phys. Rev. Lett. {\bf 105}, 247002 (2010).
\bibitem{Hashimoto}K. Hashimoto, K. Cho, T. Shibauchi, S. Kasahara, Y. Mizukami, R. Katsumata, Y. Tsuruhara, T. Terashima, H. Ikeda, M. A. Tanatar, H. Kitano, N. Salovich, R. W. Giannetta, P. Walmsley, A. Carrington, R. Prozorov, and Y. Matsuda, Science {\bf 336}, 1554
(2012).
\bibitem{Park} T. Park, F. Ronning, H. Yuan, M. Salamon, R. Movshovich, J. Sarrao, and J. Thompson, Nature (London) {\bf 440}, 65 (2006).
%
\bibitem{Moriya-1} T. Moriya and J. Kawabata, J. Phys. Soc. Jpn. {\bf 34}, 639 (1973).
\bibitem{Moriya-2} T. Moriya and J. Kawabata, J. Phys. Soc. Jpn. {\bf 35} , 669 (1973).
\bibitem{Hertz} J. A. Hertz, Phys. Rev. B  {\bf 14}, 1165 (1976).
\bibitem{Millis} A. J. Millis, Phys. Rev. B 48, 7183 (1993).
\bibitem{Holstein} T. Holstein, R. E. Norton, and P. Pincus, Phys. Rev. B {\bf 8}, 2649 (1973). 
\bibitem{Reizer} M. Y. Reizer, Phys. Rev. B {\bf 40}, 11571 (1989).
\bibitem{PLee1989} P. A. Lee, Phys. Rev. Lett. {\bf 63}, 680 (1989).
\bibitem{PLee1992} P. A. Lee and N. Nagaosa, Phys. Rev. B {\bf 46}, 5621 (1992).
\bibitem{Althshuler} B. L. Altshuler, L. B. Ioffe, and A. J. Millis, Phys. Rev. B {\bf 50}, 14048 (1994).
\bibitem{Polchinski}  J. Polchinski, Nucl. Phys. B {\bf 422}, 617 (1994).
\bibitem{Kim}  Y. B. Kim, A. Furusaki, X.-G. Wen, and P. A. Lee, Phys. Rev. B {\bf 50}, 17917 (1994).
\bibitem{Halperin} B. I. Halperin, P. A. Lee, and N. Read, Phys. Rev. B {\bf 47}, 7312 (1993).
\bibitem{Chubukov} A. V. Chubukov, Europhys. Lett. {\bf 44}, 655 (1998).
\bibitem{Abanov1} Ar. Abanov and A. V. Chubukov, Phys. Rev. Lett. {\bf 84}, 5608  (2000).
\bibitem{Abanov2} Ar. Abanov, A. V. Chubukov, and J. Schmalian, Adv. in Phys. {\bf 52}, 119 - 218 (2003).
\bibitem{Sur1} S. Sur \& S.-S. Lee, Phys. Rev. B {\bf 90}, 045121 (2014).
\bibitem{SSL1} S.-S. Lee, Phys. Rev. B {\bf 80}, 165102 (2009).
\bibitem{Metlitski1} M. A. Metlitski and S. Sachdev, Phys. Rev. B {\bf 82}, 075127 (2010).
\bibitem{Metlitski2} M. A. Metlitski and S. Sachdev, Phys. Rev. B {\bf 82}, 075128 (2010).
\bibitem{Holder15} T. Holder and W. Metzner, Phys. Rev. B {\bf 92}, 041112(R) (2015)
\bibitem{Chakravarty} S. Chakravarty, R. E. Norton, and O. F. Sylju\aa{}sen, Phys. Rev. Lett., {\bf 74}, 1423 (1995).
\bibitem{Fitzpatrick1} A. L. Fitzpatrick, S. Kachru, J. Kaplan, and S. Raghu, Phys. Rev. B {\bf 88}, 125116 (2013).
\bibitem{Fitzpatrick2} A. L. Fitzpatrick, S. Kachru, J. Kaplan, and S. Raghu, Phys. Rev. B {\bf 89}, 165114 (2014).
\bibitem{SLEE08} S.-S. Lee, Phys. Rev. B {\bf 78}, 085129 (2008).
\bibitem{Mandal} I. Mandal and S.-S. Lee, Phys. Rev. B {\bf 92} 035141,  (2015).
\bibitem{NayakNFL} C. Nayak and F. Wilczek, Nucl. Phys. B {\bf 417}, 359 (1994);
Nucl. Phys. B {\bf 430}, 534 (1994).
\bibitem{Mross} D. F. Mross, J. McGreevy, H. Liu, and T. Senthil, Phys. Rev. B {\bf 82}, 045121 (2010).
\bibitem{Senthil}  T. Senthil and R. Shankar, Phys. Rev. Lett. {\bf 102}, 046406 (2009).
\bibitem{Dalidovich} D. Dalidovich \& S.-S. Lee, Phys. Rev. B {\bf 88}, 245106 (2013).
\bibitem{Sur2} S. Sur \& S.-S. Lee, Phys. Rev. B {\bf 91}, 125136 (2015).
\bibitem{Patel} A. A. Patel, P. Strack, and S. Sachdev, Phys. Rev. B {\bf 92}, 165105 (2015).
\bibitem{Maier16} S. A. Maier and P. Strack, Phys. Rev. B {\bf 93}, 165114 (2016)
\bibitem{Taillefer1} N. D.-Leyraud and L.  Taillefer, Physica C {\bf 481}, 161 (2012).
\bibitem{Chu2009} J.-H. Chu, J. G.  Analytis, C. Kucharczyk, and I. R. Fisher, Phys. Rev. B {\bf 79}, 014506 (2009).
\bibitem{Chuang} T.-M. Chuang, M. P. Allan, Jinho Lee, Yang Xie, Ni Ni, S. L. Bud’ko, G. S. Boebinger, P. C. Canfield, and J. C. Davis, Science {\bf 327}, 181 (2010).
\bibitem{Lawler} M. J. Lawler,	K. Fujita,	Jhinhwan Lee,	A. R. Schmidt,	Y. Kohsaka,	Chung Koo Kim,	H. Eisaki,	S. Uchida,	J. C. Davis,	J. P. Sethna, and Eun-Ah Kim, Nature {\bf 466}, 347 (2010).
\bibitem{Ando} Y. Ando, K. Segawa, S. Komiya, and A. N. Lavrov, Phys. Rev. Lett. {\bf 88}, 137005 (2002).
\bibitem{Fink} J. Fink, V. Soltwisch, J.  Geck, E. Schierle, E. Weschke, and B. B\"{u}chner, Phys. Rev. B {\bf 83}, 092503 (2011).
\bibitem{Hinkov} V. Hinkov, D. Haug, B. Fauqué, P. Bourges, Y. Sidis, A. Ivanov, C. Bernhard, C. T. Lin, B. Keimer, Science {\bf 319}, 597 (2008). 
\bibitem{Kivelson} S. A. Kivelson, E. Fradkin, and V. J. Emery, Nature {\bf 393}, 550 (1998).
\bibitem{Chubukov2015} A. V. Chubukov, Springer Series in Materials Science, Volume 211, 2015, pp 255-329.
\bibitem{Chu} J.-H. Chu, J. G. Analytis, K. De Greve, P. L. McMahon, Z. Islam, Y. Yamamoto, and I. R. Fisher, Science {\bf 329}, 824 (2010).
\bibitem{Kasahara2012} S. Kasahara,	H. J. Shi, K. Hashimoto, S. Tonegawa, Y. Mizukami, T. Shibauchi,	K. Sugimoto,	T. Fukuda, T. Terashima, A. H. Nevidomskyy, and Y. Matsuda, Nature {\bf 486}, 382  (2012).
\bibitem{Zhou} R. Zhou, Z. Li, J. Yang, D.L. Sun, C. T. Lin and G.-Q. Zheng, Nature Comm. {\bf 4}, 2265 (2013).
\bibitem{Lu} X. Lu, J. T. Park, R. Zhang, H. Luo, A. H. Nevidomskyy, Q. Si, P. Dai, Science {\bf 345}, 657(2014).
\bibitem{Thewalt2015} E. Thewalt, J. P. Hinton, I. Hayes, T. Helm, D. H. Lee, James G. Analytis, and J. Orenstein, 	arXiv:1507.03981.
\bibitem{Holder14} T. Holder and W. Metzner, Phys. Rev. B {\bf 90}, 161106(R) (2014).
\bibitem{Schafer16}  T. Sch\"{a}fer, A. A. Katanin, K. Held, and A. Toschi,	 arXiv:1605.06355
\bibitem{NayakDW} C. Nayak, Phys. Rev. B {\bf 62}, 4880 (2000).
\bibitem{Burkov2011} A. A. Burkov, M. D. Hook, and Leon Balents, Phys. Rev. B  {\bf 84}, 235126 (2011).
\bibitem{Bian2016} G. Bian, T.-R. Chang, R. Sankar, S.-Y. Xu, H. Zheng, T. Neupert, C.-K. Chiu, S.-M. Huang, G. Chang, I.  Belopolski, D. S. Sanchez, M.  Neupane, N. Alidoust, C. Liu, B.  Wang, C.-C. Lee, H.-T.  Jeng, C. Zhang, Z. Yuan, S. Jia, A. Bansil, F. Chou, H. Lin, and M. Z. Hasan, Nature Communications {\bf 7}, 10556 (2016).
\bibitem{Note1} According to \eq{eq: 1PI},  we have $J_{l,0,n}^{(\lambda)}(v,\eps) = 0$ for $\lambda = v, c, g$, 
and $J_{l,m,n<2}^{(\lambda)}(v,\eps) = 0$ for $\lambda = v, g$.
Because $v$ can be renormalized only in the presence of $g$, 
one also has $J_{0,m,n}^{(v)}(v,\eps)=0$.  
Furthermore,  the full $(d+1)$-dimensional rotational invariance 
in the bosonic sub-sector implies  $J_{0,m,n}^{(c)}(v,\eps) = 0$.
\bibitem{Nakajima} M. Nakajima, T. Lianga, S. Ishidaa, Y. Tomiokab, K. Kihoub, C. H. Leeb, A. Iyob, H. Eisakib, T. Kakeshitaa, T. Itob, and S. Uchida, PNAS {\bf 108}, 12238 (2011).
\bibitem{Dusza} A. Dusza, A. Lucarelli, F. Pfuner, J.-H. Chu, I. R. Fisher, and L. Degiorgi, Europhys. Lett. {\bf 93}, 37002 (2011).
\bibitem{Tanatar} M. A. Tanatar, E. C. Blomberg, A. Kreyssig, M. G. Kim, N. Ni, A. Thaler, S. L. Bud’ko, P. C. Canfield, A. I. Goldman, I. I. Mazin, and R. Prozorov, Phys. Rev. B {\bf 81}, 184508 (2010).
\bibitem{Valenzuela} B. Valenzuela, E. Bascones, and M. J. Calderón, Phys. Rev. Lett. {\bf 105}, 207202 (2010).
\bibitem{Rech} J. Rech, C. P\'{e}pin, and A. V. Chubukov, Phys. Rev. B {\bf 74}, 195126 (2006).
\bibitem{Varma} C. M. Varma, P. B. Littlewood, S. Schmitt-Rink, E. Abrahams, and A. E. Ruckenstein, Phys. Rev. Lett. {\bf 63}, 1996 (1989).
\bibitem{Castellani} C. Castellani, C. Di Castro, and M. Grilli, Z. Phys. B Con. Mat. {\bf 103}, 137 (1996).
\bibitem{Wang} Y. Wang and A. V. Chubukov, arXiv:1507.03583.
\bibitem{Berg_MonteCarlo} E. Berg, M. A. Metlitski and S. Sachdev, Science {\bf 338}, 1606 (2012).
\bibitem{Berg15} Y. Schattner, M. H. Gerlach, S. Trebst, E. Berg, arXiv:1512.07257 
\bibitem{Varma15} C. M. Varma,  Phys. Rev. Lett. {\bf 115}, 186405 (2015).
\end{thebibliography}
\end{document}